\documentclass[
               amsfonts, amssymb, amsmath,
	       reprint, showkeys,
	       nofootinbib,
	       prd,
	       floatfix, superscriptaddress, natbib,
	      ]{revtex4-2}

\usepackage[english]{babel}
\usepackage[utf8]{inputenc}
\usepackage{multirow}
\usepackage{float}
\usepackage[pdftex, pdftitle={Article}, pdfauthor={Author}]{hyperref} 
\usepackage{amsthm}
\usepackage{mathtools}
\usepackage{physics}
\usepackage{xcolor}
\usepackage{graphicx}
\usepackage{adjustbox}
\usepackage{placeins}
\usepackage[T1]{fontenc}
\usepackage{lipsum}
\usepackage{csquotes}
\usepackage{sidecap}
\usepackage[normalem]{ulem}
\usepackage{enumitem}
\usepackage{relsize}

\setlist{nosep,leftmargin=\parindent}

\newcommand{\aap}{Astron. Astrophys.}
\newcommand{\apjl}{Astrophys. J. Lett.}
\newcommand{\apjs}{Astrophys. J. Suppl.}
\newcommand{\jcap}{J. Cosmol. Astropart. Phys.}
\newcommand{\mnras}{Mon. Not. Roy. Astron. Soc.}

\newcommand{\araa}{Annu. Rev. Astron. Astrophys.}
\newcommand{\physrep}{Phys. Rep.}

\newcommand{\pasj}{Publ. Astron. Soc. Jpn.}
\newcommand{\aj}{Astron. J.}
\newcommand{\pasp}{Publ. Astron. Soc. Pac.}
\newcommand{\baas}{Bull. Am. Astron. Soc.}

\newcommand\Mnu{\sum m_\nu}
\newcommand\Mpc{h^{-1}\text{Mpc}}
\newcommand\Mpcinv{h\text{Mpc}^{-1}}
\newcommand\kmax{k_\text{max}}

\begin{document} 

\title{Neutrino mass constraint from an Implicit Likelihood Analysis of BOSS voids}

\author{Leander~Thiele}
\email{lthiele@princeton.edu}
\affiliation{Department of Physics, Princeton University, Princeton, NJ 08544, USA}

\author{Elena~Massara}
\affiliation{Waterloo Centre for Astrophysics, University of Waterloo, 200 University Ave W, Waterloo, ON N2L 3G1, Canada}
\affiliation{Department of Physics and Astronomy, University of Waterloo, 200 University Ave W, Waterloo, ON N2L 3G1, Canada}

\author{Alice~Pisani}
\affiliation{The Cooper Union for the Advancement of Science and Art, 41 Cooper Square, New York, NY 10003, USA}
\affiliation{Center for Computational Astrophysics, Flatiron Institute, 162 5th Avenue, New York, NY 10010, USA}
\affiliation{Department of Astrophysical Sciences, Princeton University, Princeton, NJ 08544, USA}

\author{ChangHoon~Hahn}
\affiliation{Department of Astrophysical Sciences, Princeton University, Princeton, NJ 08544, USA}

\author{David~N.~Spergel}
\affiliation{Center for Computational Astrophysics, Flatiron Institute, 162 5th Avenue, New York, NY 10010, USA}

\author{Shirley~Ho}
\affiliation{Center for Computational Astrophysics, Flatiron Institute, 162 5th Avenue, New York, NY 10010, USA}

\author{Benjamin~Wandelt}
\affiliation{Institut d’Astrophysique de Paris (IAP), UMR 7095, CNRS, Sorbonne Universit\'e, Paris, France}
\affiliation{Center for Computational Astrophysics, Flatiron Institute, 162 5th Avenue, New York, NY 10010, USA}

\begin{abstract}
Cosmic voids identified in the spatial distribution of galaxies provide complementary information
to two-point statistics.
In particular, constraints on the neutrino mass sum, $\Mnu$, promise to benefit from
the inclusion of void statistics.
We perform inference on the CMASS NGC sample of SDSS-III/BOSS with the aim of constraining $\Mnu$.
We utilize the void size function, the void-galaxy cross power spectrum, and the galaxy auto power spectrum.
To extract constraints from these summary statistics we use a simulation-based approach,
specifically implicit likelihood inference.
We populate approximate gravity-only, particle neutrino cosmological simulations with an expressive
halo occupation distribution model.
With a conservative scale cut of $\kmax=0.15\,\Mpcinv$ and a \textit{Planck}-inspired $\Lambda$CDM prior,
we find upper bounds on $\Mnu$ of $0.43$ and $0.35\,\text{eV}$
from the galaxy auto power spectrum and the full data vector, respectively ($95\,\%$ credible interval).
We observe hints that the void statistics may be most effective at constraining $\Mnu$ from below.
We also substantiate the usual assumption that the void size function is Poisson distributed.
\end{abstract}

\maketitle

\section{Introduction}

The Universe's ability to provide glimpses into experimentally inaccessible conditions has a long history,
including the deduction of the laws of Gravity and the discovery of helium.
In the present day, cosmology offers a unique view on the properties of neutrinos
which are amongst the last unknowns in the standard model of particle physics.
First evidence for a non-zero neutrino mass sum, $\Mnu$, came from the solar neutrino problem~\cite{
  Bahcall1976,Wolfenstein1978,Mikheyev1985}.
Subsequently, oscillation experiments provided proof that neutrinos must have mass~\cite{
  Fukuda1998,Ahmad2002,Araki2005,Ahn2006,An2012}
and established the lower bounds of $0.06$ and $0.1\,\text{eV}$ in the normal and inverted hierarchy,
respectively.
The terrestrial experiment KATRIN currently sets an upper bound of $\geq 0.8\,\text{eV}$~\cite{Aker2022}.\footnote{
  The KATRIN bound is on $m_\beta^2 \equiv \sum_\nu |U^\text{PMNS}_{e\nu}|^2 m_\nu^2$,
  so it only equals a bound on $\Mnu$ for a special, experimentally excluded, choice of the PMNS matrix
  and the mass hierarchy. In general, the bound is weaker.}
However, the strongest upper bounds are already provided by cosmological data,
the primary CMB alone giving $0.38\,\text{eV}$~\cite{PlanckCollaboration2020}, for example.
It will be one of cosmology's primary goals in the coming decade to tighten this bound and eventually
detect neutrino mass.

One of the natural regimes to look at to constrain $\Mnu$ are extremely underdense
regions, cosmic voids~\cite{
  Icke1984,Pisani2019,Moresco2022,Schuster2023}.
As the cold dark matter (CDM) flows out of the voids and into filaments and clusters,
neutrinos are more smoothly distributed.
Thus, the neutrino/CDM ratio is higher in the voids and lower in the clusters.
These qualitative considerations have spawned considerable theoretical interest in the use of void properties
to constrain $\Mnu$.
This includes simulated data vector-level investigations~\cite{
  Massara2015,Banerjee2016,Kreisch2019,Schuster2019,Contarini2021,Verza2022}
as well as forecasts~\cite{
  Sahlen2019,Bayer2021a,Kreisch2022}.
The forecasts find promising error bars on $\Mnu$, albeit under simplifying assumptions.
Voids may be the first regime in which non-linear signatures of massive neutrinos
will be observed~\citep[also c.f.][]{Hotinli2023}.

Being large objects, voids had to wait for the era of relatively deep, large-volume surveys with 
approximately uniform selection function to be statistically usable.
While the original detections focused on individual objects~\cite{
  Gregory1978,Joeveer1978,Tully1978,Kirshner1981,deLapparent1986},
which already contain cosmological information~\cite{Sahlen2016},
we are now able to utilize catalogs of hundreds and thousands of voids~\cite{
  Hoyle2004,Pan2012,Sutter2012a,Sutter2014b,Nadathur2016,Mao2017a}
to perform precision cosmology with void shapes~\cite{
  Sutter2012b,Sutter2014a,Hamaus2016,Hamaus2017,Mao2017b,Hamaus2020,Nadathur2020,Aubert2022,Woodfinden2023}
and sizes~\cite{
  Contarini2022a,Contarini2022b}.

In this work, we use voids identified in the CMASS sample of the Sloan Digital Sky Survey~(SDSS)-III
Baryon Oscillation Spectroscopic Survey~(BOSS)~\cite{York2000,Eisenstein2011,Dawson2013}
to place constraints on $\Mnu$.
The void statistics we consider are the void size function (VSF) and the void-galaxy cross power spectrum.
We combine these with the usual galaxy auto power spectrum multipoles which by themselves already
place a tight upper bound on $\Mnu$ (through the suppression of matter power below
the neutrino free-streaming scale)~\citep[e.g.,][]{Ivanov2020b,Semenaite2023}.

Since voids can be considered anti-halos~\cite{Stopyra2021}, a popular model for the VSF
descends from Press-Schechter theory and the excursion set formalism~\cite{
  Press1974,Bond1991,Sheth1999},
with slight modifications~\cite{
  Sheth2004,Paranjape2012,Paranjape2013,Jennings2013,Pisani2015a,
  Ronconi2017,Ronconi2019,Verza2019,Contarini2019,Contarini2022c,Pelliciari2023}.
While the void-galaxy correlation function can be used for cosmological purposes without
explicit knowledge of the void profile through the Alcock-Paczynski test
and redshift space distortions~\cite{Alcock1979,Lavaux2012,Pisani2014},
the modeling of the profile itself has also been considered~\cite{
  Padilla2005,Paz2013,Hamaus2014a,Hamaus2014b,Massara2018,Kreisch2022}.

However, all analytic approaches to modeling void statistics are problematic for our purposes.
First, it is difficult to construct a consistent galaxy bias model across the different statistics
comprising the data vector.
Second, the calibration of analytic models typically did not utilize large simulations with varied neutrino mass.
Third, existing models typically apply only to an aggressively cleaned subset of the entire void catalog,
potentially leading to appreciable losses in constraining power.

Therefore, we choose to work in a simulation-based framework.
Our simulations are based on particle-neutrino, approximate gravity-only \texttt{FastPM}~\cite{Feng2016,Bayer2021b}
realizations, in which we place galaxies through an expressive
halo occupation distribution model (HOD)~\cite{Berlind2002,Cooray2002,Wechsler2018}.
We then post-process the galaxy catalogs to generate light cones incorporating survey realism.

The likelihood analysis with these simulations is a non-trivial problem.
A popular approach is to build emulators of the mean data vector and perform the analysis
under the assumption of a usually Gaussian likelihood where the covariance matrix is estimated from simulations.
However, this approach turns out to be challenging for our problem.
First, constructing an emulator in a 17-dimensional space (6 cosmology, 11 HOD) is quite difficult,
especially given a feasible number of simulations.
Second, the assumption of a Gaussian likelihood is wrong.
We demonstrate in Sec.~\ref{sec:poisson} that the VSF is very close to Poisson distributed
(as long as bins are chosen wide enough, as would be naively expected),
but modeling its covariance with the void-galaxy cross power spectrum and the galaxy auto power spectrum
is difficult.

For these reasons, we opt for an implicit-likelihood\footnote{likelihood-free, simulation-based.}
approach~\cite{Cranmer2020}.
This formalism uses neural networks to approximate functions that can be converted into posteriors.
In general structure, this work is therefore similar to the \texttt{SIMBIG} papers~\cite{Hahn2022,Hahn2023},
but it differs in almost all details (statistics, simulations, objective, HOD, code).
The resulting complementarity will therefore be useful to assess the state of implicit likelihood inference
in galaxy clustering cosmology.

The rest of this paper is structured as follows.
Sec.~\ref{sec:sims} describes our simulation pipeline.
Sec.~\ref{sec:inference} contains details on the data vector and the inference procedure.
Sec.~\ref{sec:results} collects our results and their interpretation.
We conclude in Sec.~\ref{sec:concl}.
The appendices contain additional material as well as information about data and code availability.

\section{Simulations}
\label{sec:sims}

\subsection{Cosmological prior}

Since our objective is $\Mnu$, we place a tight prior on $\Lambda$CDM.
For this, we use the posterior from the \textit{Planck}~\cite{PlanckCollaboration2020}
primary CMB analysis.\footnote{\texttt{plikHM-TTTEEE-lowl-lowE}}
Specifically, we use the chains run with fixed $\Mnu$ and measure the mean
and covariance matrix in the five ``CMB parameters'' $\omega_c$, $\omega_b$, $\log A_s$, $n_s$,
and $\theta_\text{MC}$.
In these parameters the posterior is close to Gaussian and we approximate it as such.
To ensure the robustness of our conclusions, we inflate the \textit{Planck} error bars
on cosmological parameters by a factor of two.
For $\Mnu$, we choose a flat prior between $0$ and $0.6\,\text{eV}$,
the upper boundary being motivated by preliminary tests in which we established a sensitivity
of the order $\sigma \sim 0.2\,\text{eV}$.
We assume three neutrino species with degenerate masses.

Of course, the primary CMB's information leads to some correlation between $\Mnu$ and
the CMB parameters. This correlation is not included in our prior.
However, given the sensitivity of the data used (compared to \textit{Planck}), these residual correlations
have a relatively small effect.
For example, projecting the $\Mnu$--$\omega_c$ correlation in the \textit{Planck} posterior
to our upper prior boundary of $0.6\,\text{eV}$, we obtain a shift $\Delta\omega_c/\sigma_\text{EFT} \sim 0.25$
where $\sigma_\text{EFT}$ is the error bar obtained from the EFTofLSS analysis of BOSS~\cite{Philcox2022}
(of which we only use a subset).
From these considerations, it also follows that our results do not depend strongly on the precise
choice of $\Lambda$CDM prior.

We draw from the cosmological prior using an open quasi-random sequence.
In contrast to popular sampling methods such as latin hypercube or Sobol, an open sequence
does not require initial knowledge of the total number of samples.
Our sequence is constructed by taking integer multiples of a vector whose elements are powers
of a generalized golden ratio.\footnote{
  \url{http://extremelearning.com.au/unreasonable-effectiveness-of-quasirandom-sequences}}
In hindsight it turns out that our results are insensitive to taking out random subsets from the
sequence (c.f. appendix~\ref{sec:budget}), indicating that a simple pseudo-random sampling would have been sufficient.
This is likely due to the aforementioned compactness of the prior compared to the data's sensitivity.

\subsection{Cosmological simulations}

\begin{figure}
\includegraphics[width=\linewidth]{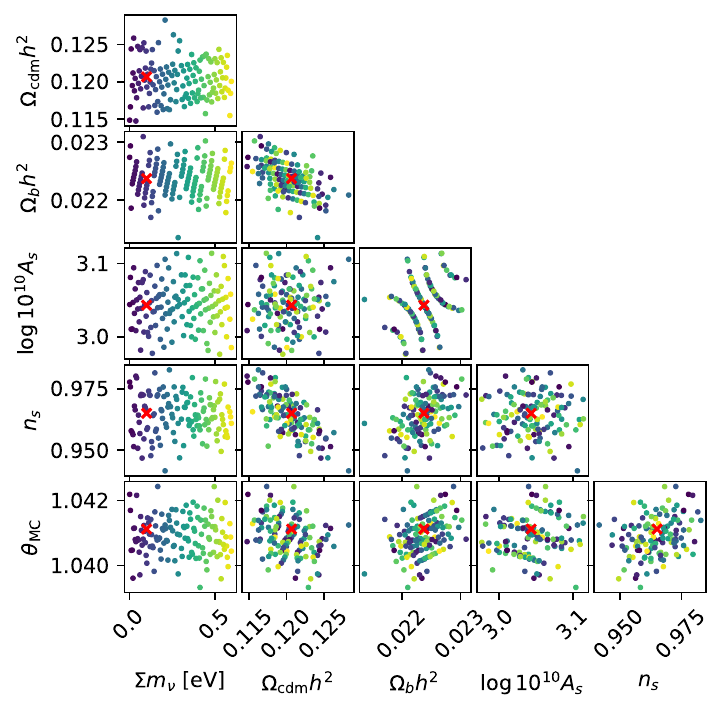}
\caption{
	 The cosmological parameter values sampled. We adopt a prior on $\Lambda$CDM that has the same
	 shape and twice the size (in $\sigma$) as the \textit{Planck} posterior.
	 We sample this prior with an open quasi-random sequence designed to yield low-variance
	 estimates of integrals.
	 The color scale correlates with $\Mnu$.
	 The red marker indicates the position of our fiducial simulations.
	}
\label{fig:sims}
\end{figure}

We run 127 simulations with varied cosmologies and 69 at a fiducial cosmology,
illustrated in Fig.~\ref{fig:sims}.\footnote{
  We attempted 130/71, but some runs failed. As discussed before, the fact that due to the failures
  we do not sample the quasi-random sequence strictly sequentially does not affect our results.}
We choose the fiducial cosmology close to the mean of the $\Lambda$CDM prior, with $\Mnu=0.1\,\text{eV}$.
The cosmo-varied simulations share the random seed. This is because the decision to adopt an
implicit-likelihood approach was only made after we encountered severe challenges with the conventional
approach, as described in the introduction.
However, as we shall see below, the simulations are large enough that sufficient quasi-independent
data vectors can be generated.

After generating a cosmological parameter vector in the CMB parameters we replace $\theta_\text{MC}$
by the Hubble constant using \texttt{CAMB}~\cite{Lewis:1999bs,Howlett:2012mh}.
We then produce power spectra at $z=99$ using \texttt{CLASS}~\cite{Blas2011}
and \texttt{REPS}~\cite{Zennaro2016,Zennaro2017}.

We run particle neutrino gravity-only simulations using the approximate \texttt{FastPM} solver.
We choose a box size of $2.5\,h^{-1}\text{Gpc}$ with $2800^3$ CDM particles,
leading to a minimum resolved halo mass of $\sim 1.3\times10^{12}\,h^{-1}M_\odot$
that is sufficient for the CMASS LRG sample.
With regard to the neutrino options in \texttt{FastPM} we follow Ref.~\cite{Bayer2021b}.\footnote{
  \texttt{every\_ncdm=4}, \texttt{n\_side=4}, \texttt{n\_shell=10}}
In particular, we follow their recommendation to increase the number of early time steps
for larger neutrino masses.
Specifically, at $\Mnu=0$ we take seven logarithmic steps (in scale factor) between
$z=99$ and $z=19$, followed by twelve linear steps until $z=0.68$. Afterwards, we take twenty steps until
$z=0.44$ during each of which we write a snapshot of CDM particles to disk.
At higher neutrino masses we insert up to twelve additional logarithmic steps before $z=79$.
The seemingly large number of twenty snapshots was established in our preliminary tests
during which we saw slight
differences between ten and twenty snapshots (in the summary statistics considered and
for a single \texttt{FastPM} run) and thus decided to err on the side of caution.

Each simulation takes about 90 minutes on 70 40-core nodes of the Tiger machine at Princeton.

\subsection{Galaxies}

We fine-tuned the method to populate CDM snapshots with galaxies through preliminary tests.
Specifically, we performed global optimization over a large and partly discrete space
of halo occupation distributions considering two objectives:
(1) the power spectrum multipoles of galaxies placed in Quijote simulations~\cite{Quijote_sims};
(2) the VSF of the CMASS data.
The first step was intended to identify degrees of freedom that are necessary to correct for potential
approximation errors in \texttt{FastPM},
while the second test was primarily meant to test our simulations' fidelity.
The optimization problems were solved with \texttt{optuna}~\cite{Akiba2019}.
In the following, we briefly describe the halo occupation distribution model (HOD)
resulting from these preliminary tests. More detail can be found in appendix~\ref{sec:hod}.

We identify halos in the CDM snapshots using \texttt{Rockstar}~\cite{Behroozi2012,Behroozi2013},
which in our preliminary tests performed better than the friends-of-friends
finder shipped with \texttt{FastPM}.

Then, galaxies are assigned stochastically to halos using an HOD.
Besides the usual five-dimensional model parameterized by
$M_\text{min}$, $\sigma_{\log M}$, $M_0$, $M_1$, $\alpha$~\cite{Zheng2005},
we introduce six additional degrees of freedom.

Although assembly bias~\cite{
  Zhu2006,Zentner2014,Pujol2014}
has been argued to be not necessary to describe the clustering of CMASS galaxies~\cite{
  Reid2014,Lin2016,Kobayashi2022},
we decide to be conservative by adding assembly bias parameterized by $P_1$ and $a_\text{bias}$.
Furthermore, we add velocity bias~\cite{Berlind2003,Yoshikawa2003,vandenBosch2005,Guo2015}
parameterized by $\eta_\text{cen}$ and $\eta_\text{sat}$.
Finally, we introduce redshift dependence to $M_\text{min}$ and $M_1$,
parameterized by $\mu(M_\text{min})$ and $\mu(M_1)$.
One advantage of having these slopes as free parameters is that we know them to be relatively close to
zero, enabling useful sanity checks on any posteriors.

The resulting 11-dimensional HOD parameterization is quite similar to, e.g., 
Refs.~\cite{Zhai2023,Hahn2022,Hahn2023}.

We populate the cosmo-varied simulations with galaxies according to HOD parameters
drawn from the priors given in appendix~\ref{sec:hod}.
For the simulations at the fiducial cosmology we only populate with a single HOD.
We choose the HOD parameters used for the fiducial mocks based on preliminary inference runs using the VSF only.
It turns out that these parameters are not very close to best-fit when considering the entire data vector;
generating new mocks closer to the best-fit point may increase the efficiency of the compression step
described below. However, the difference in HOD parameters cannot cause biases since the fiducial mocks
are not used in constructing the likelihood.

\subsection{Light cones}

We use the cuboid remapping code~\cite{Carlson2010}
to deform our simulated cubes to the CMASS NGC geometry. It turns out that there are two possible choices
of remapping and we use both (as part of the augmentation scheme discussed below).

When projecting galaxies onto the light cone, we extrapolate their positions from the snapshots
using the host halo velocities (using the stochastic galaxy velocities would weaken the correlation between centrals
and satellites). The resulting corrections are small thanks to the large number of available snapshots.

After mapping galaxies to the light cone, we apply all angular masks and approximately mimic fiber
collisions using the procedure described in Ref.~\cite{Hahn2022}.

In contrast to some other works, we downsample the galaxy field predicted by the HOD to the data's
density $n(z)$ (only, of course, if the simulation contains more galaxies at the given redshift).
This downsampling is performed iteratively in conjunction with the implementation of fiber collisions
so that both are self-consistent.
Our implementation performs any necessary downsampling regardless of host halo properties;
future work could take a prediction for stellar mass into account.

\section{Inference}
\label{sec:inference}

\subsection{Data vectors}

\begin{figure*}
\includegraphics[width=\textwidth]{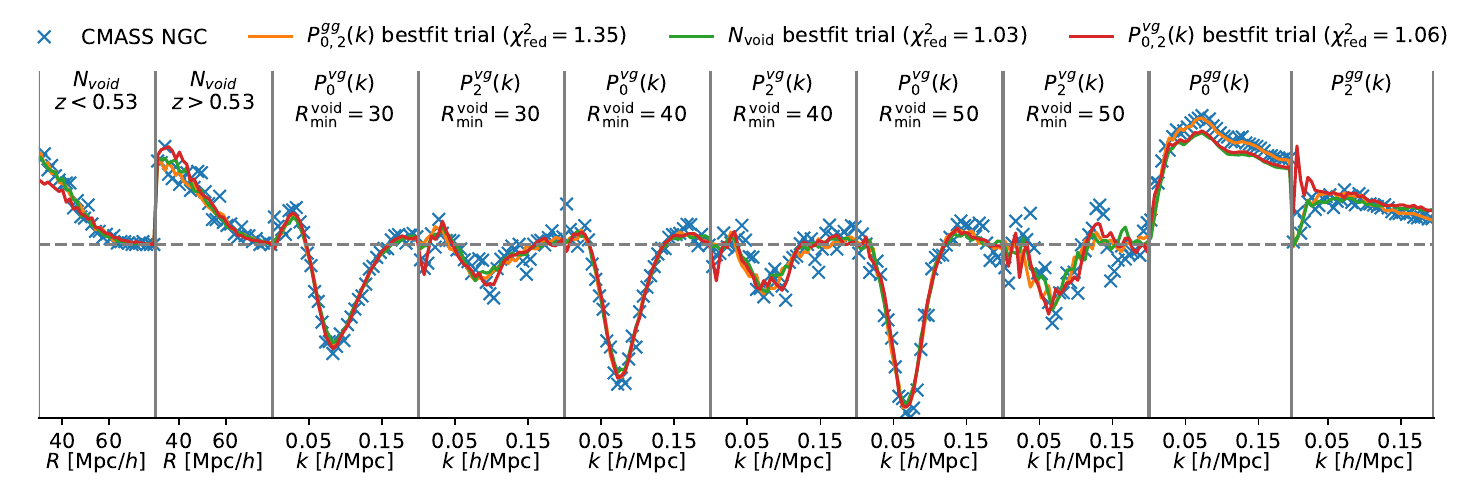}
\caption{
	 Illustration of the data vector considered.
	 The vertical axis has been scaled such that the different data vector components are
	 well visible and the power spectra are plotted as $k P(k)$.
	 The solid lines display best-fit mocks, considering separate parts of the data vector.
	 The $\chi^2$ was computed under the approximation of a Gaussian likelihood
	 and the mocks were averaged over eight augmentations
	 (no interpolation/emulation was performed).
	 The fact that, e.g., the model with best-fit VSF still reproduces the
	 other parts of the data vector reasonably well, as well as the reduced $\chi^2$ values
	 close to unity, indicate our mocks' high fidelity.
	 Note that we do not use $k<0.01$ and $k>0.15$ in our baseline analysis
	 and these scales were not included in identifying the best-fit models shown here.
	}
\label{fig:datavec}
\end{figure*}

We use the northern galactic cap (NGC) part of the CMASS sample. The southern part (SGC) is smaller and
we do not expect dramatic improvements from its inclusion. Since our focus is on better understanding the
impact of void statistics, rather than the tightest possible bounds on neutrino mass, we ignore the SGC
for simplicity.
Similarly, we do not include the LOWZ sample; its lower volume makes it less suitable for void science.

We cut galaxies into the redshift interval $0.42<z<0.7$ and map them to comoving space
using a fixed $\Omega_m=0.3439$.
Voids are identified using the \texttt{VIDE} code~\cite{Sutter2015}
which is based on \texttt{ZOBOV}~\cite{Neyrinck2008}
and works by Voronoi tessellating the galaxies and then applying a watershed algorithm to find
contiguous density minima.
We use the ``untrimmed'' catalog computed by \texttt{VIDE} as it does not require arbitrary assumptions.

While many different void finders exist~\citep[e.g.,][]{Colberg2008,vandeWeygaert2009,Cautun2018},
prior work suggests that shape-agnostic void finders such as \texttt{VIDE} yield voids with better constraining
power on $\Mnu$ than spherical finders~\cite{Kreisch2022}.
Future work could investigate the influence of void definition on signal-to-noise.

Galaxy auto power spectra $P^{gg}_\ell(k)$ and void-galaxy cross power spectra $P^{vg}_\ell(k)$
are computed using \texttt{nbodykit}~\cite{Hand2018,Hand2019}
and \texttt{pypower},\footnote{\url{https://pypower.readthedocs.io}}
reducing variance with FKP weights~\cite{Feldman1994}
and correcting for observational systematics using the provided weights
(except for fiber collisions, of course)~\cite{Reid2016}.
We only utilize the systematics weights when computing $P^{gg}$.
In the case of $P^{vg}$, we find no significant change in posteriors when using the galaxy weights,
consistent with Ref.~\cite{Hamaus2017}.
In the case of void identification, there is no guarantee that the obvious method to incorporate
the systematics weights would yield cleaner voids.
Given the relatively large void sizes considered in this work, we do not expect significant contamination
by unmodeled survey systematics, but suggest that this point may warrant future work.
Galaxy randoms are taken from the public catalogs.
Void randoms are constructed by taking a large catalog of voids from many different mocks
to choose angular positions and constructing a kernel density estimator in redshift matched to the specific
void catalog. This procedure ensures that the randoms are consistent with a given cut in void radius since
voids of different sizes have somewhat different angular distributions due to the survey mask.

We consider the VSF in 32 linearly spaced effective radius bins between $30$ and $80\,\Mpc$.
The minimum radius cut is well above the mean tracer separation and thus we expect contamination by Poisson
voids to be small~\cite{Cousinou2019,Pisani2015b}.
We split voids for the VSF into two redshift bins, separated at $z=0.53$.
This splits the CMASS sample approximately equally.

We perform analyses with $\kmax=0.15$ and $0.2$.\footnote{
  For brevity, we implicitly take all wavenumbers in units of $\Mpcinv$.}
We consider $\kmax=0.15$ the conservative baseline choice but $\kmax=0.2$ is still expected to be
reliably modeled by \texttt{FastPM} as well as the halo model~\citep[e.g.,][]{Kobayashi2022}.
We do not use power spectra on scales larger than $k_\text{min}=0.01$~\cite{Ivanov2020a}.
Since our theoretical model is simulation based, we do not deconvolve the survey window function.
This means that there is a small level of contamination by Fourier modes outside the $k$-range considered,
but we assess this effect to be negligible.

In our baseline analysis we only use the monopole $P^{gg}_0$ of the galaxy auto power spectrum multipoles.
This choice was made based on the limited information content of the quadrupole (from EFTofLSS posteriors)
and with the aim of simplicity. We discuss the effect of including the quadrupole below.
For $P^{vg}$ we use both the monopole and the quadrupole.

It is worth noting that we opt to use reciprocal space void-galaxy cross power spectra $P^{vg}$
instead of the more popular configuration space correlation function.
The correlation function has the primary advantage that one can rescale its argument on a void-by-void
basis by the respective radius and thus sharpen the resulting void profile
(this is also possible in reciprocal space but computationally expensive,
future work could explore this point).
We believe, however, that the mixing of Fourier modes in the correlation function
could lead to problems with approximate solvers like \texttt{FastPM} whose domain of validity
is better localized in reciprocal space.
In order to optimize signal-to-noise, we consider $P^{vg}$ computed with three different choices of minimum
void radius, $30$, $40$, and $50\,\Mpc$.

An illustration of the data vector is given in Fig.~\ref{fig:datavec}.

\subsection{Implicit likelihood inference}

As discussed in the introduction, the standard emulator-based approach is difficult in 17 dimensions.
The main reason is that the training objective for an emulator does not directly map
to the ultimate goal of accurate posteriors, implying that the optimum needs to be very sharp (which requires many simulations).

Combined with the unknown likelihood function, we believe implicit likelihood inference (ILI) to be the appropriate tool for our task.

We opt for neural ratio estimation~(NRE)~\cite{Cranmer2015,Hermans2019,Durkan2020,Delaunoy2022,Miller2022}
which recasts inference as a classification problem.
The choice of an amortized instead of a sequential method was made based on the hierarchical structure
of our simulations; we then opt for NRE because of its simplicity.
In its original and simplest form, NRE works with pairs of parameter vectors $\theta, \theta'$
drawn from the prior $p(\theta)$. We then consider a data vector $x$ drawn from the likelihood $p(x|\theta)$,
where the simulation process described above approximates this draw.
A neural network $f$ maps the pairs $(x,\theta)$, $(x,\theta')$ to scalars $y$, $y'$.
If we now choose a classification loss function $L(y,y')$, e.g., binary cross entropy
\begin{equation}
L(y,y') = -\log(y) - \log(1-y')\,,
\end{equation}
it is easy to show that the functional optimization problem
\begin{equation}
f^* = \underset{f}{\text{argmin}} \int d\theta d\theta' dx\,p(\theta)p(\theta')p(x|\theta)\,L
\label{eq:argmin}
\end{equation}
has the solution
\begin{equation}
\frac{p(x|\theta)}{p(x)} = \frac{f^*(x,\theta)}{1 - f^*(x,\theta)}\,.
\end{equation}
In other words, a neural network trained to distinguish between samples from $p(x,\theta)$ and samples from $p(x)p(\theta)$
approximates the likelihood-to-evidence ratio at optimum.
Posteriors can then be obtained through usual Monte Carlo Markov Chain sampling
which we perform with \texttt{emcee}~\cite{Foreman-Mackey2013a,Foreman-Mackey2013b}.
In practice, this general idea of approximating $p(x|\theta)/p(x)$ through a classifier
works better in the multi-class version ``NRE-B''~\cite{Durkan2020}.
We use the implementation provided in the \texttt{sbi} package~\cite{tejero-cantero2020sbi}.

In the above, it is actually not necessary for the parameter vectors $\theta$ to be drawn independently from
the prior $p(\theta)$. In fact, all that is required is that a sum over the simulated parameter vectors
approximates the integral in Eq.~\eqref{eq:argmin}. For this reason, it is correct for us to populate each
of the 127 cosmo-varied simulations with multiple draws from the HOD prior ($\sim 230$).
For each HOD draw we compute 8 augmentations as described below, yielding $\sim 1.7\times 10^5$ training samples.

The ILI framework allows implicit marginalization over nuisance parameters.
This is one of its primary benefits in high dimensional parameter spaces.\footnote{
  Consider computation of a high-dimensional integral over $f(\mathbf{x})$ given samples $f(\mathbf{x}_i)$.
  Interpolating $f(\mathbf{x})$ using these samples and then performing quadrature is a more difficult
  problem than using the Monte Carlo estimator.}
In principle, we could take $\theta = \{ \Mnu \}$ as one dimensional.
In practice, it is likely better to include a subset of the nuisance parameters in $\theta$.
This is because we have intuition for the posteriors expected for some nuisance parameters and thus
making them explicit allows useful checks.
We opt to include $\log M_\text{min}$ and $\mu(M_\text{min})$ in the parameter vector.
For the former we know that the data should provide a constraint considerably tighter than the prior,
while for the latter we expect a result close to zero.
The posterior on $\Mnu$ is unaffected by this choice of $\theta$ and the extra computational cost
in training and sampling the neural network is marginal when making two nuisance parameters explicit
(however, making all nuisance parameters explicit would complicate the training unnecessarily).

We parameterize the classifier $f$ as a residual neural network.
Hyperparameter optimization was performed considering the
loss on a validation set of 13 cosmologies (i.e., $\sim 2.4\times 10^4$ mocks).\footnote{
  It is actually important to separate training and validation sets by cosmologies.
  Initial trials that mixed simulations exhibited hidden overfitting.}
We converged at a relatively large network with $1.7\times 10^7$ trainable parameters but high dropout rates.

High-dimensional data vectors $x$ are often problematic for ILI, our problem being no exception.
This necessitates a compression step before the data vector is passed to the neural network.
Since we expect our likelihood to be close to Gaussian/Poissonian, we use the linear score compression
\texttt{MOPED}~\cite{Heavens2000}
to 17 compressed statistics.
Indeed, \texttt{MOPED} is locally optimal both for a pure Gaussian and a pure Poissonian likelihood.
We also experimented with the nuisance-hardened generalization~\cite{Alsing2019}
to five and ten dimensions, obtaining consistent but slightly wider posteriors.
In order to construct the \texttt{MOPED} compression matrix, estimates for the covariance and derivatives
of the data vector are required.
We construct the covariance matrix from our fiducial mocks using the usual estimator.
For the derivatives, we generate $\sim 10^5$ additional mocks ($10^3$ parameter vectors, each with $96$ augmentations)
in a small ball around the fiducial model. We then perform linear regression and read off the derivatives.
Simple tests indicate that the dependence on parameters is close to linear in the region considered.

\subsection{Augmentations}

\begin{figure}
\includegraphics[width=\linewidth]{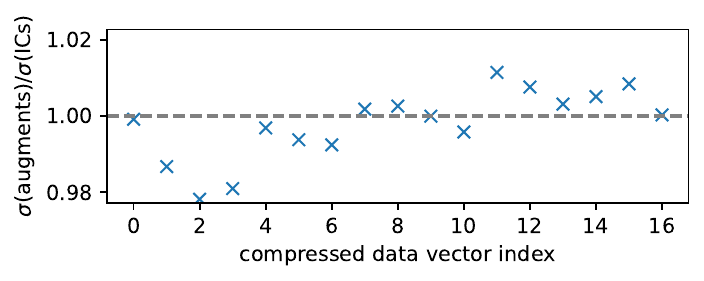}
\includegraphics[width=\linewidth]{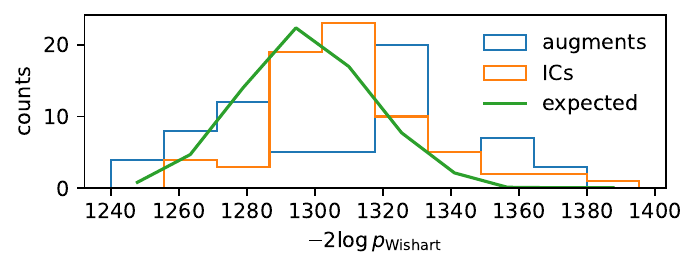}
\caption{
	 Checks for statistical independence of augmentations, in the compressed space.
	 The data labeled ``augments'' are obtained by marginalizing over augmentations, while those
	 labeled ``ICs'' are obtained by marginalizing over initial conditions.
	 \emph{Top}: ratio of average standard deviations, difference with unity not exceeding two percent.
	             This check can also be performed with the uncompressed data vector,
		     leading to similar conclusions.
	 \emph{Bottom}: distributions of log-likelihoods of covariance matrices under the Wishart distribution with
	                our fiducial covariance matrix.
			The distributions have large overlap, again indicating that the augmentations
			are very close to statistical independence.
	}
\label{fig:augments}
\end{figure}

As discussed before, our cosmo-varied simulations share the random seed. This fact ostensibly makes them
unsuitable for the ILI approach discussed before since the integral in Eq.~\eqref{eq:argmin} requires
a sampling of initial conditions.

However, as we shall discuss in this section, it is possible to generate many quasi-independent realizations
from a single simulation.
As mentioned before, we do not require independent identically distributed realizations,
so this is in fact enough to approximate Eq.~\eqref{eq:argmin} with sufficient accuracy.

For a single $2.5\,h^{-1}\text{Gpc}$ simulation box populated with galaxies, we can take the product of the
following transformations: 2 cuboid remappings, 8 reflections, 6 axis transpositions.
This results in 96 augmentations.
In principle, many more augmentations can be generated through translations, but we expect these to be
more correlated.

The crucial question now is whether these 96 augmentations approximate the distribution over initial conditions.
We can answer this question by considering our fiducial simulations which have 69 different random seeds.
Given the fiducial parameter vector, we generate a matrix $D_{\mu a}$ whose elements are data vector-valued
and where $\mu=1\ldots69$, $a=1\ldots96$ index the initial conditions and augmentations, respectively.
We can perform statistical tests by computing marginals over $\mu$ and $a$ individually or jointly.
In order to simplify the statistical interpretation, we restrict $a$ to $69$ randomly chosen indices.

In the upper panel of Fig.~\ref{fig:augments}, we compare the diagonals of covariance matrices
in the \texttt{MOPED} compressed space.
We see that the standard deviations are almost identical for marginalization over $\mu$ and $a$.
This test can also be performed for the uncompressed data vector, yielding consistent results
and no systematic differences between different summary statistics or scales.

In the lower panel, we perform a test considering the entire content of the covariance matrices.
We construct the Wishart distribution given the covariance matrix $C_\text{joint}$ obtained by marginalizing over $\mu$ and $a$
jointly and then compute the log-likelihood of the individually marginalized covariance matrices.
If these covariance matrices were drawn from the Wishart distribution sourced by $C_\text{joint}$,
their log-likelihoods would be distributed as indicated by the green line.
We see that the distributions are somewhat different but still have large overlap.
In conclusion, the 96 augmentations reproduce the distribution over initial conditions reasonably
well.
Since $96 \gg 1$, we expect the augmentations to provide a good approximation to the integral in Eq.~\eqref{eq:argmin}.

Why does this augmentation procedure work?
First, our simulation boxes are about $5.7\times$ larger than the survey volume. 
Second, the augmentations alter the redshift direction.
Third, galaxies and the survey mask interact.
Fourth, galaxies are captured at different times so their peculiar motions alter their real space positions.
All these points need to be seen relative to the specific survey and simulation configuration;
the described augmentation procedure is certainly not expected to work universally.

\section{Results}
\label{sec:results}

In this section, we first present our main posteriors on $\Mnu$
from the CMASS NGC data, taking various combinations of the summary statistics VSF $N_v$,
void-galaxy cross power spectrum $P^{vg}$, and galaxy auto power spectrum $P^{gg}$.

We present most of our posteriors in their cumulative form. This is because at the current level
of precision, no neutrino mass detection is expected and upper bounds are the main objective.
The cumulative posterior is the most direct visualization of upper bounds.
In all plots we include a diagonal dashed line indicating the prior.

In the following, we will occasionally compare with results obtained with the EFTofLSS~\cite{
  Ivanov2020a,dAmico2020,Chen2022}.
The EFTofLSS allows for the analysis of the full-shape galaxy auto power spectrum
(as well as other statistics we will not consider here).
We use the window-less full-shape likelihood~\cite{
  Philcox2020,Philcox2021,Philcox2022}\footnote{\url{https://github.com/oliverphilcox/full_shape_likelihoods}}
and the $\texttt{CLASS-PT}$ code~\cite{
  Chudaykin2020}.\footnote{\url{https://github.com/michalychforever/CLASS-PT}}
We restrict the data included in the likelihood to the NGC high-$z$ sample,
approximately equal to the data we use for our analysis.
Furthermore, we impose the same $\Lambda$CDM prior while keeping the nuisance parameter priors equal to those
implemented in the public likelihood code.
In any comparison with our results we use identical $\kmax$.
Likewise, we usually only use the monopole $P^{gg}_0$, consistent with our simulation-based analysis.
The likelihood part termed ``Alcock-Paczynski'' in the EFTofLSS likelihood is included, since our method
also effectively includes this term. On the other hand, we do not include the BAO reconstruction or real space
likelihoods.

We emphasize that a comparison between EFTofLSS and HOD methods is beyond the scope of this work.
Therefore, we will use the EFTofLSS posteriors to provide intuition, show that our posteriors are at
least qualitatively reasonable, and for an interesting observation about the quadrupole $P^{gg}_2$ later on.

\subsection{Neutrino mass posterior}

\begin{figure}
\includegraphics[width=\linewidth]{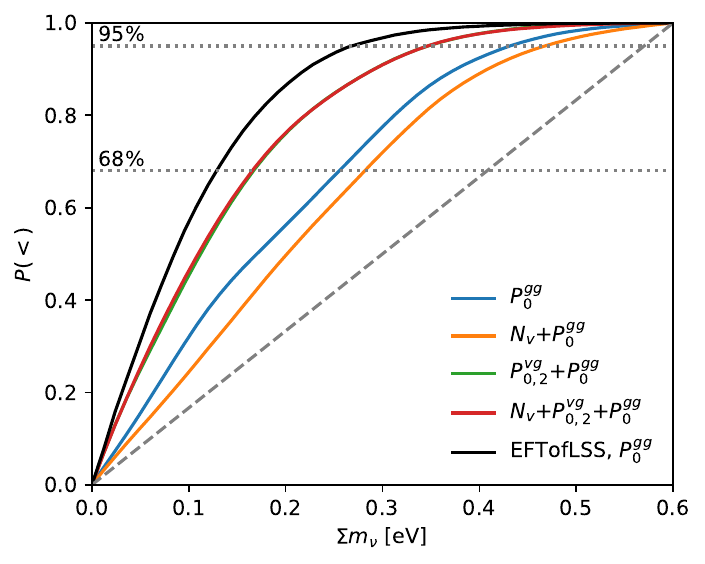}
\caption{
	 Cumulative posteriors on $\Mnu$ from different data vector combinations,
	 with our baseline choice of $\kmax = 0.15$.
	 The addition of void-galaxy cross power spectra improves the constraint compared to the
	 galaxy auto power spectrum, whereas the VSF has negligible impact.
	}
\label{fig:mainposterior}
\end{figure}

In Fig.~\ref{fig:mainposterior}, we show the baseline posterior on $\Mnu$, with $\kmax=0.15$.
The galaxy auto power spectrum gives a $95\,\%$ credible interval constraint of $\Mnu < 0.43\,\text{eV}$.
Upon inclusion of the VSF, the posterior broadens somewhat.
Including the void-galaxy cross power spectrum tightens the posterior to $\Mnu < 0.35\,\text{eV}$,
a $\sim 20\,\%$ improvement.
Further adding the VSF does not lead to any appreciable change.
Posteriors are generally wider than the EFTofLSS result.

\begin{figure}
\includegraphics[width=\linewidth]{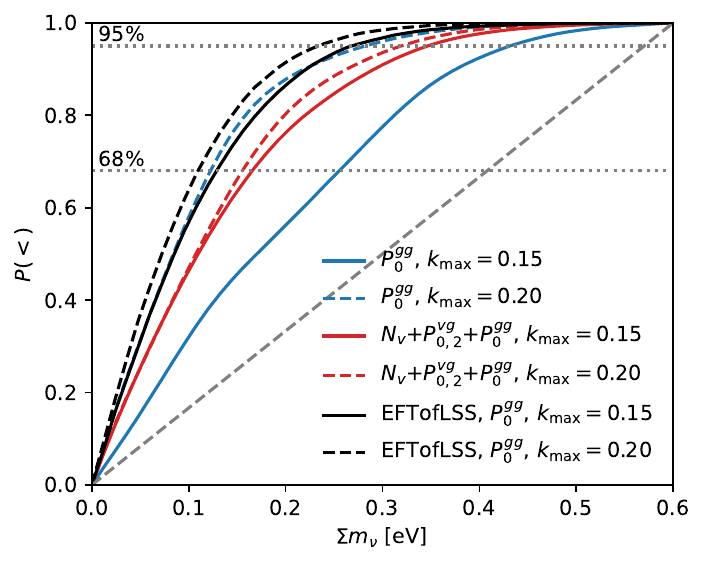}
\caption{
	 Effect of increasing $\kmax$ to $0.2$. The posterior shrinks, as expected,
	 but remains broader than the EFTofLSS result. The impact of adding voids to the data vector
	 is now reversed.
	}
\label{fig:kmax}
\end{figure}

In Fig.~\ref{fig:kmax}, we show a similar set of posteriors obtained with $\kmax=0.2$.
We believe that our simulated model should still have a high level of fidelity at these somewhat smaller scales.
We observe that including smaller scales tightens the posterior, as expected.
However, adding void statistics to $P^{gg}$ now slightly broadens the posterior.
Most of the remainder of this section will be devoted to better understanding the observations
from Figs.~\ref{fig:mainposterior},~\ref{fig:kmax}.

\subsection{Validation}

\begin{figure}
\includegraphics[width=\linewidth]{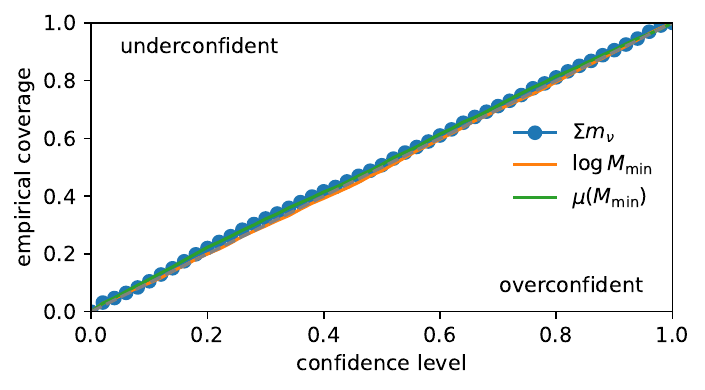}
\caption{
	 Coverage (q-q) plot, demonstrating that when performing inference on mocks the posteriors
	 are well calibrated.
	}
\label{fig:coverage}
\end{figure}

Any simulation-based, and especially implicit-likelihood, inference necessitates rigorous validation
of the simulated model, the likelihood approximation, and the resulting posteriors.
In this section, we present three tests verifying different aspects of our pipeline.

First, in Fig.~\ref{fig:coverage}, we present the usual coverage (or q-q) plot~\cite{Hermans2021}.
For this diagnostic, we perform inference on mocks drawn from the prior;
in particular, we use the $\sim 2.4\times10^4$ validation mocks discussed before.
We use the $N_v+P^{vg}+P^{gg}$ likelihood with $\kmax=0.15$.
For each resulting chain, we compute the marginal distributions of the explicit parameters and
then the confidence level at which the true input parameter is located.
In Fig.~\ref{fig:coverage}, we show the cumulative histograms of these confidence levels.
If the posterior is well-calibrated, these CDFs should coincide with the diagonal.
As can be seen, for all parameters considered this is the case.
This diagnostic is a powerful internal consistency check and verifies that the neural network is well-trained.

\begin{figure}
\includegraphics[width=\linewidth]{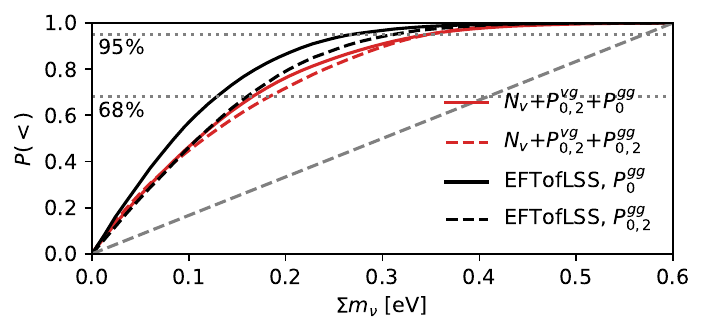}
\caption{
	 Effect of adding the quadrupole $P_2^{gg}$ to the data vector.
	 Consistent with the EFTofLSS result, the $\Mnu$ posterior broadens slightly.
	 This effect is likely a statistical fluctuation.
	 However, it provides a useful cross-check for how well our simulations model redshift space
	 distortions.
	}
\label{fig:quadrupole}
\end{figure}

Second, in Fig.~\ref{fig:quadrupole}, we show an interesting observation concerning the galaxy auto power spectrum
quadrupole $P^{gg}_2$. As discussed before, this summary statistic has limited constraining power and we do not use
it for our main posteriors.
As can be seen in Fig.~\ref{fig:quadrupole}, adding the quadrupole to the data vector slightly broadens the posterior.
This happens consistently in our analysis and in the EFTofLSS.
We believe that this observation increases confidence in the validity of our simulation model,
in particular the modeling of redshift space distortions.

\begin{figure}
\includegraphics[width=\linewidth]{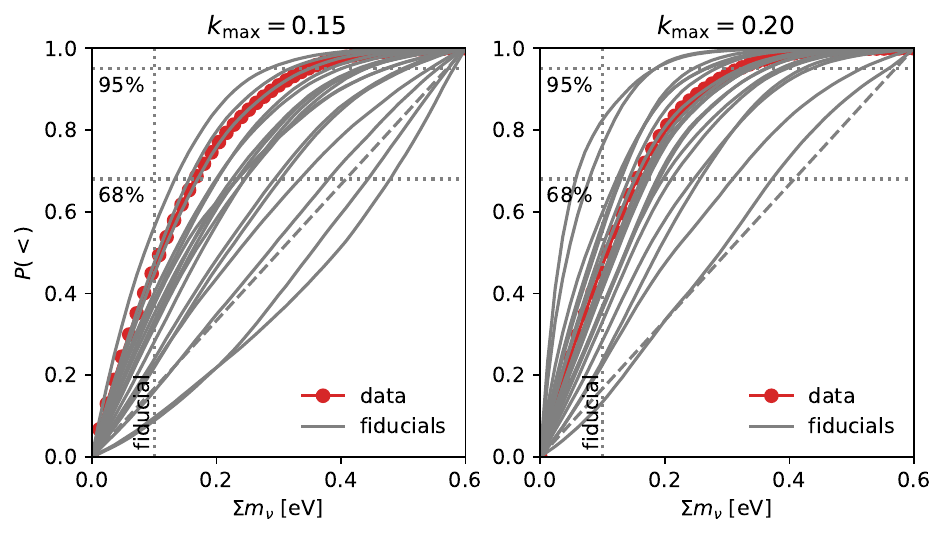}
\caption{
	 Comparison of the data posterior with posteriors obtained on fiducial mocks,
	 using the $N_v+P^{vg}_{0,2}+P^{gg}_0$ data vector.
	 Note that the HOD used in generating these mocks is somewhat different from the best-fit.
	 While the data posterior at $\kmax=0.15$ is slightly unusual, at $\kmax=0.20$
	 it becomes more typical (compared to the distribution of mock posteriors).
	}
\label{fig:cdffid}
\end{figure}

Third, in Fig.~\ref{fig:cdffid} we compare the data posteriors with posteriors obtained by running inference
on randomly chosen mocks generated at the fiducial point.
We remind the reader that the fiducial HOD is rather far from best-fit which somewhat complicates the interpretation.
We observe that at $\kmax=0.15$ the data posterior is tighter than most of the mock ones.
If the cosmological simulations were to blame for this,
the naive expectation would be for the discrepancy to become more severe
as smaller scales are included in the analysis.
However, this appears not to be the case: at $\kmax=0.20$ the data posterior becomes more typical.
We conclude that even though we observe hints of differences between data and simulations,
the evidence is not conclusive and the data posterior could well be consistent with the observed distribution.
It should also be noted that the real $\Mnu$ may be less than the choice $0.1\,\text{eV}$ with
which the fiducials were run, potentially leading to a tighter data posterior.

\subsection{Broadening of posteriors}

One peculiar observation is that inclusion of void statistics can broaden the posterior on $\Mnu$.
We do not fully understand this phenomenon and can only provide some suggestive results.
These are more comprehensively described in appendix~\ref{sec:broadening};
here we only provide a summary.

We observe similar broadening on fiducial mocks and thus propose that we are in fact observing a generic
phenomenon.
Therefore, we suggest that void statistics are most effective at constraining the neutrino mass sum
from below.
A further test using artificially enlarged volumes supports this theory.

For a potential physical explanation, we consider the free streaming length.
At $z=0.5$, $\lambda_\text{fs} = 90\,\Mpc\,(0.3\,\text{eV}/\Mnu)$ for degenerate masses.
This length scale is comparable to the diameter of the voids that seem to contribute most
(c.f. Sec.~\ref{sec:voidphys}).
Thus, it may be that $\Mnu$ at the upper end of the posterior is ``invisible'' to voids.
However, we identify voids using tracers of small-scale fluctuations,
so the full picture is much more complicated and could be a subject for further study.

\subsection{Void radius}
\label{sec:voidphys}

\begin{figure}
\includegraphics[width=\linewidth]{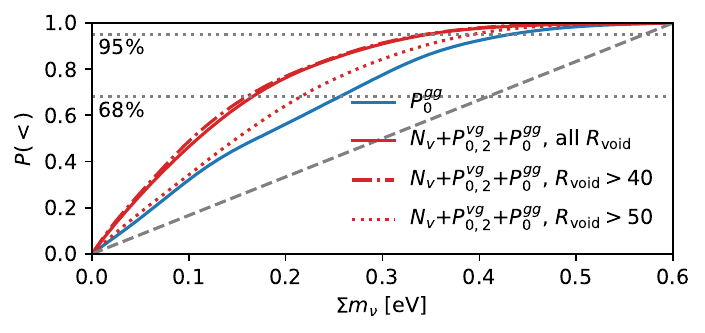}
\caption{
	 Effect of considering voids above different radius cuts.
	 Voids with radii between $40$ and $50\,\Mpc$ contribute the majority of
	 the observed tightening of the posterior relative to the $P^{gg}$-only result.
	}
\label{fig:largevoids}
\end{figure}

In Fig.~\ref{fig:largevoids}, we show posteriors obtained with the $N_v+P^{vg}+P^{gg}$ likelihood,
concentrating on void size.
Cuts on effective radius are performed both in the VSF and $P^{vg}$ parts of the data vector.
We observe that the posteriors are almost identical regardless of whether we cut at $30$
(the baseline analysis) or $40\,\Mpc$.
On the other hand, further increasing the minimum radius to $50\,\Mpc$ removes much of
the effect of voids on the posterior.
Fig.~\ref{fig:largevoids} indicates that at least for the present analysis voids with effective radii
between $40$ and $50\,\Mpc$ are the most constraining.
Smaller voids might be contaminated by spurious Poisson voids and perhaps also due to their shallower
density profile less affected by neutrinos.
Larger voids presumably suffer from their low abundance.

\begin{figure}
\includegraphics[width=\linewidth]{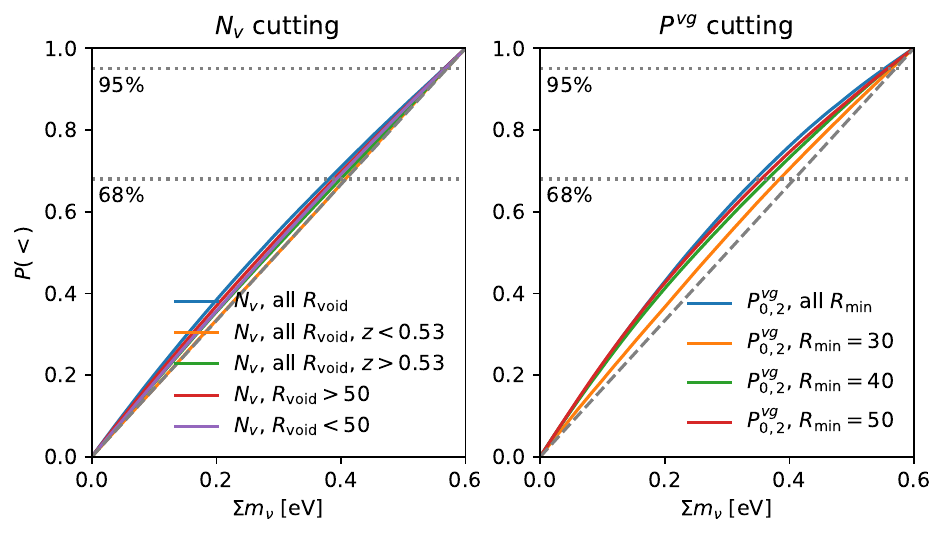}
\caption{
	 Different void-only statistics combinations.
	 Consistent with Fig.~\ref{fig:largevoids}, it appears that larger voids carry more neutrino mass signal,
	 although the posteriors are quite close to the prior.
	}
\label{fig:voidcuts}
\end{figure}

In Fig.~\ref{fig:voidcuts} we show posteriors obtained from void statistics only.
We show them mostly for completeness; in the present analysis these are entirely prior dominated.
However, even in this plot we see the previously mentioned observation that larger voids appear to
carry more signal.

\subsection{Poissonian void size function}
\label{sec:poisson}

\begin{figure}
\includegraphics[width=\linewidth]{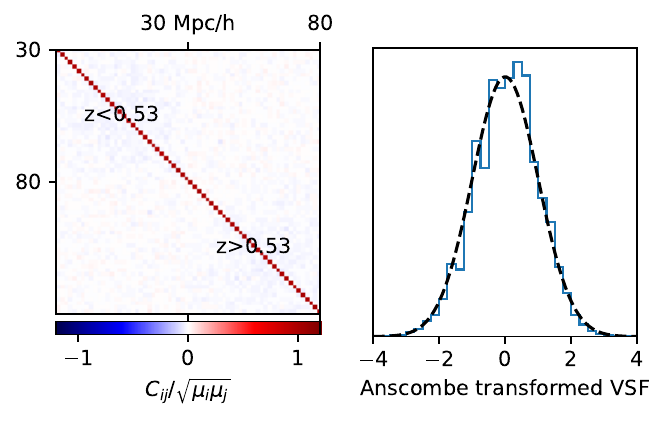}
\caption{
	 Checks for Poissonian nature of the VSF.
	 In the left panel, we show a rescaled covariance matrix obtained from our fiducial simulations
	 of the VSF part of the data vector.
	 For an exactly Poissonian distribution, this matrix would be the identity.
	 In the right panel, we show the distribution of Anscombe transformed VSF counts.
	 Overplotted is a standard normal.
	 Both tests indicate a distribution that is very close to Poissonian.
	}
\label{fig:poisson}
\end{figure}

As a final point of this section, we substantiate the previous claim that the VSF is very close to Poissonian distributed.
While this seems to be a natural assumption, void exclusion makes it non-trivial.
Indeed, previous works have assumed Poisson likelihoods~\cite{Sahlen2019,Contarini2022a};
our simulations enable us to check this assumption.

In Fig.~\ref{fig:poisson}, we show two checks performed with our fiducial mocks.
The left panel shows the covariance matrix divided by the outer square of Poissonian standard deviations;
the result is close to the identity.
The right panel shows a check using the variance-stabilizing Anscombe transform~\cite{Anscombe1948}.
For each mock data vector~$c^{(\alpha)}$ and bin~$i$, we compute the transformed VSF count
\begin{equation}
\tilde c^{(\alpha)}_i = 2\left( \sqrt{c^{(\alpha)}_i+\frac{3}{8}} - \sqrt{\langle c_i \rangle+\frac{3}{8}} \right)
                  + \frac{1}{4\sqrt{\langle c_i \rangle}}\,.
\end{equation}
In the limit of large counts the distribution of these transformed counts converges to the standard normal 
if the counts themselves are Poissonian.
As can be seen, the agreement with the standard normal is quite good indeed.
These tests demonstrate that deviations from Poissonian distribution are small for the VSF,
at least for the choice of binning considered here.

\section{Conclusions}
\label{sec:concl}

We have performed inference on galaxy clustering in the BOSS CMASS northern sample,
combining the void size function, the void-galaxy cross power spectrum, and the galaxy auto power spectrum.
Our primary target was the neutrino mass sum, $\Mnu$; thus, we imposed a tight prior on $\Lambda$CDM informed
by primary CMB data.

We argued that analytic models for the considered void statistics are not mature enough
and unsuitable for our specific problem, necessitating a simulation-based approach.
To this end, we ran approximate gravity-only simulations and populated them with galaxies
using an expressive halo occupation distribution.
Several factors motivated the use of implicit likelihood inference.

In our baseline analysis, we find $\Mnu < 0.43\,\text{eV}$ from the galaxy auto power spectrum alone,
and $\Mnu < 0.35\,\text{eV}$ with the void statistics included ($95\,\%$ credible interval).
We performed several tests to confirm statistical and systematic validity of our likelihood approximation.

We performed a short investigation of the impact of voids on the neutrino mass posterior.
It appears that the void statistics may be most effective in constraining $\Mnu$ from below.
This result would imply that future analyses aiming at \emph{measuring}
$\Mnu$ may benefit from including void statistics.

Our results suggest that larger voids with effective radii $>40\,\Mpc$ carry most of
the signal despite their lower abundance.
This has interesting implications for future analyses, since voids of this size should be detectable
in photometric catalogs with relatively low redshift error, such as the one expected for Rubin/LSST~\cite{Ivezic2019}. 
Of course, spectroscopic surveys such as
DESI~\cite{Dey2019}, Euclid~\cite{Laureijs2011}, SPHEREx~\cite{Dore2014}, PFS~\cite{Takada2014}, and Roman~\cite{Spergel2015}
will continue to be cornerstones of void science.
The trade-off between volume, galaxy number density, and redshift precision warrants further investigation.

We also demonstrate that the void size function is very close to Poisson distributed,
a feature that had been assumed in previous analyses but never explicitly confirmed.

Future work could improve upon our analysis in multiple ways.
First, the cosmo-varied simulations should be run with different random seeds
(we decided for a fixed seed in anticipation of an emulator-based analysis which ultimately
turned out to be very difficult).
Second, it may be beneficial to normalize the void-galaxy cross power spectrum by void number.
Although in principle this would contain the same information as our data vector once the VSF is included,
the necessary transformation is non-linear and thus potentially invisible to our data compression.
Third, the HOD modeling could be improved. Some of our priors may not be optimal, and our $n(z)$ downsampling
is simplistic. The CMASS sample's completeness is quite well known and could be used to put a prior
on the downsampling.
Fourth, it turns out that the cosmological simulations did not dominate compute cost.
It may therefore be economical to increase accuracy in \texttt{FastPM} or switch to a different solver. 

Our results point toward a complicated picture with regard to the relationship between massive neutrinos
and voids.
Future data sets, both spectroscopic as well as photometric, promise to bring tight cosmological constraints
from void science,
since it scales well with number.

\acknowledgments

We thank Sofia Contarini, Adrian Bayer, Jia Liu, Jo Dunkley, Masahiro Takada
for useful discussions.
We thank Oliver Philcox for explaining the EFTofLSS likelihood.
The work of LT is supported by the NSF grant AST~2108078.
The authors are pleased to acknowledge that the work reported on in this paper was substantially performed using the Princeton Research Computing resources at Princeton University which is a consortium of groups led by the Princeton Institute for Computational Science and Engineering (PICSciE) and Office of Information Technology's Research Computing.

\appendix

\section{Halo occupation distribution}
\label{sec:hod}

In this appendix, we provide a more detailed description of the adopted HOD model.

First, for reference, the baseline five-parameter model only depends on halo mass $M$
and has mean occupations
\begin{equation}
\overline N_\text{cen} = \frac{1}{2}\left[
  1 + \text{erf}\left( \frac{\log M - \log M_\text{min}}{\sigma_{\log M}} \right)
  \right]
\end{equation}
for the central galaxy and
\begin{equation}
\overline N_\text{sat}  = \overline N_\text{cen}
  \left( \frac{M - M_0}{M_1} \right)^\alpha
\end{equation}
for the satellites.
A central is placed with probability $\overline N_\text{cen}$ and the number of
satellites is drawn from a Poisson distribution with mean $\overline N_\text{sat}$.
The central is placed at the halo's center and assigned the halo velocity.
The satellites are distributed isotropically with an NFW profile~\cite{Navarro1997}
and the concentration model of Ref.~\cite{Duffy2008},
using the analytic solution for the inverse NFW CDF from Ref.~\cite{Robotham2018}.
Satellite velocities are drawn from a distribution assuming virialization.

On top of this baseline model, we implement assembly bias using the
decorated HOD~\cite{Hearin2016a,Hearin2016b} with the ratio of kinetic to potential energy
$r \equiv T/U$ as proxy for assembly history.
In our preliminary tests $T/U$ outperformed halo concentration, possibly due to limited resolution
within the \texttt{FastPM} halos.
The decoration works by splitting halos into two groups according to $r$.
In order to reduce the effect of any potential evolution of $r$ with halo mass,
we do this splitting separately within $64$ groups containing equal numbers of halos.
The fraction $P_1$ of halos with lowest $r$ is assigned type 1 (2) for positive (negative) $a_\text{bias}$,
while the rest is assigned type 2 (1).
Then, the mean occupations are modified as
\begin{align}
\Delta \overline N_\text{cen}& = |a_\text{bias}|
  \text{min}\left[1-\overline N_\text{cen}, \frac{1-P_1}{P_1} \overline N_\text{cen} \right]\,\\
\Delta \overline N_\text{sat}& = |a_\text{bias}|
  \frac{1-P_1}{P_1} \overline N_\text{sat}
\end{align}
for type 1 and
\begin{align}
\Delta \overline N_\text{cen}& = |a_\text{bias}|
  \text{max}\left[-\overline N_\text{cen}, \frac{1-P_1}{P_1} (\overline N_\text{cen} - 1) \right]\,\\
\Delta \overline N_\text{sat}& = |a_\text{bias}|
  (-\overline N_\text{sat})
\end{align}
for type 2.

Velocity bias for the centrals is implemented by adding $\eta_\text{cen} V_\text{vir} n$
where $n \sim N(0,1)$.
For the satellites, the velocity difference from the host halo is scaled by $\eta_\text{sat}$.

Redshift dependence for $M_\text{min}$, $M_1$ is approximated as linear in scale factor,
such that
\begin{equation}
\Delta \log M_i = \mu(M_i) ( a - a_0 )
\end{equation}
with $a_0=1/(1+0.53)$.

We adopt flat priors $12.5<\log M_\text{min}<13.2$, $0.1<\sigma_{\log M}<0.8$, $12.5<\log M_{0,1}<15.5$,
$0.2<\alpha<1.5$, $-3<P_1'<3$, $-1<a_\text{bias}<1$, $5<\eta_\text{cen}'<10$, $-1<\eta_\text{sat}'<1$,
$-20<\mu(M_\text{min})<20$, $-40<\mu(M_1)<40$.
Here, all masses are in $h^{-1}M_\odot$, and the primed parameters are defined as
$2 P_1=(1+\tanh P_1')$, $\eta_\text{sat}=\exp(\eta_\text{sat}')$, $\eta_\text{cen}=\exp(-10+\eta_\text{cen}')$.
The above intervals were found during preliminary inference runs.
Note in particular the small values of $M_\text{min}$, compared to other analyses.
This is partly explained by systematically lower halo masses in \texttt{FastPM},
and partly by the $n(z)$ downsampling described below.
We picked the transformations given by the primed parameters based on the intuition
that strictly (mathematically) bounded intervals often indicate that a uniform prior
in a transformed quantity is a better choice.

\section{Broadening of posteriors}
\label{sec:broadening}

We have seen in our main posteriors that adding void statistics to the data vector sometimes broadens
the posterior on $\Mnu$.
In this appendix, we attempt to better understand this observation, focusing on $\kmax=0.20$.
For this, we will rely on inference on our fiducial mocks.

\begin{figure}
\includegraphics[width=\linewidth]{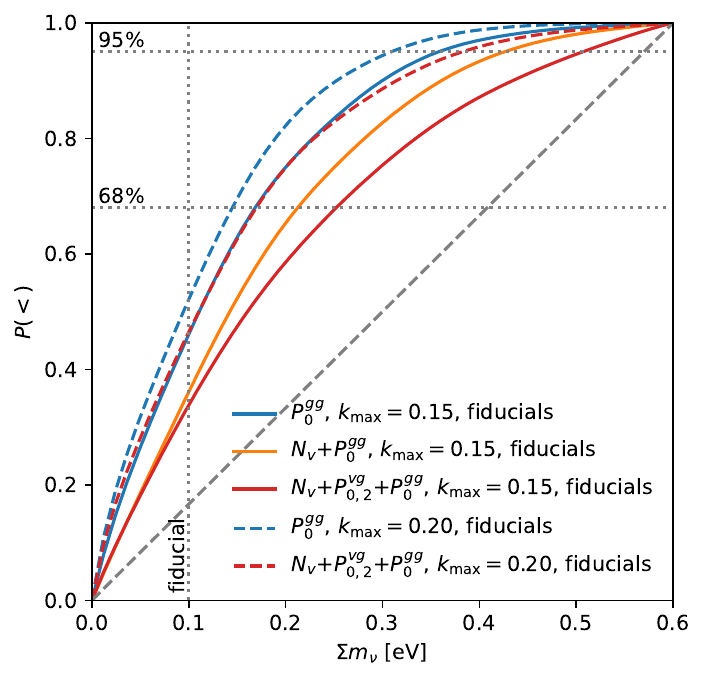}
\caption{
	 Averaged posteriors on $\sim 20$ randomly chosen fiducial mocks.
	 A widening of the $\Mnu$ posterior upon including
	 void statistics is also observed in simulations.
	}
\label{fig:fid1}
\end{figure}

The first possible explanation could be a statistical fluctuation,
and we cannot definitely exclude this hypothesis.
One way, however, to test it is to look at \emph{average} posteriors on our fiducial mocks.
We perform inference on $\sim 20$ randomly chosen mocks and plot the CDFs of concatenated chains
in Fig.~\ref{fig:fid1}.
There, we observe that the expected, average behavior is for the posterior to broaden once void statistics
are added to the data vector.

The second possible explanation could be that once void statistics are added our compression procedure
becomes less efficient. This could certainly be the case if at linear order the void statistics appear more
constraining than they are globally, thus $P^{gg}$ would be unnecessarily downweighted.
This hypothesis appears unlikely in light of the full posteriors presented
in Figs.~\ref{fig:hodposterior},~\ref{fig:moreparams}.
In these posteriors, we observe that for the parameters that are actually constrained (like $M_\text{min}$)
adding void statistics generically tightens the posteriors.
It appears unlikely that $\Mnu$ should be an exception.

\begin{figure}
\includegraphics[width=\linewidth]{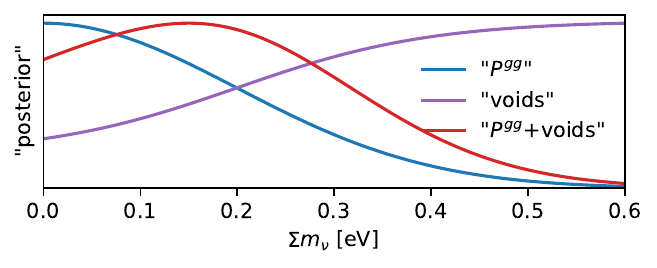}
\caption{
	 Schematic illustration of the proposed mechanism explaining the broadening of posteriors
	 when voids are included.
	 This figure is only meant as a guide and not as a literal depiction of posteriors.
	 Correlations and nuisance parameters complicate the picture.
	}
\label{fig:schematic}
\end{figure}

Having found these two hypotheses unsatisfactory, we arrive at the third one:
\emph{void statistics tend to constrain $\Mnu$ from below}.
We illustrate this theory qualitatively in Fig.~\ref{fig:schematic}, which should not be interpreted as a literal depiction.
In fact, in Sec.~\ref{sec:voidphys} we show that void statistics alone yield posteriors close to the prior.
Fig.~\ref{fig:schematic} provides merely an effective depiction.

\begin{figure}
\includegraphics[width=\linewidth]{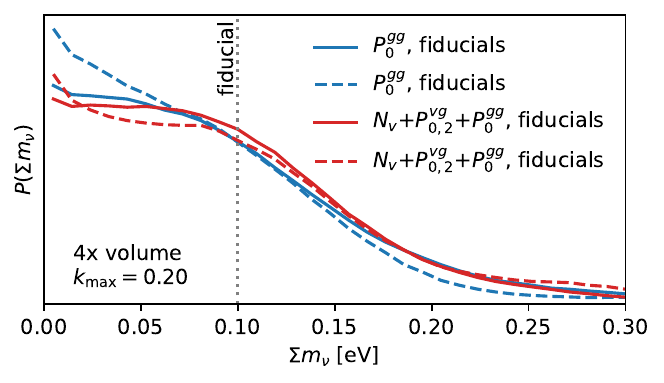}
\caption{
	 Posteriors on joint analyses of four randomly chosen fiducial mocks,
	 averaged over $\sim 30$ groups.
	 The solid and dashed lines correspond to likelihoods with two different sets of five nuisance
	 parameters kept explicit.
	 We see that the posteriors where void statistics are included have a slightly more pronounced
	 bump at the true value $\Mnu=0.1\,\text{eV}$, consistent with the speculative picture
	 in Fig.~\ref{fig:schematic}.
	}
\label{fig:product}
\end{figure}

We can investigate this hypothesis further by performing the following test.
In order to increase signal-to-noise, we perform inference on four fiducial mocks at the same time,
shown in Fig.~\ref{fig:product}.
For this, we use a different set of neural nets in which we leave five nuisance parameters explicit.
The reason is that all implicitly marginalized nuisance parameters are effectively assumed to be different
for each of the four mocks, an effect we would like to minimize.
Of course, increasing the number of explicit nuisance parameters complicates the training and we have less
confidence in the precise calibration of the posteriors.
For this reason, our baseline results were obtained with only two explicit nuisance parameters.
For reference, the real data posteriors obtained with these alternative neural nets are shown in
Fig.~\ref{fig:moreparams}.
We perform this test with two different sets of nuisance parameters kept explicit in order to gauge robustness
(corresponding to the solid and dashed lines in Fig.~\ref{fig:product}).
Similar to Fig.~\ref{fig:fid1}, we average posteriors over $\sim 30$ randomly chosen groups of four mocks in order to decrease
sample variance.
We observe that, consistent with our theory, the posteriors that include void statistics show a more pronounced
hint of a bump at the true $\Mnu$.
In principle, one could increase the simulated volume further by combining more mocks,
but our neural nets are not calibrated at the required level of precision and thus the resulting posteriors
would not be robust.

In summary, the more mundane ideas to explain the observed broadening of posteriors appear questionable
given the tests presented.
On the other hand, the idea that void statistics are most effective at constraining $\Mnu$ from below
receives support from our experiments.
A more in-depth examination of this issue would constitute a great starting point for future work.

\section{Corner plots}

\begin{figure}
\includegraphics[width=\linewidth]{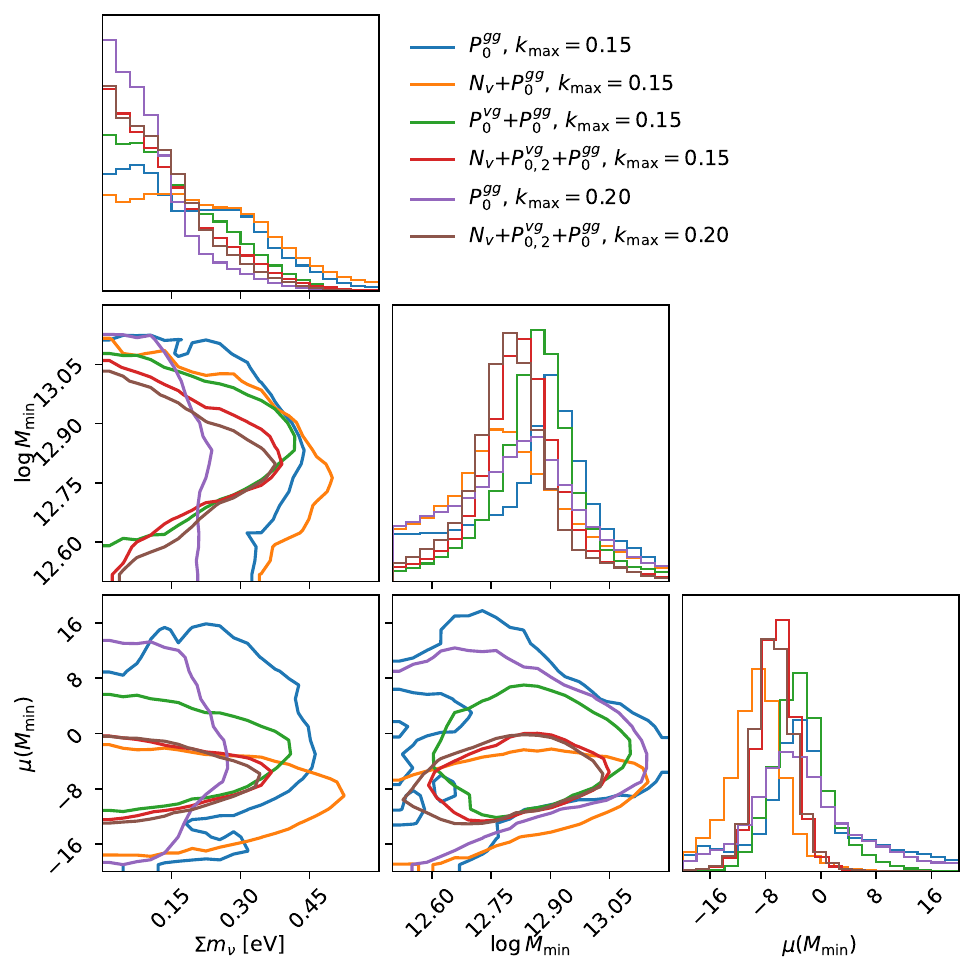}
\caption{
	 Posteriors in the full parameter space considered, including the two HOD parameters
	 we choose to keep explicit.
	 The HOD posteriors are consistent between different analysis choices.
	}
\label{fig:hodposterior}
\end{figure}

\begin{figure*}
\includegraphics[width=0.48\linewidth]{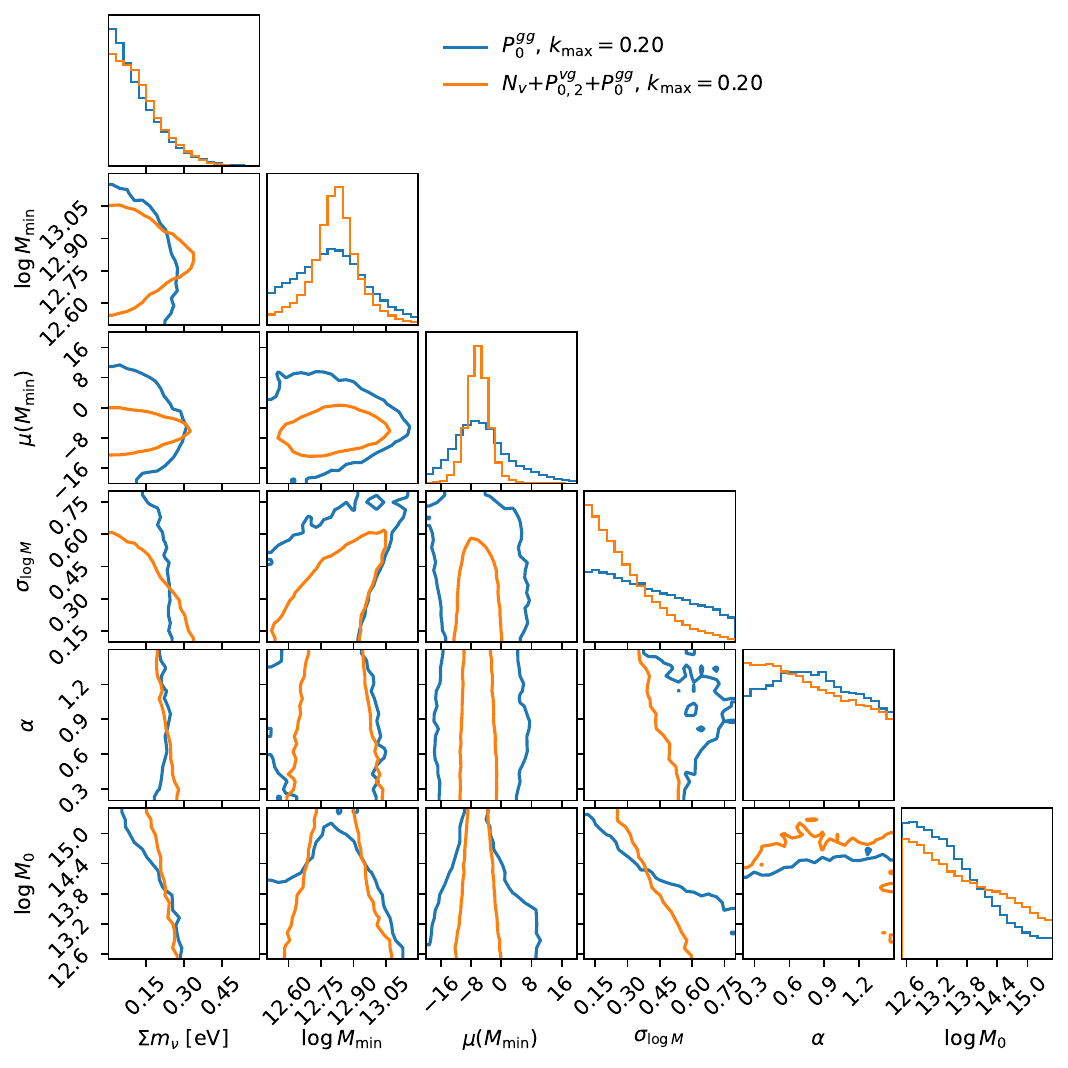}
\includegraphics[width=0.48\linewidth]{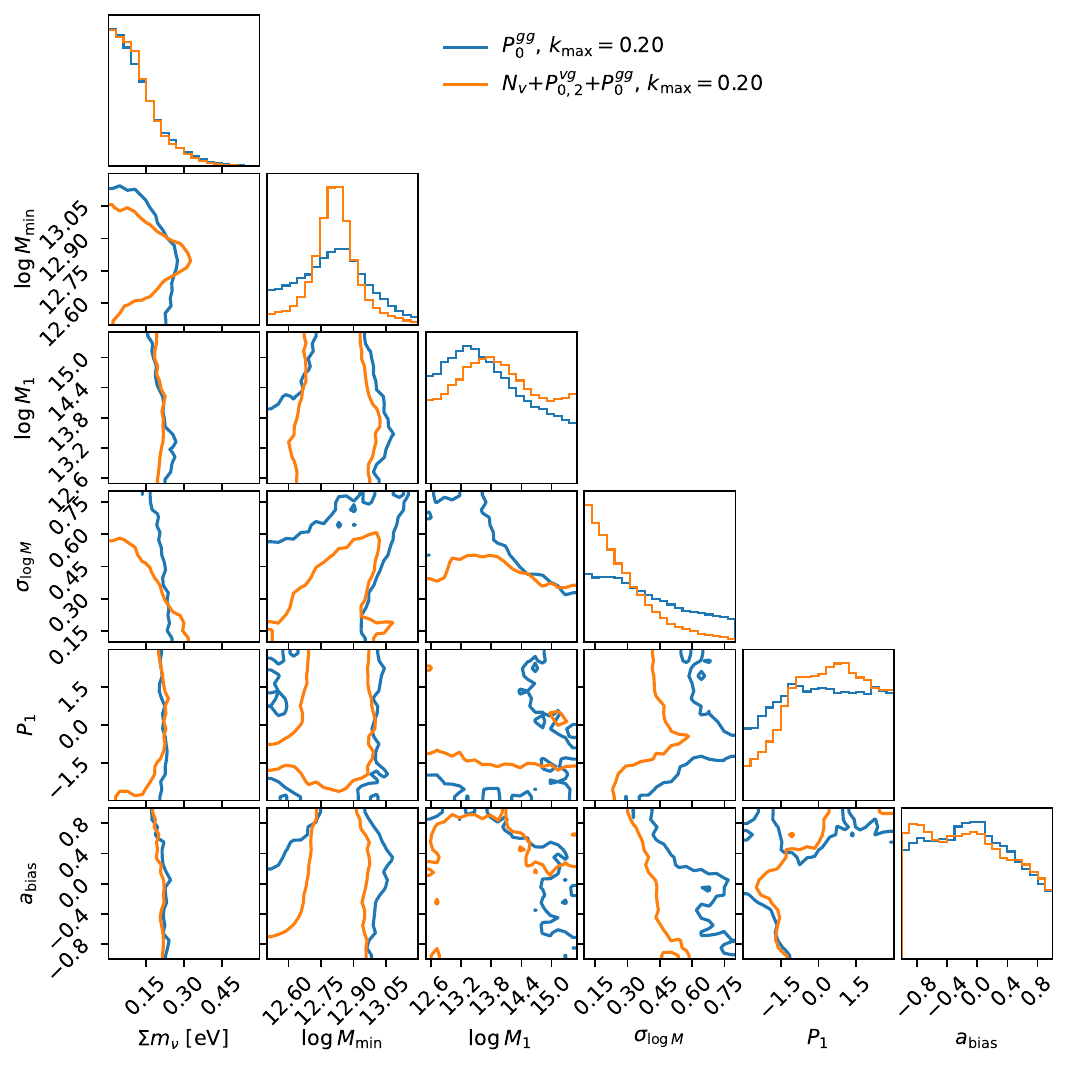}
\caption{
         Posteriors with different sets of nuisance parameters kept explicit.
	 Since no hyperparameter optimization was performed when training the corresponding networks
	 and the larger parameter space makes training more difficult,
	 we assess these posteriors as less robust than our baseline results.
	 The neutrino mass posteriors, however, are quite consistent.
	 The left panel corresponds to the solid lines in Fig.~\ref{fig:fid1},
	 the right panel to the dashed lines.
	}
\label{fig:moreparams}
\end{figure*}

\begin{figure}
\includegraphics[width=\linewidth]{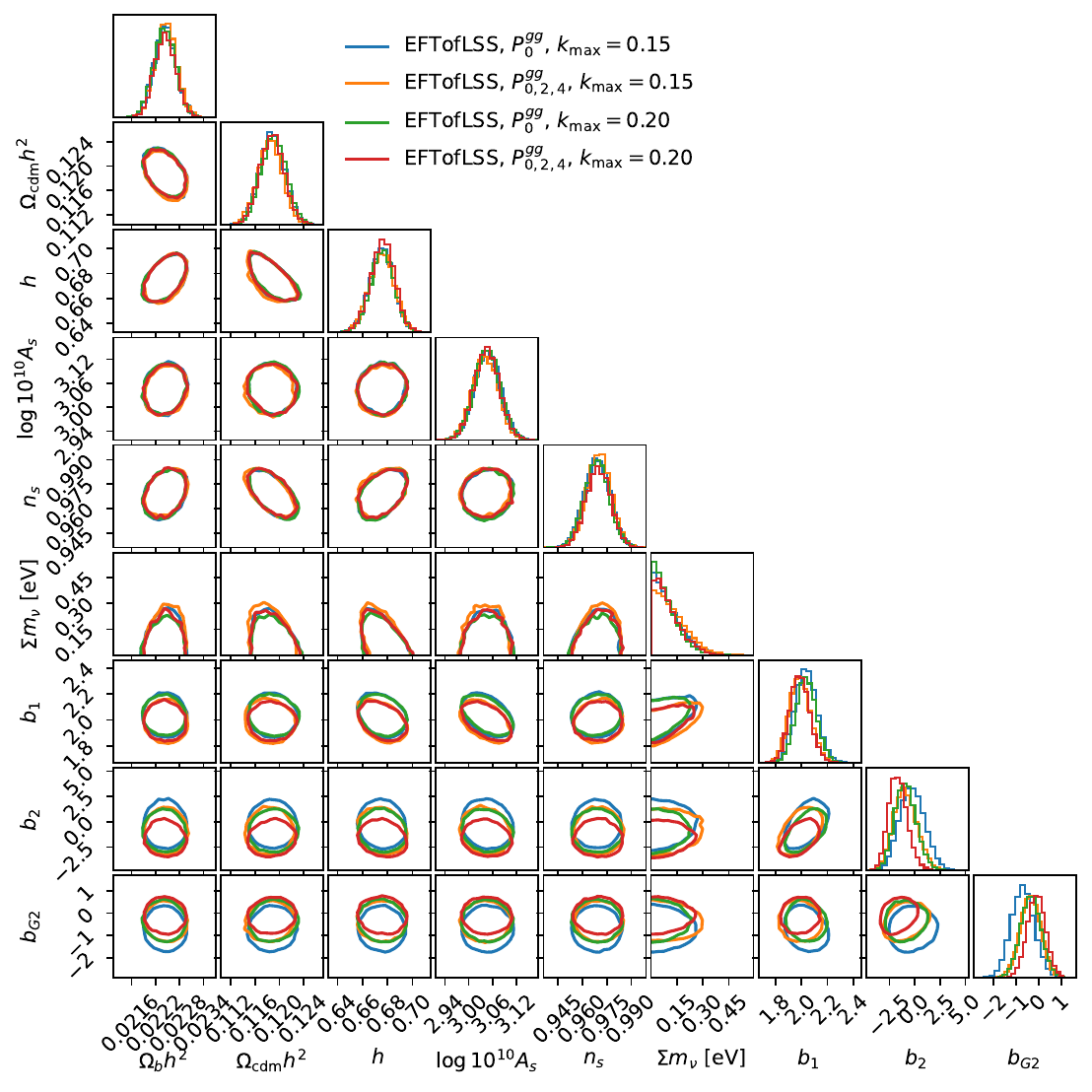}
\caption{
	 For reference, we show the full EFTofLSS posteriors.
	}
\label{fig:eft}
\end{figure}

This appendix collects posteriors in the full parameter spaces considered.
Fig.~\ref{fig:hodposterior} shows the baseline parameter space with two explicit nuisance parameters.
Fig.~\ref{fig:moreparams} shows larger sections of parameter space
(it should be mentioned, however, that the corresponding neural networks were trained without further
hyperparameter optimization, implying a somewhat lower level of confidence in the validity of these posteriors).
Fig.~\ref{fig:eft} shows our EFTofLSS posteriors, demonstrating that the $\Lambda$CDM part of the parameter
space is prior-dominated.

\section{Simulation budget}
\label{sec:budget}

\begin{figure}
\includegraphics[width=\linewidth]{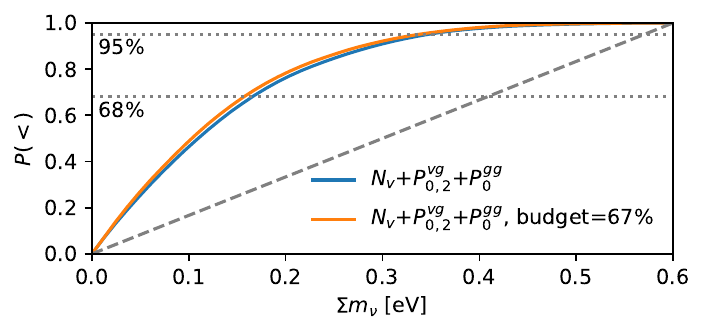}
\caption{
	 Comparison of posteriors obtained from neural networks trained with the full training set
	 and with 2/3 of the training set.
	 We observe very good agreement between these posteriors.
	}
\label{fig:budget}
\end{figure}

One might worry that the 127 cosmo-varied simulations are not enough to properly sample the cosmological prior.
We test this by discarding a third of the simulations and training on the rest.
The resulting posterior, compared to our baseline result, is shown in Fig.~\ref{fig:budget}.
Agreement between the two posteriors is almost perfect, demonstrating that our simulations cover the cosmological
prior sufficiently well.

\section{Simulation data}

About 50TB of halo catalogs, light cones, void catalogs, and summary statistics have been saved
(at 20 times between $z=0.44$ and $z=0.68$ in 127 different massive-neutrino cosmologies with various HODs
and 69 different initial conditions with a fiducial model).
We are currently finalizing how to make this data set publicly available.

\vfill\eject

\section{Code}

In terms of new code, we have written
C\nolinebreak[4]\hspace{-.05em}\raisebox{.4ex}{\relsize{-3}{\textbf{++}}}
code to populate halo catalogs with galaxies
and to generate light cones including survey realism.

We have also written a C implementation of the quasi-random sampling scheme for uniform
and Gaussian priors.

This work necessitated several small modifications to public codes:
\begin{itemize}
\item \texttt{REPS}: read files generated by the current \texttt{CLASS} version;
                     write output in a user-defined directory.
\item \texttt{FastPM}: do not write neutrinos to disk.
\item \texttt{bigfile}: support for half-precision floats.
\item \texttt{Rockstar}: native reading of the \texttt{bigfile} snapshots generated by \texttt{FastPM}
                         (using the file chunking to read in distributed fashion since \texttt{Rockstar}
			  does not use \texttt{MPI}).
\item \texttt{Rockstar/find\_parents}: output to \texttt{bigfile} with lower priority fields in half precision.
\item \texttt{cuboidremap}: support for velocities.
\item \texttt{sbi}: custom splitting into training and validation data.
\end{itemize}

Since all these items are relatively obscure, we do not provide documentation.
However, we are happy to share any of these with interested researchers.
A repository with most of the code is available at \url{https://github.com/leanderthiele/nuvoid_production}.

\newpage
\ \\
\newpage


\begin{thebibliography}{146}%
\makeatletter
\providecommand \@ifxundefined [1]{%
 \@ifx{#1\undefined}
}%
\providecommand \@ifnum [1]{%
 \ifnum #1\expandafter \@firstoftwo
 \else \expandafter \@secondoftwo
 \fi
}%
\providecommand \@ifx [1]{%
 \ifx #1\expandafter \@firstoftwo
 \else \expandafter \@secondoftwo
 \fi
}%
\providecommand \natexlab [1]{#1}%
\providecommand \enquote  [1]{``#1''}%
\providecommand \bibnamefont  [1]{#1}%
\providecommand \bibfnamefont [1]{#1}%
\providecommand \citenamefont [1]{#1}%
\providecommand \href@noop [0]{\@secondoftwo}%
\providecommand \href [0]{\begingroup \@sanitize@url \@href}%
\providecommand \@href[1]{\@@startlink{#1}\@@href}%
\providecommand \@@href[1]{\endgroup#1\@@endlink}%
\providecommand \@sanitize@url [0]{\catcode `\\12\catcode `\$12\catcode
  `\&12\catcode `\#12\catcode `\^12\catcode `\_12\catcode `\%12\relax}%
\providecommand \@@startlink[1]{}%
\providecommand \@@endlink[0]{}%
\providecommand \url  [0]{\begingroup\@sanitize@url \@url }%
\providecommand \@url [1]{\endgroup\@href {#1}{\urlprefix }}%
\providecommand \urlprefix  [0]{URL }%
\providecommand \Eprint [0]{\href }%
\providecommand \doibase [0]{https://doi.org/}%
\providecommand \selectlanguage [0]{\@gobble}%
\providecommand \bibinfo  [0]{\@secondoftwo}%
\providecommand \bibfield  [0]{\@secondoftwo}%
\providecommand \translation [1]{[#1]}%
\providecommand \BibitemOpen [0]{}%
\providecommand \bibitemStop [0]{}%
\providecommand \bibitemNoStop [0]{.\EOS\space}%
\providecommand \EOS [0]{\spacefactor3000\relax}%
\providecommand \BibitemShut  [1]{\csname bibitem#1\endcsname}%
\let\auto@bib@innerbib\@empty
\bibitem [{\citenamefont {{Bahcall}}\ and\ \citenamefont {{Davis,
  Jr.}}(1976)}]{Bahcall1976}%
  \BibitemOpen
  \bibfield  {author} {\bibinfo {author} {\bibfnamefont {J.~N.}\ \bibnamefont
  {{Bahcall}}}\ and\ \bibinfo {author} {\bibfnamefont {R.}~\bibnamefont
  {{Davis, Jr.}}},\ }\href {https://doi.org/10.1126/science.191.4224.264}
  {\bibfield  {journal} {\bibinfo  {journal} {Science}\ }\textbf {\bibinfo
  {volume} {191}},\ \bibinfo {pages} {264} (\bibinfo {year}
  {1976})}\BibitemShut {NoStop}%
\bibitem [{\citenamefont {{Wolfenstein}}(1978)}]{Wolfenstein1978}%
  \BibitemOpen
  \bibfield  {author} {\bibinfo {author} {\bibfnamefont {L.}~\bibnamefont
  {{Wolfenstein}}},\ }\href {https://doi.org/10.1103/PhysRevD.17.2369}
  {\bibfield  {journal} {\bibinfo  {journal} {\prd}\ }\textbf {\bibinfo
  {volume} {17}},\ \bibinfo {pages} {2369} (\bibinfo {year}
  {1978})}\BibitemShut {NoStop}%
\bibitem [{\citenamefont {{Mikheyev}}\ and\ \citenamefont
  {{Smirnov}}(1985)}]{Mikheyev1985}%
  \BibitemOpen
  \bibfield  {author} {\bibinfo {author} {\bibfnamefont {S.~P.}\ \bibnamefont
  {{Mikheyev}}}\ and\ \bibinfo {author} {\bibfnamefont {A.~Y.}\ \bibnamefont
  {{Smirnov}}},\ }\href@noop {} {\bibfield  {journal} {\bibinfo  {journal}
  {Yadernaya Fizika}\ }\textbf {\bibinfo {volume} {42}},\ \bibinfo {pages}
  {1441} (\bibinfo {year} {1985})}\BibitemShut {NoStop}%
\bibitem [{\citenamefont {{Super-Kamiokande Collaboration}}\ \emph
  {et~al.}(1998)\citenamefont {{Super-Kamiokande Collaboration}}, \citenamefont
  {{Fukuda}} \emph {et~al.}}]{Fukuda1998}%
  \BibitemOpen
  \bibfield  {author} {\bibinfo {author} {\bibnamefont {{Super-Kamiokande
  Collaboration}}}, \bibinfo {author} {\bibfnamefont {Y.}~\bibnamefont
  {{Fukuda}}}, \emph {et~al.},\ }\href
  {https://doi.org/10.1103/PhysRevLett.81.1562} {\bibfield  {journal} {\bibinfo
   {journal} {\prl}\ }\textbf {\bibinfo {volume} {81}},\ \bibinfo {pages}
  {1562} (\bibinfo {year} {1998})},\ \Eprint
  {https://arxiv.org/abs/hep-ex/9807003} {arXiv:hep-ex/9807003 [hep-ex]}
  \BibitemShut {NoStop}%
\bibitem [{\citenamefont {{SNO Collaboration}}\ \emph
  {et~al.}(2002)\citenamefont {{SNO Collaboration}}, \citenamefont {{Ahmad}}
  \emph {et~al.}}]{Ahmad2002}%
  \BibitemOpen
  \bibfield  {author} {\bibinfo {author} {\bibnamefont {{SNO Collaboration}}},
  \bibinfo {author} {\bibfnamefont {Q.~R.}\ \bibnamefont {{Ahmad}}}, \emph
  {et~al.},\ }\href {https://doi.org/10.1103/PhysRevLett.89.011301} {\bibfield
  {journal} {\bibinfo  {journal} {\prl}\ }\textbf {\bibinfo {volume} {89}},\
  \bibinfo {eid} {011301} (\bibinfo {year} {2002})},\ \Eprint
  {https://arxiv.org/abs/nucl-ex/0204008} {arXiv:nucl-ex/0204008 [nucl-ex]}
  \BibitemShut {NoStop}%
\bibitem [{\citenamefont {{KamLAND Collaboration}}\ \emph
  {et~al.}(2005)\citenamefont {{KamLAND Collaboration}}, \citenamefont
  {{Araki}} \emph {et~al.}}]{Araki2005}%
  \BibitemOpen
  \bibfield  {author} {\bibinfo {author} {\bibnamefont {{KamLAND
  Collaboration}}}, \bibinfo {author} {\bibfnamefont {T.}~\bibnamefont
  {{Araki}}}, \emph {et~al.},\ }\href
  {https://doi.org/10.1103/PhysRevLett.94.081801} {\bibfield  {journal}
  {\bibinfo  {journal} {\prl}\ }\textbf {\bibinfo {volume} {94}},\ \bibinfo
  {eid} {081801} (\bibinfo {year} {2005})},\ \Eprint
  {https://arxiv.org/abs/hep-ex/0406035} {arXiv:hep-ex/0406035 [hep-ex]}
  \BibitemShut {NoStop}%
\bibitem [{\citenamefont {{K2K Collaboration}}\ \emph
  {et~al.}(2006)\citenamefont {{K2K Collaboration}}, \citenamefont {{Ahn}}
  \emph {et~al.}}]{Ahn2006}%
  \BibitemOpen
  \bibfield  {author} {\bibinfo {author} {\bibnamefont {{K2K Collaboration}}},
  \bibinfo {author} {\bibfnamefont {M.~H.}\ \bibnamefont {{Ahn}}}, \emph
  {et~al.},\ }\href {https://doi.org/10.1103/PhysRevD.74.072003} {\bibfield
  {journal} {\bibinfo  {journal} {\prd}\ }\textbf {\bibinfo {volume} {74}},\
  \bibinfo {eid} {072003} (\bibinfo {year} {2006})},\ \Eprint
  {https://arxiv.org/abs/hep-ex/0606032} {arXiv:hep-ex/0606032 [hep-ex]}
  \BibitemShut {NoStop}%
\bibitem [{\citenamefont {{Daya Bay Collaboration}}\ \emph
  {et~al.}(2012)\citenamefont {{Daya Bay Collaboration}}, \citenamefont {{An}}
  \emph {et~al.}}]{An2012}%
  \BibitemOpen
  \bibfield  {author} {\bibinfo {author} {\bibnamefont {{Daya Bay
  Collaboration}}}, \bibinfo {author} {\bibfnamefont {F.~P.}\ \bibnamefont
  {{An}}}, \emph {et~al.},\ }\href
  {https://doi.org/10.1103/PhysRevLett.108.171803} {\bibfield  {journal}
  {\bibinfo  {journal} {\prl}\ }\textbf {\bibinfo {volume} {108}},\ \bibinfo
  {eid} {171803} (\bibinfo {year} {2012})},\ \Eprint
  {https://arxiv.org/abs/1203.1669} {arXiv:1203.1669 [hep-ex]} \BibitemShut
  {NoStop}%
\bibitem [{\citenamefont {{KATRIN Collaboration}}\ \emph
  {et~al.}(2022)\citenamefont {{KATRIN Collaboration}}, \citenamefont {{Aker}}
  \emph {et~al.}}]{Aker2022}%
  \BibitemOpen
  \bibfield  {author} {\bibinfo {author} {\bibnamefont {{KATRIN
  Collaboration}}}, \bibinfo {author} {\bibfnamefont {M.}~\bibnamefont
  {{Aker}}}, \emph {et~al.},\ }\href {https://doi.org/10.1088/1361-6471/ac834e}
  {\bibfield  {journal} {\bibinfo  {journal} {Journal of Physics G Nuclear
  Physics}\ }\textbf {\bibinfo {volume} {49}},\ \bibinfo {eid} {100501}
  (\bibinfo {year} {2022})},\ \Eprint {https://arxiv.org/abs/2203.08059}
  {arXiv:2203.08059 [nucl-ex]} \BibitemShut {NoStop}%
\bibitem [{\citenamefont {{Planck Collaboration}}\ \emph
  {et~al.}(2020)\citenamefont {{Planck Collaboration}}, \citenamefont
  {{Aghanim}} \emph {et~al.}}]{PlanckCollaboration2020}%
  \BibitemOpen
  \bibfield  {author} {\bibinfo {author} {\bibnamefont {{Planck
  Collaboration}}}, \bibinfo {author} {\bibfnamefont {N.}~\bibnamefont
  {{Aghanim}}}, \emph {et~al.},\ }\href
  {https://doi.org/10.1051/0004-6361/201833910} {\bibfield  {journal} {\bibinfo
   {journal} {\aap}\ }\textbf {\bibinfo {volume} {641}},\ \bibinfo {eid} {A6}
  (\bibinfo {year} {2020})},\ \Eprint {https://arxiv.org/abs/1807.06209}
  {arXiv:1807.06209 [astro-ph.CO]} \BibitemShut {NoStop}%
\bibitem [{\citenamefont {{Icke}}(1984)}]{Icke1984}%
  \BibitemOpen
  \bibfield  {author} {\bibinfo {author} {\bibfnamefont {V.}~\bibnamefont
  {{Icke}}},\ }\href {https://doi.org/10.1093/mnras/206.1.1P} {\bibfield
  {journal} {\bibinfo  {journal} {\mnras}\ }\textbf {\bibinfo {volume} {206}},\
  \bibinfo {pages} {1P} (\bibinfo {year} {1984})}\BibitemShut {NoStop}%
\bibitem [{\citenamefont {{Pisani}}\ \emph {et~al.}(2019)\citenamefont
  {{Pisani}}, \citenamefont {{Massara}}, \citenamefont {{Spergel}} \emph
  {et~al.}}]{Pisani2019}%
  \BibitemOpen
  \bibfield  {author} {\bibinfo {author} {\bibfnamefont {A.}~\bibnamefont
  {{Pisani}}}, \bibinfo {author} {\bibfnamefont {E.}~\bibnamefont {{Massara}}},
  \bibinfo {author} {\bibfnamefont {D.~N.}\ \bibnamefont {{Spergel}}}, \emph
  {et~al.},\ }\href {https://doi.org/10.48550/arXiv.1903.05161} {\bibfield
  {journal} {\bibinfo  {journal} {\baas}\ }\textbf {\bibinfo {volume} {51}},\
  \bibinfo {eid} {40} (\bibinfo {year} {2019})},\ \Eprint
  {https://arxiv.org/abs/1903.05161} {arXiv:1903.05161 [astro-ph.CO]}
  \BibitemShut {NoStop}%
\bibitem [{\citenamefont {{Moresco}}\ \emph {et~al.}(2022)\citenamefont
  {{Moresco}} \emph {et~al.}}]{Moresco2022}%
  \BibitemOpen
  \bibfield  {author} {\bibinfo {author} {\bibfnamefont {M.}~\bibnamefont
  {{Moresco}}} \emph {et~al.},\ }\href
  {https://doi.org/10.1007/s41114-022-00040-z} {\bibfield  {journal} {\bibinfo
  {journal} {Living Reviews in Relativity}\ }\textbf {\bibinfo {volume} {25}},\
  \bibinfo {eid} {6} (\bibinfo {year} {2022})},\ \Eprint
  {https://arxiv.org/abs/2201.07241} {arXiv:2201.07241 [astro-ph.CO]}
  \BibitemShut {NoStop}%
\bibitem [{\citenamefont {{Schuster}}\ \emph {et~al.}(2023)\citenamefont
  {{Schuster}}, \citenamefont {{Hamaus}}, \citenamefont {{Dolag}},\ and\
  \citenamefont {{Weller}}}]{Schuster2023}%
  \BibitemOpen
  \bibfield  {author} {\bibinfo {author} {\bibfnamefont {N.}~\bibnamefont
  {{Schuster}}}, \bibinfo {author} {\bibfnamefont {N.}~\bibnamefont
  {{Hamaus}}}, \bibinfo {author} {\bibfnamefont {K.}~\bibnamefont {{Dolag}}},\
  and\ \bibinfo {author} {\bibfnamefont {J.}~\bibnamefont {{Weller}}},\ }\href
  {https://doi.org/10.1088/1475-7516/2023/05/031} {\bibfield  {journal}
  {\bibinfo  {journal} {\jcap}\ }\textbf {\bibinfo {volume} {2023}},\ \bibinfo
  {eid} {031} (\bibinfo {year} {2023})},\ \Eprint
  {https://arxiv.org/abs/2210.02457} {arXiv:2210.02457 [astro-ph.CO]}
  \BibitemShut {NoStop}%
\bibitem [{\citenamefont {{Massara}}\ \emph {et~al.}(2015)\citenamefont
  {{Massara}}, \citenamefont {{Villaescusa-Navarro}}, \citenamefont {{Viel}},\
  and\ \citenamefont {{Sutter}}}]{Massara2015}%
  \BibitemOpen
  \bibfield  {author} {\bibinfo {author} {\bibfnamefont {E.}~\bibnamefont
  {{Massara}}}, \bibinfo {author} {\bibfnamefont {F.}~\bibnamefont
  {{Villaescusa-Navarro}}}, \bibinfo {author} {\bibfnamefont {M.}~\bibnamefont
  {{Viel}}},\ and\ \bibinfo {author} {\bibfnamefont {P.~M.}\ \bibnamefont
  {{Sutter}}},\ }\href {https://doi.org/10.1088/1475-7516/2015/11/018}
  {\bibfield  {journal} {\bibinfo  {journal} {\jcap}\ }\textbf {\bibinfo
  {volume} {2015}},\ \bibinfo {pages} {018} (\bibinfo {year} {2015})},\ \Eprint
  {https://arxiv.org/abs/1506.03088} {arXiv:1506.03088 [astro-ph.CO]}
  \BibitemShut {NoStop}%
\bibitem [{\citenamefont {{Banerjee}}\ and\ \citenamefont
  {{Dalal}}(2016)}]{Banerjee2016}%
  \BibitemOpen
  \bibfield  {author} {\bibinfo {author} {\bibfnamefont {A.}~\bibnamefont
  {{Banerjee}}}\ and\ \bibinfo {author} {\bibfnamefont {N.}~\bibnamefont
  {{Dalal}}},\ }\href {https://doi.org/10.1088/1475-7516/2016/11/015}
  {\bibfield  {journal} {\bibinfo  {journal} {\jcap}\ }\textbf {\bibinfo
  {volume} {2016}},\ \bibinfo {eid} {015} (\bibinfo {year} {2016})},\ \Eprint
  {https://arxiv.org/abs/1606.06167} {arXiv:1606.06167 [astro-ph.CO]}
  \BibitemShut {NoStop}%
\bibitem [{\citenamefont {{Kreisch}}\ \emph {et~al.}(2019)\citenamefont
  {{Kreisch}}, \citenamefont {{Pisani}}, \citenamefont {{Carbone}},
  \citenamefont {{Liu}}, \citenamefont {{Hawken}}, \citenamefont {{Massara}},
  \citenamefont {{Spergel}},\ and\ \citenamefont {{Wandelt}}}]{Kreisch2019}%
  \BibitemOpen
  \bibfield  {author} {\bibinfo {author} {\bibfnamefont {C.~D.}\ \bibnamefont
  {{Kreisch}}}, \bibinfo {author} {\bibfnamefont {A.}~\bibnamefont {{Pisani}}},
  \bibinfo {author} {\bibfnamefont {C.}~\bibnamefont {{Carbone}}}, \bibinfo
  {author} {\bibfnamefont {J.}~\bibnamefont {{Liu}}}, \bibinfo {author}
  {\bibfnamefont {A.~J.}\ \bibnamefont {{Hawken}}}, \bibinfo {author}
  {\bibfnamefont {E.}~\bibnamefont {{Massara}}}, \bibinfo {author}
  {\bibfnamefont {D.~N.}\ \bibnamefont {{Spergel}}},\ and\ \bibinfo {author}
  {\bibfnamefont {B.~D.}\ \bibnamefont {{Wandelt}}},\ }\href
  {https://doi.org/10.1093/mnras/stz1944} {\bibfield  {journal} {\bibinfo
  {journal} {\mnras}\ }\textbf {\bibinfo {volume} {488}},\ \bibinfo {pages}
  {4413} (\bibinfo {year} {2019})},\ \Eprint {https://arxiv.org/abs/1808.07464}
  {arXiv:1808.07464 [astro-ph.CO]} \BibitemShut {NoStop}%
\bibitem [{\citenamefont {{Schuster}}\ \emph {et~al.}(2019)\citenamefont
  {{Schuster}}, \citenamefont {{Hamaus}}, \citenamefont {{Pisani}},
  \citenamefont {{Carbone}}, \citenamefont {{Kreisch}}, \citenamefont
  {{Pollina}},\ and\ \citenamefont {{Weller}}}]{Schuster2019}%
  \BibitemOpen
  \bibfield  {author} {\bibinfo {author} {\bibfnamefont {N.}~\bibnamefont
  {{Schuster}}}, \bibinfo {author} {\bibfnamefont {N.}~\bibnamefont
  {{Hamaus}}}, \bibinfo {author} {\bibfnamefont {A.}~\bibnamefont {{Pisani}}},
  \bibinfo {author} {\bibfnamefont {C.}~\bibnamefont {{Carbone}}}, \bibinfo
  {author} {\bibfnamefont {C.~D.}\ \bibnamefont {{Kreisch}}}, \bibinfo {author}
  {\bibfnamefont {G.}~\bibnamefont {{Pollina}}},\ and\ \bibinfo {author}
  {\bibfnamefont {J.}~\bibnamefont {{Weller}}},\ }\href
  {https://doi.org/10.1088/1475-7516/2019/12/055} {\bibfield  {journal}
  {\bibinfo  {journal} {\jcap}\ }\textbf {\bibinfo {volume} {2019}},\ \bibinfo
  {eid} {055} (\bibinfo {year} {2019})},\ \Eprint
  {https://arxiv.org/abs/1905.00436} {arXiv:1905.00436 [astro-ph.CO]}
  \BibitemShut {NoStop}%
\bibitem [{\citenamefont {{Contarini}}\ \emph {et~al.}(2021)\citenamefont
  {{Contarini}}, \citenamefont {{Marulli}}, \citenamefont {{Moscardini}},
  \citenamefont {{Veropalumbo}}, \citenamefont {{Giocoli}},\ and\ \citenamefont
  {{Baldi}}}]{Contarini2021}%
  \BibitemOpen
  \bibfield  {author} {\bibinfo {author} {\bibfnamefont {S.}~\bibnamefont
  {{Contarini}}}, \bibinfo {author} {\bibfnamefont {F.}~\bibnamefont
  {{Marulli}}}, \bibinfo {author} {\bibfnamefont {L.}~\bibnamefont
  {{Moscardini}}}, \bibinfo {author} {\bibfnamefont {A.}~\bibnamefont
  {{Veropalumbo}}}, \bibinfo {author} {\bibfnamefont {C.}~\bibnamefont
  {{Giocoli}}},\ and\ \bibinfo {author} {\bibfnamefont {M.}~\bibnamefont
  {{Baldi}}},\ }\href {https://doi.org/10.1093/mnras/stab1112} {\bibfield
  {journal} {\bibinfo  {journal} {\mnras}\ }\textbf {\bibinfo {volume} {504}},\
  \bibinfo {pages} {5021} (\bibinfo {year} {2021})},\ \Eprint
  {https://arxiv.org/abs/2009.03309} {arXiv:2009.03309 [astro-ph.CO]}
  \BibitemShut {NoStop}%
\bibitem [{\citenamefont {{Verza}}\ \emph {et~al.}(2022)\citenamefont
  {{Verza}}, \citenamefont {{Carbone}}, \citenamefont {{Pisani}},\ and\
  \citenamefont {{Renzi}}}]{Verza2022}%
  \BibitemOpen
  \bibfield  {author} {\bibinfo {author} {\bibfnamefont {G.}~\bibnamefont
  {{Verza}}}, \bibinfo {author} {\bibfnamefont {C.}~\bibnamefont {{Carbone}}},
  \bibinfo {author} {\bibfnamefont {A.}~\bibnamefont {{Pisani}}},\ and\
  \bibinfo {author} {\bibfnamefont {A.}~\bibnamefont {{Renzi}}},\ }\href
  {https://doi.org/10.48550/arXiv.2212.09740} {\bibfield  {journal} {\bibinfo
  {journal} {arXiv e-prints}\ ,\ \bibinfo {eid} {arXiv:2212.09740}} (\bibinfo
  {year} {2022})},\ \Eprint {https://arxiv.org/abs/2212.09740}
  {arXiv:2212.09740 [astro-ph.CO]} \BibitemShut {NoStop}%
\bibitem [{\citenamefont {{Sahl{\'e}n}}(2019)}]{Sahlen2019}%
  \BibitemOpen
  \bibfield  {author} {\bibinfo {author} {\bibfnamefont {M.}~\bibnamefont
  {{Sahl{\'e}n}}},\ }\href {https://doi.org/10.1103/PhysRevD.99.063525}
  {\bibfield  {journal} {\bibinfo  {journal} {\prd}\ }\textbf {\bibinfo
  {volume} {99}},\ \bibinfo {eid} {063525} (\bibinfo {year} {2019})},\ \Eprint
  {https://arxiv.org/abs/1807.02470} {arXiv:1807.02470 [astro-ph.CO]}
  \BibitemShut {NoStop}%
\bibitem [{\citenamefont {{Bayer}}\ \emph
  {et~al.}(2021{\natexlab{a}})\citenamefont {{Bayer}}, \citenamefont
  {{Villaescusa-Navarro}}, \citenamefont {{Massara}}, \citenamefont {{Liu}},
  \citenamefont {{Spergel}}, \citenamefont {{Verde}}, \citenamefont
  {{Wandelt}}, \citenamefont {{Viel}},\ and\ \citenamefont
  {{Ho}}}]{Bayer2021a}%
  \BibitemOpen
  \bibfield  {author} {\bibinfo {author} {\bibfnamefont {A.~E.}\ \bibnamefont
  {{Bayer}}}, \bibinfo {author} {\bibfnamefont {F.}~\bibnamefont
  {{Villaescusa-Navarro}}}, \bibinfo {author} {\bibfnamefont {E.}~\bibnamefont
  {{Massara}}}, \bibinfo {author} {\bibfnamefont {J.}~\bibnamefont {{Liu}}},
  \bibinfo {author} {\bibfnamefont {D.~N.}\ \bibnamefont {{Spergel}}}, \bibinfo
  {author} {\bibfnamefont {L.}~\bibnamefont {{Verde}}}, \bibinfo {author}
  {\bibfnamefont {B.~D.}\ \bibnamefont {{Wandelt}}}, \bibinfo {author}
  {\bibfnamefont {M.}~\bibnamefont {{Viel}}},\ and\ \bibinfo {author}
  {\bibfnamefont {S.}~\bibnamefont {{Ho}}},\ }\href
  {https://doi.org/10.3847/1538-4357/ac0e91} {\bibfield  {journal} {\bibinfo
  {journal} {\apj}\ }\textbf {\bibinfo {volume} {919}},\ \bibinfo {eid} {24}
  (\bibinfo {year} {2021}{\natexlab{a}})},\ \Eprint
  {https://arxiv.org/abs/2102.05049} {arXiv:2102.05049 [astro-ph.CO]}
  \BibitemShut {NoStop}%
\bibitem [{\citenamefont {{Kreisch}}\ \emph {et~al.}(2022)\citenamefont
  {{Kreisch}}, \citenamefont {{Pisani}}, \citenamefont {{Villaescusa-Navarro}},
  \citenamefont {{Spergel}}, \citenamefont {{Wandelt}}, \citenamefont
  {{Hamaus}},\ and\ \citenamefont {{Bayer}}}]{Kreisch2022}%
  \BibitemOpen
  \bibfield  {author} {\bibinfo {author} {\bibfnamefont {C.~D.}\ \bibnamefont
  {{Kreisch}}}, \bibinfo {author} {\bibfnamefont {A.}~\bibnamefont {{Pisani}}},
  \bibinfo {author} {\bibfnamefont {F.}~\bibnamefont {{Villaescusa-Navarro}}},
  \bibinfo {author} {\bibfnamefont {D.~N.}\ \bibnamefont {{Spergel}}}, \bibinfo
  {author} {\bibfnamefont {B.~D.}\ \bibnamefont {{Wandelt}}}, \bibinfo {author}
  {\bibfnamefont {N.}~\bibnamefont {{Hamaus}}},\ and\ \bibinfo {author}
  {\bibfnamefont {A.~E.}\ \bibnamefont {{Bayer}}},\ }\href
  {https://doi.org/10.3847/1538-4357/ac7d4b} {\bibfield  {journal} {\bibinfo
  {journal} {\apj}\ }\textbf {\bibinfo {volume} {935}},\ \bibinfo {eid} {100}
  (\bibinfo {year} {2022})},\ \Eprint {https://arxiv.org/abs/2107.02304}
  {arXiv:2107.02304 [astro-ph.CO]} \BibitemShut {NoStop}%
\bibitem [{\citenamefont {{Hotinli}}\ \emph {et~al.}(2023)\citenamefont
  {{Hotinli}}, \citenamefont {{Sabti}}, \citenamefont {{North}},\ and\
  \citenamefont {{Kamionkowski}}}]{Hotinli2023}%
  \BibitemOpen
  \bibfield  {author} {\bibinfo {author} {\bibfnamefont {S.~C.}\ \bibnamefont
  {{Hotinli}}}, \bibinfo {author} {\bibfnamefont {N.}~\bibnamefont {{Sabti}}},
  \bibinfo {author} {\bibfnamefont {J.}~\bibnamefont {{North}}},\ and\ \bibinfo
  {author} {\bibfnamefont {M.}~\bibnamefont {{Kamionkowski}}},\ }\href
  {https://doi.org/10.48550/arXiv.2306.15715} {\bibfield  {journal} {\bibinfo
  {journal} {arXiv e-prints}\ ,\ \bibinfo {eid} {arXiv:2306.15715}} (\bibinfo
  {year} {2023})},\ \Eprint {https://arxiv.org/abs/2306.15715}
  {arXiv:2306.15715 [astro-ph.CO]} \BibitemShut {NoStop}%
\bibitem [{\citenamefont {{Gregory}}\ and\ \citenamefont
  {{Thompson}}(1978)}]{Gregory1978}%
  \BibitemOpen
  \bibfield  {author} {\bibinfo {author} {\bibfnamefont {S.~A.}\ \bibnamefont
  {{Gregory}}}\ and\ \bibinfo {author} {\bibfnamefont {L.~A.}\ \bibnamefont
  {{Thompson}}},\ }\href {https://doi.org/10.1086/156198} {\bibfield  {journal}
  {\bibinfo  {journal} {\apj}\ }\textbf {\bibinfo {volume} {222}},\ \bibinfo
  {pages} {784} (\bibinfo {year} {1978})}\BibitemShut {NoStop}%
\bibitem [{\citenamefont {{J{\~o}eveer}}\ \emph {et~al.}(1978)\citenamefont
  {{J{\~o}eveer}}, \citenamefont {{Einasto}},\ and\ \citenamefont
  {{Tago}}}]{Joeveer1978}%
  \BibitemOpen
  \bibfield  {author} {\bibinfo {author} {\bibfnamefont {M.}~\bibnamefont
  {{J{\~o}eveer}}}, \bibinfo {author} {\bibfnamefont {J.}~\bibnamefont
  {{Einasto}}},\ and\ \bibinfo {author} {\bibfnamefont {E.}~\bibnamefont
  {{Tago}}},\ }\href {https://doi.org/10.1093/mnras/185.2.357} {\bibfield
  {journal} {\bibinfo  {journal} {\mnras}\ }\textbf {\bibinfo {volume} {185}},\
  \bibinfo {pages} {357} (\bibinfo {year} {1978})}\BibitemShut {NoStop}%
\bibitem [{\citenamefont {{Tully}}\ and\ \citenamefont
  {{Fisher}}(1978)}]{Tully1978}%
  \BibitemOpen
  \bibfield  {author} {\bibinfo {author} {\bibfnamefont {R.~B.}\ \bibnamefont
  {{Tully}}}\ and\ \bibinfo {author} {\bibfnamefont {J.~R.}\ \bibnamefont
  {{Fisher}}},\ }in\ \href@noop {} {\emph {\bibinfo {booktitle} {Large Scale
  Structures in the Universe}}},\ Vol.~\bibinfo {volume} {79},\ \bibinfo
  {editor} {edited by\ \bibinfo {editor} {\bibfnamefont {M.~S.}\ \bibnamefont
  {{Longair}}}\ and\ \bibinfo {editor} {\bibfnamefont {J.}~\bibnamefont
  {{Einasto}}}}\ (\bibinfo {year} {1978})\ p.~\bibinfo {pages} {31}\BibitemShut
  {NoStop}%
\bibitem [{\citenamefont {{Kirshner}}\ \emph {et~al.}(1981)\citenamefont
  {{Kirshner}}, \citenamefont {{Oemler}}, \citenamefont {{Schechter}},\ and\
  \citenamefont {{Shectman}}}]{Kirshner1981}%
  \BibitemOpen
  \bibfield  {author} {\bibinfo {author} {\bibfnamefont {R.~P.}\ \bibnamefont
  {{Kirshner}}}, \bibinfo {author} {\bibfnamefont {J.}~\bibnamefont {{Oemler}},
  \bibfnamefont {A.}}, \bibinfo {author} {\bibfnamefont {P.~L.}\ \bibnamefont
  {{Schechter}}},\ and\ \bibinfo {author} {\bibfnamefont {S.~A.}\ \bibnamefont
  {{Shectman}}},\ }\href {https://doi.org/10.1086/183623} {\bibfield  {journal}
  {\bibinfo  {journal} {\apjl}\ }\textbf {\bibinfo {volume} {248}},\ \bibinfo
  {pages} {L57} (\bibinfo {year} {1981})}\BibitemShut {NoStop}%
\bibitem [{\citenamefont {{de Lapparent}}\ \emph {et~al.}(1986)\citenamefont
  {{de Lapparent}}, \citenamefont {{Geller}},\ and\ \citenamefont
  {{Huchra}}}]{deLapparent1986}%
  \BibitemOpen
  \bibfield  {author} {\bibinfo {author} {\bibfnamefont {V.}~\bibnamefont {{de
  Lapparent}}}, \bibinfo {author} {\bibfnamefont {M.~J.}\ \bibnamefont
  {{Geller}}},\ and\ \bibinfo {author} {\bibfnamefont {J.~P.}\ \bibnamefont
  {{Huchra}}},\ }\href {https://doi.org/10.1086/184625} {\bibfield  {journal}
  {\bibinfo  {journal} {\apjl}\ }\textbf {\bibinfo {volume} {302}},\ \bibinfo
  {pages} {L1} (\bibinfo {year} {1986})}\BibitemShut {NoStop}%
\bibitem [{\citenamefont {{Sahl{\'e}n}}\ \emph {et~al.}(2016)\citenamefont
  {{Sahl{\'e}n}}, \citenamefont {{Zubeld{\'\i}a}},\ and\ \citenamefont
  {{Silk}}}]{Sahlen2016}%
  \BibitemOpen
  \bibfield  {author} {\bibinfo {author} {\bibfnamefont {M.}~\bibnamefont
  {{Sahl{\'e}n}}}, \bibinfo {author} {\bibfnamefont {{\'I}.}~\bibnamefont
  {{Zubeld{\'\i}a}}},\ and\ \bibinfo {author} {\bibfnamefont {J.}~\bibnamefont
  {{Silk}}},\ }\href {https://doi.org/10.3847/2041-8205/820/1/L7} {\bibfield
  {journal} {\bibinfo  {journal} {\apjl}\ }\textbf {\bibinfo {volume} {820}},\
  \bibinfo {eid} {L7} (\bibinfo {year} {2016})},\ \Eprint
  {https://arxiv.org/abs/1511.04075} {arXiv:1511.04075 [astro-ph.CO]}
  \BibitemShut {NoStop}%
\bibitem [{\citenamefont {{Hoyle}}\ and\ \citenamefont
  {{Vogeley}}(2004)}]{Hoyle2004}%
  \BibitemOpen
  \bibfield  {author} {\bibinfo {author} {\bibfnamefont {F.}~\bibnamefont
  {{Hoyle}}}\ and\ \bibinfo {author} {\bibfnamefont {M.~S.}\ \bibnamefont
  {{Vogeley}}},\ }\href {https://doi.org/10.1086/386279} {\bibfield  {journal}
  {\bibinfo  {journal} {\apj}\ }\textbf {\bibinfo {volume} {607}},\ \bibinfo
  {pages} {751} (\bibinfo {year} {2004})},\ \Eprint
  {https://arxiv.org/abs/astro-ph/0312533} {arXiv:astro-ph/0312533 [astro-ph]}
  \BibitemShut {NoStop}%
\bibitem [{\citenamefont {{Pan}}\ \emph {et~al.}(2012)\citenamefont {{Pan}},
  \citenamefont {{Vogeley}}, \citenamefont {{Hoyle}}, \citenamefont {{Choi}},\
  and\ \citenamefont {{Park}}}]{Pan2012}%
  \BibitemOpen
  \bibfield  {author} {\bibinfo {author} {\bibfnamefont {D.~C.}\ \bibnamefont
  {{Pan}}}, \bibinfo {author} {\bibfnamefont {M.~S.}\ \bibnamefont
  {{Vogeley}}}, \bibinfo {author} {\bibfnamefont {F.}~\bibnamefont {{Hoyle}}},
  \bibinfo {author} {\bibfnamefont {Y.-Y.}\ \bibnamefont {{Choi}}},\ and\
  \bibinfo {author} {\bibfnamefont {C.}~\bibnamefont {{Park}}},\ }\href
  {https://doi.org/10.1111/j.1365-2966.2011.20197.x} {\bibfield  {journal}
  {\bibinfo  {journal} {\mnras}\ }\textbf {\bibinfo {volume} {421}},\ \bibinfo
  {pages} {926} (\bibinfo {year} {2012})},\ \Eprint
  {https://arxiv.org/abs/1103.4156} {arXiv:1103.4156 [astro-ph.CO]}
  \BibitemShut {NoStop}%
\bibitem [{\citenamefont {{Sutter}}\ \emph
  {et~al.}(2012{\natexlab{a}})\citenamefont {{Sutter}}, \citenamefont
  {{Lavaux}}, \citenamefont {{Wandelt}},\ and\ \citenamefont
  {{Weinberg}}}]{Sutter2012a}%
  \BibitemOpen
  \bibfield  {author} {\bibinfo {author} {\bibfnamefont {P.~M.}\ \bibnamefont
  {{Sutter}}}, \bibinfo {author} {\bibfnamefont {G.}~\bibnamefont {{Lavaux}}},
  \bibinfo {author} {\bibfnamefont {B.~D.}\ \bibnamefont {{Wandelt}}},\ and\
  \bibinfo {author} {\bibfnamefont {D.~H.}\ \bibnamefont {{Weinberg}}},\ }\href
  {https://doi.org/10.1088/0004-637X/761/1/44} {\bibfield  {journal} {\bibinfo
  {journal} {\apj}\ }\textbf {\bibinfo {volume} {761}},\ \bibinfo {eid} {44}
  (\bibinfo {year} {2012}{\natexlab{a}})},\ \Eprint
  {https://arxiv.org/abs/1207.2524} {arXiv:1207.2524 [astro-ph.CO]}
  \BibitemShut {NoStop}%
\bibitem [{\citenamefont {{Sutter}}\ \emph
  {et~al.}(2014{\natexlab{a}})\citenamefont {{Sutter}}, \citenamefont
  {{Lavaux}}, \citenamefont {{Wandelt}}, \citenamefont {{Weinberg}},
  \citenamefont {{Warren}},\ and\ \citenamefont {{Pisani}}}]{Sutter2014b}%
  \BibitemOpen
  \bibfield  {author} {\bibinfo {author} {\bibfnamefont {P.~M.}\ \bibnamefont
  {{Sutter}}}, \bibinfo {author} {\bibfnamefont {G.}~\bibnamefont {{Lavaux}}},
  \bibinfo {author} {\bibfnamefont {B.~D.}\ \bibnamefont {{Wandelt}}}, \bibinfo
  {author} {\bibfnamefont {D.~H.}\ \bibnamefont {{Weinberg}}}, \bibinfo
  {author} {\bibfnamefont {M.~S.}\ \bibnamefont {{Warren}}},\ and\ \bibinfo
  {author} {\bibfnamefont {A.}~\bibnamefont {{Pisani}}},\ }\href
  {https://doi.org/10.1093/mnras/stu1094} {\bibfield  {journal} {\bibinfo
  {journal} {\mnras}\ }\textbf {\bibinfo {volume} {442}},\ \bibinfo {pages}
  {3127} (\bibinfo {year} {2014}{\natexlab{a}})},\ \Eprint
  {https://arxiv.org/abs/1310.7155} {arXiv:1310.7155 [astro-ph.CO]}
  \BibitemShut {NoStop}%
\bibitem [{\citenamefont {{Nadathur}}(2016)}]{Nadathur2016}%
  \BibitemOpen
  \bibfield  {author} {\bibinfo {author} {\bibfnamefont {S.}~\bibnamefont
  {{Nadathur}}},\ }\href {https://doi.org/10.1093/mnras/stw1340} {\bibfield
  {journal} {\bibinfo  {journal} {\mnras}\ }\textbf {\bibinfo {volume} {461}},\
  \bibinfo {pages} {358} (\bibinfo {year} {2016})},\ \Eprint
  {https://arxiv.org/abs/1602.04752} {arXiv:1602.04752 [astro-ph.CO]}
  \BibitemShut {NoStop}%
\bibitem [{\citenamefont {{Mao}}\ \emph
  {et~al.}(2017{\natexlab{a}})\citenamefont {{Mao}}, \citenamefont {{Berlind}},
  \citenamefont {{Scherrer}}, \citenamefont {{Neyrinck}}, \citenamefont
  {{Scoccimarro}}, \citenamefont {{Tinker}}, \citenamefont {{McBride}},
  \citenamefont {{Schneider}}, \citenamefont {{Pan}}, \citenamefont
  {{Bizyaev}}, \citenamefont {{Malanushenko}},\ and\ \citenamefont
  {{Malanushenko}}}]{Mao2017a}%
  \BibitemOpen
  \bibfield  {author} {\bibinfo {author} {\bibfnamefont {Q.}~\bibnamefont
  {{Mao}}}, \bibinfo {author} {\bibfnamefont {A.~A.}\ \bibnamefont
  {{Berlind}}}, \bibinfo {author} {\bibfnamefont {R.~J.}\ \bibnamefont
  {{Scherrer}}}, \bibinfo {author} {\bibfnamefont {M.~C.}\ \bibnamefont
  {{Neyrinck}}}, \bibinfo {author} {\bibfnamefont {R.}~\bibnamefont
  {{Scoccimarro}}}, \bibinfo {author} {\bibfnamefont {J.~L.}\ \bibnamefont
  {{Tinker}}}, \bibinfo {author} {\bibfnamefont {C.~K.}\ \bibnamefont
  {{McBride}}}, \bibinfo {author} {\bibfnamefont {D.~P.}\ \bibnamefont
  {{Schneider}}}, \bibinfo {author} {\bibfnamefont {K.}~\bibnamefont {{Pan}}},
  \bibinfo {author} {\bibfnamefont {D.}~\bibnamefont {{Bizyaev}}}, \bibinfo
  {author} {\bibfnamefont {E.}~\bibnamefont {{Malanushenko}}},\ and\ \bibinfo
  {author} {\bibfnamefont {V.}~\bibnamefont {{Malanushenko}}},\ }\href
  {https://doi.org/10.3847/1538-4357/835/2/161} {\bibfield  {journal} {\bibinfo
   {journal} {\apj}\ }\textbf {\bibinfo {volume} {835}},\ \bibinfo {eid} {161}
  (\bibinfo {year} {2017}{\natexlab{a}})},\ \Eprint
  {https://arxiv.org/abs/1602.02771} {arXiv:1602.02771 [astro-ph.CO]}
  \BibitemShut {NoStop}%
\bibitem [{\citenamefont {{Sutter}}\ \emph
  {et~al.}(2012{\natexlab{b}})\citenamefont {{Sutter}}, \citenamefont
  {{Lavaux}}, \citenamefont {{Wandelt}},\ and\ \citenamefont
  {{Weinberg}}}]{Sutter2012b}%
  \BibitemOpen
  \bibfield  {author} {\bibinfo {author} {\bibfnamefont {P.~M.}\ \bibnamefont
  {{Sutter}}}, \bibinfo {author} {\bibfnamefont {G.}~\bibnamefont {{Lavaux}}},
  \bibinfo {author} {\bibfnamefont {B.~D.}\ \bibnamefont {{Wandelt}}},\ and\
  \bibinfo {author} {\bibfnamefont {D.~H.}\ \bibnamefont {{Weinberg}}},\ }\href
  {https://doi.org/10.1088/0004-637X/761/2/187} {\bibfield  {journal} {\bibinfo
   {journal} {\apj}\ }\textbf {\bibinfo {volume} {761}},\ \bibinfo {eid} {187}
  (\bibinfo {year} {2012}{\natexlab{b}})},\ \Eprint
  {https://arxiv.org/abs/1208.1058} {arXiv:1208.1058 [astro-ph.CO]}
  \BibitemShut {NoStop}%
\bibitem [{\citenamefont {{Sutter}}\ \emph
  {et~al.}(2014{\natexlab{b}})\citenamefont {{Sutter}}, \citenamefont
  {{Pisani}}, \citenamefont {{Wandelt}},\ and\ \citenamefont
  {{Weinberg}}}]{Sutter2014a}%
  \BibitemOpen
  \bibfield  {author} {\bibinfo {author} {\bibfnamefont {P.~M.}\ \bibnamefont
  {{Sutter}}}, \bibinfo {author} {\bibfnamefont {A.}~\bibnamefont {{Pisani}}},
  \bibinfo {author} {\bibfnamefont {B.~D.}\ \bibnamefont {{Wandelt}}},\ and\
  \bibinfo {author} {\bibfnamefont {D.~H.}\ \bibnamefont {{Weinberg}}},\ }\href
  {https://doi.org/10.1093/mnras/stu1392} {\bibfield  {journal} {\bibinfo
  {journal} {\mnras}\ }\textbf {\bibinfo {volume} {443}},\ \bibinfo {pages}
  {2983} (\bibinfo {year} {2014}{\natexlab{b}})},\ \Eprint
  {https://arxiv.org/abs/1404.5618} {arXiv:1404.5618 [astro-ph.CO]}
  \BibitemShut {NoStop}%
\bibitem [{\citenamefont {{Hamaus}}\ \emph {et~al.}(2016)\citenamefont
  {{Hamaus}}, \citenamefont {{Pisani}}, \citenamefont {{Sutter}}, \citenamefont
  {{Lavaux}}, \citenamefont {{Escoffier}}, \citenamefont {{Wandelt}},\ and\
  \citenamefont {{Weller}}}]{Hamaus2016}%
  \BibitemOpen
  \bibfield  {author} {\bibinfo {author} {\bibfnamefont {N.}~\bibnamefont
  {{Hamaus}}}, \bibinfo {author} {\bibfnamefont {A.}~\bibnamefont {{Pisani}}},
  \bibinfo {author} {\bibfnamefont {P.~M.}\ \bibnamefont {{Sutter}}}, \bibinfo
  {author} {\bibfnamefont {G.}~\bibnamefont {{Lavaux}}}, \bibinfo {author}
  {\bibfnamefont {S.}~\bibnamefont {{Escoffier}}}, \bibinfo {author}
  {\bibfnamefont {B.~D.}\ \bibnamefont {{Wandelt}}},\ and\ \bibinfo {author}
  {\bibfnamefont {J.}~\bibnamefont {{Weller}}},\ }\href
  {https://doi.org/10.1103/PhysRevLett.117.091302} {\bibfield  {journal}
  {\bibinfo  {journal} {\prl}\ }\textbf {\bibinfo {volume} {117}},\ \bibinfo
  {eid} {091302} (\bibinfo {year} {2016})},\ \Eprint
  {https://arxiv.org/abs/1602.01784} {arXiv:1602.01784 [astro-ph.CO]}
  \BibitemShut {NoStop}%
\bibitem [{\citenamefont {{Hamaus}}\ \emph {et~al.}(2017)\citenamefont
  {{Hamaus}}, \citenamefont {{Cousinou}}, \citenamefont {{Pisani}},
  \citenamefont {{Aubert}}, \citenamefont {{Escoffier}},\ and\ \citenamefont
  {{Weller}}}]{Hamaus2017}%
  \BibitemOpen
  \bibfield  {author} {\bibinfo {author} {\bibfnamefont {N.}~\bibnamefont
  {{Hamaus}}}, \bibinfo {author} {\bibfnamefont {M.-C.}\ \bibnamefont
  {{Cousinou}}}, \bibinfo {author} {\bibfnamefont {A.}~\bibnamefont
  {{Pisani}}}, \bibinfo {author} {\bibfnamefont {M.}~\bibnamefont {{Aubert}}},
  \bibinfo {author} {\bibfnamefont {S.}~\bibnamefont {{Escoffier}}},\ and\
  \bibinfo {author} {\bibfnamefont {J.}~\bibnamefont {{Weller}}},\ }\href
  {https://doi.org/10.1088/1475-7516/2017/07/014} {\bibfield  {journal}
  {\bibinfo  {journal} {\jcap}\ }\textbf {\bibinfo {volume} {2017}},\ \bibinfo
  {eid} {014} (\bibinfo {year} {2017})},\ \Eprint
  {https://arxiv.org/abs/1705.05328} {arXiv:1705.05328 [astro-ph.CO]}
  \BibitemShut {NoStop}%
\bibitem [{\citenamefont {{Mao}}\ \emph
  {et~al.}(2017{\natexlab{b}})\citenamefont {{Mao}}, \citenamefont {{Berlind}},
  \citenamefont {{Scherrer}}, \citenamefont {{Neyrinck}}, \citenamefont
  {{Scoccimarro}}, \citenamefont {{Tinker}}, \citenamefont {{McBride}},\ and\
  \citenamefont {{Schneider}}}]{Mao2017b}%
  \BibitemOpen
  \bibfield  {author} {\bibinfo {author} {\bibfnamefont {Q.}~\bibnamefont
  {{Mao}}}, \bibinfo {author} {\bibfnamefont {A.~A.}\ \bibnamefont
  {{Berlind}}}, \bibinfo {author} {\bibfnamefont {R.~J.}\ \bibnamefont
  {{Scherrer}}}, \bibinfo {author} {\bibfnamefont {M.~C.}\ \bibnamefont
  {{Neyrinck}}}, \bibinfo {author} {\bibfnamefont {R.}~\bibnamefont
  {{Scoccimarro}}}, \bibinfo {author} {\bibfnamefont {J.~L.}\ \bibnamefont
  {{Tinker}}}, \bibinfo {author} {\bibfnamefont {C.~K.}\ \bibnamefont
  {{McBride}}},\ and\ \bibinfo {author} {\bibfnamefont {D.~P.}\ \bibnamefont
  {{Schneider}}},\ }\href {https://doi.org/10.3847/1538-4357/835/2/160}
  {\bibfield  {journal} {\bibinfo  {journal} {\apj}\ }\textbf {\bibinfo
  {volume} {835}},\ \bibinfo {eid} {160} (\bibinfo {year}
  {2017}{\natexlab{b}})},\ \Eprint {https://arxiv.org/abs/1602.06306}
  {arXiv:1602.06306 [astro-ph.CO]} \BibitemShut {NoStop}%
\bibitem [{\citenamefont {{Hamaus}}\ \emph {et~al.}(2020)\citenamefont
  {{Hamaus}}, \citenamefont {{Pisani}}, \citenamefont {{Choi}}, \citenamefont
  {{Lavaux}}, \citenamefont {{Wandelt}},\ and\ \citenamefont
  {{Weller}}}]{Hamaus2020}%
  \BibitemOpen
  \bibfield  {author} {\bibinfo {author} {\bibfnamefont {N.}~\bibnamefont
  {{Hamaus}}}, \bibinfo {author} {\bibfnamefont {A.}~\bibnamefont {{Pisani}}},
  \bibinfo {author} {\bibfnamefont {J.-A.}\ \bibnamefont {{Choi}}}, \bibinfo
  {author} {\bibfnamefont {G.}~\bibnamefont {{Lavaux}}}, \bibinfo {author}
  {\bibfnamefont {B.~D.}\ \bibnamefont {{Wandelt}}},\ and\ \bibinfo {author}
  {\bibfnamefont {J.}~\bibnamefont {{Weller}}},\ }\href
  {https://doi.org/10.1088/1475-7516/2020/12/023} {\bibfield  {journal}
  {\bibinfo  {journal} {\jcap}\ }\textbf {\bibinfo {volume} {2020}},\ \bibinfo
  {eid} {023} (\bibinfo {year} {2020})},\ \Eprint
  {https://arxiv.org/abs/2007.07895} {arXiv:2007.07895 [astro-ph.CO]}
  \BibitemShut {NoStop}%
\bibitem [{\citenamefont {{Nadathur}}\ \emph {et~al.}(2020)\citenamefont
  {{Nadathur}}, \citenamefont {{Woodfinden}}, \citenamefont {{Percival}} \emph
  {et~al.}}]{Nadathur2020}%
  \BibitemOpen
  \bibfield  {author} {\bibinfo {author} {\bibfnamefont {S.}~\bibnamefont
  {{Nadathur}}}, \bibinfo {author} {\bibfnamefont {A.}~\bibnamefont
  {{Woodfinden}}}, \bibinfo {author} {\bibfnamefont {W.~J.}\ \bibnamefont
  {{Percival}}}, \emph {et~al.},\ }\href
  {https://doi.org/10.1093/mnras/staa3074} {\bibfield  {journal} {\bibinfo
  {journal} {\mnras}\ }\textbf {\bibinfo {volume} {499}},\ \bibinfo {pages}
  {4140} (\bibinfo {year} {2020})},\ \Eprint {https://arxiv.org/abs/2008.06060}
  {arXiv:2008.06060 [astro-ph.CO]} \BibitemShut {NoStop}%
\bibitem [{\citenamefont {{Aubert}}\ \emph {et~al.}(2022)\citenamefont
  {{Aubert}}, \citenamefont {{Cousinou}}, \citenamefont {{Escoffier}},
  \citenamefont {{Hawken}}, \citenamefont {{Nadathur}} \emph
  {et~al.}}]{Aubert2022}%
  \BibitemOpen
  \bibfield  {author} {\bibinfo {author} {\bibfnamefont {M.}~\bibnamefont
  {{Aubert}}}, \bibinfo {author} {\bibfnamefont {M.-C.}\ \bibnamefont
  {{Cousinou}}}, \bibinfo {author} {\bibfnamefont {S.}~\bibnamefont
  {{Escoffier}}}, \bibinfo {author} {\bibfnamefont {A.~J.}\ \bibnamefont
  {{Hawken}}}, \bibinfo {author} {\bibfnamefont {S.}~\bibnamefont
  {{Nadathur}}}, \emph {et~al.},\ }\href
  {https://doi.org/10.1093/mnras/stac828} {\bibfield  {journal} {\bibinfo
  {journal} {\mnras}\ }\textbf {\bibinfo {volume} {513}},\ \bibinfo {pages}
  {186} (\bibinfo {year} {2022})},\ \Eprint {https://arxiv.org/abs/2007.09013}
  {arXiv:2007.09013 [astro-ph.CO]} \BibitemShut {NoStop}%
\bibitem [{\citenamefont {{Woodfinden}}\ \emph {et~al.}(2023)\citenamefont
  {{Woodfinden}}, \citenamefont {{Percival}}, \citenamefont {{Nadathur}},
  \citenamefont {{Winther}}, \citenamefont {{Fraser}}, \citenamefont
  {{Massara}}, \citenamefont {{Paillas}},\ and\ \citenamefont
  {{Radinovic}}}]{Woodfinden2023}%
  \BibitemOpen
  \bibfield  {author} {\bibinfo {author} {\bibfnamefont {A.}~\bibnamefont
  {{Woodfinden}}}, \bibinfo {author} {\bibfnamefont {W.~J.}\ \bibnamefont
  {{Percival}}}, \bibinfo {author} {\bibfnamefont {S.}~\bibnamefont
  {{Nadathur}}}, \bibinfo {author} {\bibfnamefont {H.~A.}\ \bibnamefont
  {{Winther}}}, \bibinfo {author} {\bibfnamefont {T.~S.}\ \bibnamefont
  {{Fraser}}}, \bibinfo {author} {\bibfnamefont {E.}~\bibnamefont {{Massara}}},
  \bibinfo {author} {\bibfnamefont {E.}~\bibnamefont {{Paillas}}},\ and\
  \bibinfo {author} {\bibfnamefont {S.}~\bibnamefont {{Radinovic}}},\
  }\bibfield  {journal} {\bibinfo  {journal} {\mnras}\ }\href
  {https://doi.org/10.1093/mnras/stad1725} {10.1093/mnras/stad1725} (\bibinfo
  {year} {2023}),\ \Eprint {https://arxiv.org/abs/2303.06143} {arXiv:2303.06143
  [astro-ph.CO]} \BibitemShut {NoStop}%
\bibitem [{\citenamefont {{Contarini}}\ \emph
  {et~al.}(2022{\natexlab{a}})\citenamefont {{Contarini}}, \citenamefont
  {{Pisani}}, \citenamefont {{Hamaus}}, \citenamefont {{Marulli}},
  \citenamefont {{Moscardini}},\ and\ \citenamefont
  {{Baldi}}}]{Contarini2022a}%
  \BibitemOpen
  \bibfield  {author} {\bibinfo {author} {\bibfnamefont {S.}~\bibnamefont
  {{Contarini}}}, \bibinfo {author} {\bibfnamefont {A.}~\bibnamefont
  {{Pisani}}}, \bibinfo {author} {\bibfnamefont {N.}~\bibnamefont {{Hamaus}}},
  \bibinfo {author} {\bibfnamefont {F.}~\bibnamefont {{Marulli}}}, \bibinfo
  {author} {\bibfnamefont {L.}~\bibnamefont {{Moscardini}}},\ and\ \bibinfo
  {author} {\bibfnamefont {M.}~\bibnamefont {{Baldi}}},\ }\href
  {https://doi.org/10.48550/arXiv.2212.03873} {\bibfield  {journal} {\bibinfo
  {journal} {arXiv e-prints}\ ,\ \bibinfo {eid} {arXiv:2212.03873}} (\bibinfo
  {year} {2022}{\natexlab{a}})},\ \Eprint {https://arxiv.org/abs/2212.03873}
  {arXiv:2212.03873 [astro-ph.CO]} \BibitemShut {NoStop}%
\bibitem [{\citenamefont {{Contarini}}\ \emph
  {et~al.}(2022{\natexlab{b}})\citenamefont {{Contarini}}, \citenamefont
  {{Pisani}}, \citenamefont {{Hamaus}}, \citenamefont {{Marulli}},
  \citenamefont {{Moscardini}},\ and\ \citenamefont
  {{Baldi}}}]{Contarini2022b}%
  \BibitemOpen
  \bibfield  {author} {\bibinfo {author} {\bibfnamefont {S.}~\bibnamefont
  {{Contarini}}}, \bibinfo {author} {\bibfnamefont {A.}~\bibnamefont
  {{Pisani}}}, \bibinfo {author} {\bibfnamefont {N.}~\bibnamefont {{Hamaus}}},
  \bibinfo {author} {\bibfnamefont {F.}~\bibnamefont {{Marulli}}}, \bibinfo
  {author} {\bibfnamefont {L.}~\bibnamefont {{Moscardini}}},\ and\ \bibinfo
  {author} {\bibfnamefont {M.}~\bibnamefont {{Baldi}}},\ }\href
  {https://doi.org/10.48550/arXiv.2212.07438} {\bibfield  {journal} {\bibinfo
  {journal} {arXiv e-prints}\ ,\ \bibinfo {eid} {arXiv:2212.07438}} (\bibinfo
  {year} {2022}{\natexlab{b}})},\ \Eprint {https://arxiv.org/abs/2212.07438}
  {arXiv:2212.07438 [astro-ph.CO]} \BibitemShut {NoStop}%
\bibitem [{\citenamefont {{SDSS Collaboration}}\ \emph
  {et~al.}(2000)\citenamefont {{SDSS Collaboration}}, \citenamefont {{York}}
  \emph {et~al.}}]{York2000}%
  \BibitemOpen
  \bibfield  {author} {\bibinfo {author} {\bibnamefont {{SDSS Collaboration}}},
  \bibinfo {author} {\bibfnamefont {D.~G.}\ \bibnamefont {{York}}}, \emph
  {et~al.},\ }\href {https://doi.org/10.1086/301513} {\bibfield  {journal}
  {\bibinfo  {journal} {\aj}\ }\textbf {\bibinfo {volume} {120}},\ \bibinfo
  {pages} {1579} (\bibinfo {year} {2000})},\ \Eprint
  {https://arxiv.org/abs/astro-ph/0006396} {arXiv:astro-ph/0006396 [astro-ph]}
  \BibitemShut {NoStop}%
\bibitem [{\citenamefont {{SDSS Collaboration}}\ \emph
  {et~al.}(2011)\citenamefont {{SDSS Collaboration}}, \citenamefont
  {{Eisenstein}}, \citenamefont {{Weinberg}} \emph {et~al.}}]{Eisenstein2011}%
  \BibitemOpen
  \bibfield  {author} {\bibinfo {author} {\bibnamefont {{SDSS Collaboration}}},
  \bibinfo {author} {\bibfnamefont {D.~J.}\ \bibnamefont {{Eisenstein}}},
  \bibinfo {author} {\bibfnamefont {D.~H.}\ \bibnamefont {{Weinberg}}}, \emph
  {et~al.},\ }\href {https://doi.org/10.1088/0004-6256/142/3/72} {\bibfield
  {journal} {\bibinfo  {journal} {\aj}\ }\textbf {\bibinfo {volume} {142}},\
  \bibinfo {eid} {72} (\bibinfo {year} {2011})},\ \Eprint
  {https://arxiv.org/abs/1101.1529} {arXiv:1101.1529 [astro-ph.IM]}
  \BibitemShut {NoStop}%
\bibitem [{\citenamefont {{SDSS Collaboration}}\ \emph
  {et~al.}(2013)\citenamefont {{SDSS Collaboration}}, \citenamefont {{Dawson}},
  \citenamefont {{Schlegel}} \emph {et~al.}}]{Dawson2013}%
  \BibitemOpen
  \bibfield  {author} {\bibinfo {author} {\bibnamefont {{SDSS Collaboration}}},
  \bibinfo {author} {\bibfnamefont {K.~S.}\ \bibnamefont {{Dawson}}}, \bibinfo
  {author} {\bibfnamefont {D.~J.}\ \bibnamefont {{Schlegel}}}, \emph {et~al.},\
  }\href {https://doi.org/10.1088/0004-6256/145/1/10} {\bibfield  {journal}
  {\bibinfo  {journal} {\aj}\ }\textbf {\bibinfo {volume} {145}},\ \bibinfo
  {eid} {10} (\bibinfo {year} {2013})},\ \Eprint
  {https://arxiv.org/abs/1208.0022} {arXiv:1208.0022 [astro-ph.CO]}
  \BibitemShut {NoStop}%
\bibitem [{\citenamefont {{Ivanov}}\ \emph
  {et~al.}(2020{\natexlab{a}})\citenamefont {{Ivanov}}, \citenamefont
  {{Simonovi{\'c}}},\ and\ \citenamefont {{Zaldarriaga}}}]{Ivanov2020b}%
  \BibitemOpen
  \bibfield  {author} {\bibinfo {author} {\bibfnamefont {M.~M.}\ \bibnamefont
  {{Ivanov}}}, \bibinfo {author} {\bibfnamefont {M.}~\bibnamefont
  {{Simonovi{\'c}}}},\ and\ \bibinfo {author} {\bibfnamefont {M.}~\bibnamefont
  {{Zaldarriaga}}},\ }\href {https://doi.org/10.1103/PhysRevD.101.083504}
  {\bibfield  {journal} {\bibinfo  {journal} {\prd}\ }\textbf {\bibinfo
  {volume} {101}},\ \bibinfo {eid} {083504} (\bibinfo {year}
  {2020}{\natexlab{a}})},\ \Eprint {https://arxiv.org/abs/1912.08208}
  {arXiv:1912.08208 [astro-ph.CO]} \BibitemShut {NoStop}%
\bibitem [{\citenamefont {{Semenaite}}\ \emph {et~al.}(2023)\citenamefont
  {{Semenaite}}, \citenamefont {{S{\'a}nchez}}, \citenamefont {{Pezzotta}},
  \citenamefont {{Hou}}, \citenamefont {{Eggemeier}}, \citenamefont {{Crocce}},
  \citenamefont {{Zhao}}, \citenamefont {{Brownstein}}, \citenamefont
  {{Rossi}},\ and\ \citenamefont {{Schneider}}}]{Semenaite2023}%
  \BibitemOpen
  \bibfield  {author} {\bibinfo {author} {\bibfnamefont {A.}~\bibnamefont
  {{Semenaite}}}, \bibinfo {author} {\bibfnamefont {A.~G.}\ \bibnamefont
  {{S{\'a}nchez}}}, \bibinfo {author} {\bibfnamefont {A.}~\bibnamefont
  {{Pezzotta}}}, \bibinfo {author} {\bibfnamefont {J.}~\bibnamefont {{Hou}}},
  \bibinfo {author} {\bibfnamefont {A.}~\bibnamefont {{Eggemeier}}}, \bibinfo
  {author} {\bibfnamefont {M.}~\bibnamefont {{Crocce}}}, \bibinfo {author}
  {\bibfnamefont {C.}~\bibnamefont {{Zhao}}}, \bibinfo {author} {\bibfnamefont
  {J.~R.}\ \bibnamefont {{Brownstein}}}, \bibinfo {author} {\bibfnamefont
  {G.}~\bibnamefont {{Rossi}}},\ and\ \bibinfo {author} {\bibfnamefont {D.~P.}\
  \bibnamefont {{Schneider}}},\ }\href {https://doi.org/10.1093/mnras/stad849}
  {\bibfield  {journal} {\bibinfo  {journal} {\mnras}\ }\textbf {\bibinfo
  {volume} {521}},\ \bibinfo {pages} {5013} (\bibinfo {year} {2023})},\ \Eprint
  {https://arxiv.org/abs/2210.07304} {arXiv:2210.07304 [astro-ph.CO]}
  \BibitemShut {NoStop}%
\bibitem [{\citenamefont {{Stopyra}}\ \emph {et~al.}(2021)\citenamefont
  {{Stopyra}}, \citenamefont {{Peiris}},\ and\ \citenamefont
  {{Pontzen}}}]{Stopyra2021}%
  \BibitemOpen
  \bibfield  {author} {\bibinfo {author} {\bibfnamefont {S.}~\bibnamefont
  {{Stopyra}}}, \bibinfo {author} {\bibfnamefont {H.~V.}\ \bibnamefont
  {{Peiris}}},\ and\ \bibinfo {author} {\bibfnamefont {A.}~\bibnamefont
  {{Pontzen}}},\ }\href {https://doi.org/10.1093/mnras/staa3587} {\bibfield
  {journal} {\bibinfo  {journal} {\mnras}\ }\textbf {\bibinfo {volume} {500}},\
  \bibinfo {pages} {4173} (\bibinfo {year} {2021})},\ \Eprint
  {https://arxiv.org/abs/2007.14395} {arXiv:2007.14395 [astro-ph.CO]}
  \BibitemShut {NoStop}%
\bibitem [{\citenamefont {{Press}}\ and\ \citenamefont
  {{Schechter}}(1974)}]{Press1974}%
  \BibitemOpen
  \bibfield  {author} {\bibinfo {author} {\bibfnamefont {W.~H.}\ \bibnamefont
  {{Press}}}\ and\ \bibinfo {author} {\bibfnamefont {P.}~\bibnamefont
  {{Schechter}}},\ }\href {https://doi.org/10.1086/152650} {\bibfield
  {journal} {\bibinfo  {journal} {\apj}\ }\textbf {\bibinfo {volume} {187}},\
  \bibinfo {pages} {425} (\bibinfo {year} {1974})}\BibitemShut {NoStop}%
\bibitem [{\citenamefont {{Bond}}\ \emph {et~al.}(1991)\citenamefont {{Bond}},
  \citenamefont {{Cole}}, \citenamefont {{Efstathiou}},\ and\ \citenamefont
  {{Kaiser}}}]{Bond1991}%
  \BibitemOpen
  \bibfield  {author} {\bibinfo {author} {\bibfnamefont {J.~R.}\ \bibnamefont
  {{Bond}}}, \bibinfo {author} {\bibfnamefont {S.}~\bibnamefont {{Cole}}},
  \bibinfo {author} {\bibfnamefont {G.}~\bibnamefont {{Efstathiou}}},\ and\
  \bibinfo {author} {\bibfnamefont {N.}~\bibnamefont {{Kaiser}}},\ }\href
  {https://doi.org/10.1086/170520} {\bibfield  {journal} {\bibinfo  {journal}
  {\apj}\ }\textbf {\bibinfo {volume} {379}},\ \bibinfo {pages} {440} (\bibinfo
  {year} {1991})}\BibitemShut {NoStop}%
\bibitem [{\citenamefont {{Sheth}}\ and\ \citenamefont
  {{Tormen}}(1999)}]{Sheth1999}%
  \BibitemOpen
  \bibfield  {author} {\bibinfo {author} {\bibfnamefont {R.~K.}\ \bibnamefont
  {{Sheth}}}\ and\ \bibinfo {author} {\bibfnamefont {G.}~\bibnamefont
  {{Tormen}}},\ }\href {https://doi.org/10.1046/j.1365-8711.1999.02692.x}
  {\bibfield  {journal} {\bibinfo  {journal} {\mnras}\ }\textbf {\bibinfo
  {volume} {308}},\ \bibinfo {pages} {119} (\bibinfo {year} {1999})},\ \Eprint
  {https://arxiv.org/abs/astro-ph/9901122} {arXiv:astro-ph/9901122 [astro-ph]}
  \BibitemShut {NoStop}%
\bibitem [{\citenamefont {{Sheth}}\ and\ \citenamefont {{van de
  Weygaert}}(2004)}]{Sheth2004}%
  \BibitemOpen
  \bibfield  {author} {\bibinfo {author} {\bibfnamefont {R.~K.}\ \bibnamefont
  {{Sheth}}}\ and\ \bibinfo {author} {\bibfnamefont {R.}~\bibnamefont {{van de
  Weygaert}}},\ }\href {https://doi.org/10.1111/j.1365-2966.2004.07661.x}
  {\bibfield  {journal} {\bibinfo  {journal} {\mnras}\ }\textbf {\bibinfo
  {volume} {350}},\ \bibinfo {pages} {517} (\bibinfo {year} {2004})},\ \Eprint
  {https://arxiv.org/abs/astro-ph/0311260} {arXiv:astro-ph/0311260 [astro-ph]}
  \BibitemShut {NoStop}%
\bibitem [{\citenamefont {{Paranjape}}\ \emph {et~al.}(2012)\citenamefont
  {{Paranjape}}, \citenamefont {{Lam}},\ and\ \citenamefont
  {{Sheth}}}]{Paranjape2012}%
  \BibitemOpen
  \bibfield  {author} {\bibinfo {author} {\bibfnamefont {A.}~\bibnamefont
  {{Paranjape}}}, \bibinfo {author} {\bibfnamefont {T.~Y.}\ \bibnamefont
  {{Lam}}},\ and\ \bibinfo {author} {\bibfnamefont {R.~K.}\ \bibnamefont
  {{Sheth}}},\ }\href {https://doi.org/10.1111/j.1365-2966.2011.20128.x}
  {\bibfield  {journal} {\bibinfo  {journal} {\mnras}\ }\textbf {\bibinfo
  {volume} {420}},\ \bibinfo {pages} {1429} (\bibinfo {year} {2012})},\ \Eprint
  {https://arxiv.org/abs/1105.1990} {arXiv:1105.1990 [astro-ph.CO]}
  \BibitemShut {NoStop}%
\bibitem [{\citenamefont {{Paranjape}}\ \emph {et~al.}(2013)\citenamefont
  {{Paranjape}}, \citenamefont {{Sheth}},\ and\ \citenamefont
  {{Desjacques}}}]{Paranjape2013}%
  \BibitemOpen
  \bibfield  {author} {\bibinfo {author} {\bibfnamefont {A.}~\bibnamefont
  {{Paranjape}}}, \bibinfo {author} {\bibfnamefont {R.~K.}\ \bibnamefont
  {{Sheth}}},\ and\ \bibinfo {author} {\bibfnamefont {V.}~\bibnamefont
  {{Desjacques}}},\ }\href {https://doi.org/10.1093/mnras/stt267} {\bibfield
  {journal} {\bibinfo  {journal} {\mnras}\ }\textbf {\bibinfo {volume} {431}},\
  \bibinfo {pages} {1503} (\bibinfo {year} {2013})},\ \Eprint
  {https://arxiv.org/abs/1210.1483} {arXiv:1210.1483 [astro-ph.CO]}
  \BibitemShut {NoStop}%
\bibitem [{\citenamefont {{Jennings}}\ \emph {et~al.}(2013)\citenamefont
  {{Jennings}}, \citenamefont {{Li}},\ and\ \citenamefont
  {{Hu}}}]{Jennings2013}%
  \BibitemOpen
  \bibfield  {author} {\bibinfo {author} {\bibfnamefont {E.}~\bibnamefont
  {{Jennings}}}, \bibinfo {author} {\bibfnamefont {Y.}~\bibnamefont {{Li}}},\
  and\ \bibinfo {author} {\bibfnamefont {W.}~\bibnamefont {{Hu}}},\ }\href
  {https://doi.org/10.1093/mnras/stt1169} {\bibfield  {journal} {\bibinfo
  {journal} {\mnras}\ }\textbf {\bibinfo {volume} {434}},\ \bibinfo {pages}
  {2167} (\bibinfo {year} {2013})},\ \Eprint {https://arxiv.org/abs/1304.6087}
  {arXiv:1304.6087 [astro-ph.CO]} \BibitemShut {NoStop}%
\bibitem [{\citenamefont {{Pisani}}\ \emph
  {et~al.}(2015{\natexlab{a}})\citenamefont {{Pisani}}, \citenamefont
  {{Sutter}}, \citenamefont {{Hamaus}}, \citenamefont {{Alizadeh}},
  \citenamefont {{Biswas}}, \citenamefont {{Wandelt}},\ and\ \citenamefont
  {{Hirata}}}]{Pisani2015a}%
  \BibitemOpen
  \bibfield  {author} {\bibinfo {author} {\bibfnamefont {A.}~\bibnamefont
  {{Pisani}}}, \bibinfo {author} {\bibfnamefont {P.~M.}\ \bibnamefont
  {{Sutter}}}, \bibinfo {author} {\bibfnamefont {N.}~\bibnamefont {{Hamaus}}},
  \bibinfo {author} {\bibfnamefont {E.}~\bibnamefont {{Alizadeh}}}, \bibinfo
  {author} {\bibfnamefont {R.}~\bibnamefont {{Biswas}}}, \bibinfo {author}
  {\bibfnamefont {B.~D.}\ \bibnamefont {{Wandelt}}},\ and\ \bibinfo {author}
  {\bibfnamefont {C.~M.}\ \bibnamefont {{Hirata}}},\ }\href
  {https://doi.org/10.1103/PhysRevD.92.083531} {\bibfield  {journal} {\bibinfo
  {journal} {\prd}\ }\textbf {\bibinfo {volume} {92}},\ \bibinfo {eid} {083531}
  (\bibinfo {year} {2015}{\natexlab{a}})},\ \Eprint
  {https://arxiv.org/abs/1503.07690} {arXiv:1503.07690 [astro-ph.CO]}
  \BibitemShut {NoStop}%
\bibitem [{\citenamefont {{Ronconi}}\ and\ \citenamefont
  {{Marulli}}(2017)}]{Ronconi2017}%
  \BibitemOpen
  \bibfield  {author} {\bibinfo {author} {\bibfnamefont {T.}~\bibnamefont
  {{Ronconi}}}\ and\ \bibinfo {author} {\bibfnamefont {F.}~\bibnamefont
  {{Marulli}}},\ }\href {https://doi.org/10.1051/0004-6361/201730852}
  {\bibfield  {journal} {\bibinfo  {journal} {\aap}\ }\textbf {\bibinfo
  {volume} {607}},\ \bibinfo {eid} {A24} (\bibinfo {year} {2017})},\ \Eprint
  {https://arxiv.org/abs/1703.07848} {arXiv:1703.07848 [astro-ph.CO]}
  \BibitemShut {NoStop}%
\bibitem [{\citenamefont {{Ronconi}}\ \emph {et~al.}(2019)\citenamefont
  {{Ronconi}}, \citenamefont {{Contarini}}, \citenamefont {{Marulli}},
  \citenamefont {{Baldi}},\ and\ \citenamefont {{Moscardini}}}]{Ronconi2019}%
  \BibitemOpen
  \bibfield  {author} {\bibinfo {author} {\bibfnamefont {T.}~\bibnamefont
  {{Ronconi}}}, \bibinfo {author} {\bibfnamefont {S.}~\bibnamefont
  {{Contarini}}}, \bibinfo {author} {\bibfnamefont {F.}~\bibnamefont
  {{Marulli}}}, \bibinfo {author} {\bibfnamefont {M.}~\bibnamefont {{Baldi}}},\
  and\ \bibinfo {author} {\bibfnamefont {L.}~\bibnamefont {{Moscardini}}},\
  }\href {https://doi.org/10.1093/mnras/stz2115} {\bibfield  {journal}
  {\bibinfo  {journal} {\mnras}\ }\textbf {\bibinfo {volume} {488}},\ \bibinfo
  {pages} {5075} (\bibinfo {year} {2019})},\ \Eprint
  {https://arxiv.org/abs/1902.04585} {arXiv:1902.04585 [astro-ph.CO]}
  \BibitemShut {NoStop}%
\bibitem [{\citenamefont {{Verza}}\ \emph {et~al.}(2019)\citenamefont
  {{Verza}}, \citenamefont {{Pisani}}, \citenamefont {{Carbone}}, \citenamefont
  {{Hamaus}},\ and\ \citenamefont {{Guzzo}}}]{Verza2019}%
  \BibitemOpen
  \bibfield  {author} {\bibinfo {author} {\bibfnamefont {G.}~\bibnamefont
  {{Verza}}}, \bibinfo {author} {\bibfnamefont {A.}~\bibnamefont {{Pisani}}},
  \bibinfo {author} {\bibfnamefont {C.}~\bibnamefont {{Carbone}}}, \bibinfo
  {author} {\bibfnamefont {N.}~\bibnamefont {{Hamaus}}},\ and\ \bibinfo
  {author} {\bibfnamefont {L.}~\bibnamefont {{Guzzo}}},\ }\href
  {https://doi.org/10.1088/1475-7516/2019/12/040} {\bibfield  {journal}
  {\bibinfo  {journal} {\jcap}\ }\textbf {\bibinfo {volume} {2019}},\ \bibinfo
  {eid} {040} (\bibinfo {year} {2019})},\ \Eprint
  {https://arxiv.org/abs/1906.00409} {arXiv:1906.00409 [astro-ph.CO]}
  \BibitemShut {NoStop}%
\bibitem [{\citenamefont {{Contarini}}\ \emph {et~al.}(2019)\citenamefont
  {{Contarini}}, \citenamefont {{Ronconi}}, \citenamefont {{Marulli}},
  \citenamefont {{Moscardini}}, \citenamefont {{Veropalumbo}},\ and\
  \citenamefont {{Baldi}}}]{Contarini2019}%
  \BibitemOpen
  \bibfield  {author} {\bibinfo {author} {\bibfnamefont {S.}~\bibnamefont
  {{Contarini}}}, \bibinfo {author} {\bibfnamefont {T.}~\bibnamefont
  {{Ronconi}}}, \bibinfo {author} {\bibfnamefont {F.}~\bibnamefont
  {{Marulli}}}, \bibinfo {author} {\bibfnamefont {L.}~\bibnamefont
  {{Moscardini}}}, \bibinfo {author} {\bibfnamefont {A.}~\bibnamefont
  {{Veropalumbo}}},\ and\ \bibinfo {author} {\bibfnamefont {M.}~\bibnamefont
  {{Baldi}}},\ }\href {https://doi.org/10.1093/mnras/stz1989} {\bibfield
  {journal} {\bibinfo  {journal} {\mnras}\ }\textbf {\bibinfo {volume} {488}},\
  \bibinfo {pages} {3526} (\bibinfo {year} {2019})},\ \Eprint
  {https://arxiv.org/abs/1904.01022} {arXiv:1904.01022 [astro-ph.CO]}
  \BibitemShut {NoStop}%
\bibitem [{\citenamefont {{Contarini}}\ \emph
  {et~al.}(2022{\natexlab{c}})\citenamefont {{Contarini}}, \citenamefont
  {{Verza}}, \citenamefont {{Pisani}}, \citenamefont {{Hamaus}}, \citenamefont
  {{Sahl{\'e}n}} \emph {et~al.}}]{Contarini2022c}%
  \BibitemOpen
  \bibfield  {author} {\bibinfo {author} {\bibfnamefont {S.}~\bibnamefont
  {{Contarini}}}, \bibinfo {author} {\bibfnamefont {G.}~\bibnamefont
  {{Verza}}}, \bibinfo {author} {\bibfnamefont {A.}~\bibnamefont {{Pisani}}},
  \bibinfo {author} {\bibfnamefont {N.}~\bibnamefont {{Hamaus}}}, \bibinfo
  {author} {\bibfnamefont {M.}~\bibnamefont {{Sahl{\'e}n}}}, \emph {et~al.},\
  }\href {https://doi.org/10.1051/0004-6361/202244095} {\bibfield  {journal}
  {\bibinfo  {journal} {\aap}\ }\textbf {\bibinfo {volume} {667}},\ \bibinfo
  {eid} {A162} (\bibinfo {year} {2022}{\natexlab{c}})},\ \Eprint
  {https://arxiv.org/abs/2205.11525} {arXiv:2205.11525 [astro-ph.CO]}
  \BibitemShut {NoStop}%
\bibitem [{\citenamefont {{Pelliciari}}\ \emph {et~al.}(2023)\citenamefont
  {{Pelliciari}}, \citenamefont {{Contarini}}, \citenamefont {{Marulli}},
  \citenamefont {{Moscardini}}, \citenamefont {{Giocoli}}, \citenamefont
  {{Lesci}},\ and\ \citenamefont {{Dolag}}}]{Pelliciari2023}%
  \BibitemOpen
  \bibfield  {author} {\bibinfo {author} {\bibfnamefont {D.}~\bibnamefont
  {{Pelliciari}}}, \bibinfo {author} {\bibfnamefont {S.}~\bibnamefont
  {{Contarini}}}, \bibinfo {author} {\bibfnamefont {F.}~\bibnamefont
  {{Marulli}}}, \bibinfo {author} {\bibfnamefont {L.}~\bibnamefont
  {{Moscardini}}}, \bibinfo {author} {\bibfnamefont {C.}~\bibnamefont
  {{Giocoli}}}, \bibinfo {author} {\bibfnamefont {G.~F.}\ \bibnamefont
  {{Lesci}}},\ and\ \bibinfo {author} {\bibfnamefont {K.}~\bibnamefont
  {{Dolag}}},\ }\href {https://doi.org/10.1093/mnras/stad956} {\bibfield
  {journal} {\bibinfo  {journal} {\mnras}\ }\textbf {\bibinfo {volume} {522}},\
  \bibinfo {pages} {152} (\bibinfo {year} {2023})},\ \Eprint
  {https://arxiv.org/abs/2210.07248} {arXiv:2210.07248 [astro-ph.CO]}
  \BibitemShut {NoStop}%
\bibitem [{\citenamefont {{Alcock}}\ and\ \citenamefont
  {{Paczynski}}(1979)}]{Alcock1979}%
  \BibitemOpen
  \bibfield  {author} {\bibinfo {author} {\bibfnamefont {C.}~\bibnamefont
  {{Alcock}}}\ and\ \bibinfo {author} {\bibfnamefont {B.}~\bibnamefont
  {{Paczynski}}},\ }\href {https://doi.org/10.1038/281358a0} {\bibfield
  {journal} {\bibinfo  {journal} {\nat}\ }\textbf {\bibinfo {volume} {281}},\
  \bibinfo {pages} {358} (\bibinfo {year} {1979})}\BibitemShut {NoStop}%
\bibitem [{\citenamefont {{Lavaux}}\ and\ \citenamefont
  {{Wandelt}}(2012)}]{Lavaux2012}%
  \BibitemOpen
  \bibfield  {author} {\bibinfo {author} {\bibfnamefont {G.}~\bibnamefont
  {{Lavaux}}}\ and\ \bibinfo {author} {\bibfnamefont {B.~D.}\ \bibnamefont
  {{Wandelt}}},\ }\href {https://doi.org/10.1088/0004-637X/754/2/109}
  {\bibfield  {journal} {\bibinfo  {journal} {\apj}\ }\textbf {\bibinfo
  {volume} {754}},\ \bibinfo {eid} {109} (\bibinfo {year} {2012})},\ \Eprint
  {https://arxiv.org/abs/1110.0345} {arXiv:1110.0345 [astro-ph.CO]}
  \BibitemShut {NoStop}%
\bibitem [{\citenamefont {{Pisani}}\ \emph {et~al.}(2014)\citenamefont
  {{Pisani}}, \citenamefont {{Lavaux}}, \citenamefont {{Sutter}},\ and\
  \citenamefont {{Wandelt}}}]{Pisani2014}%
  \BibitemOpen
  \bibfield  {author} {\bibinfo {author} {\bibfnamefont {A.}~\bibnamefont
  {{Pisani}}}, \bibinfo {author} {\bibfnamefont {G.}~\bibnamefont {{Lavaux}}},
  \bibinfo {author} {\bibfnamefont {P.~M.}\ \bibnamefont {{Sutter}}},\ and\
  \bibinfo {author} {\bibfnamefont {B.~D.}\ \bibnamefont {{Wandelt}}},\ }\href
  {https://doi.org/10.1093/mnras/stu1399} {\bibfield  {journal} {\bibinfo
  {journal} {\mnras}\ }\textbf {\bibinfo {volume} {443}},\ \bibinfo {pages}
  {3238} (\bibinfo {year} {2014})},\ \Eprint {https://arxiv.org/abs/1306.3052}
  {arXiv:1306.3052 [astro-ph.CO]} \BibitemShut {NoStop}%
\bibitem [{\citenamefont {{Padilla}}\ \emph {et~al.}(2005)\citenamefont
  {{Padilla}}, \citenamefont {{Ceccarelli}},\ and\ \citenamefont
  {{Lambas}}}]{Padilla2005}%
  \BibitemOpen
  \bibfield  {author} {\bibinfo {author} {\bibfnamefont {N.~D.}\ \bibnamefont
  {{Padilla}}}, \bibinfo {author} {\bibfnamefont {L.}~\bibnamefont
  {{Ceccarelli}}},\ and\ \bibinfo {author} {\bibfnamefont {D.~G.}\ \bibnamefont
  {{Lambas}}},\ }\href {https://doi.org/10.1111/j.1365-2966.2005.09500.x}
  {\bibfield  {journal} {\bibinfo  {journal} {\mnras}\ }\textbf {\bibinfo
  {volume} {363}},\ \bibinfo {pages} {977} (\bibinfo {year} {2005})},\ \Eprint
  {https://arxiv.org/abs/astro-ph/0508297} {arXiv:astro-ph/0508297 [astro-ph]}
  \BibitemShut {NoStop}%
\bibitem [{\citenamefont {{Paz}}\ \emph {et~al.}(2013)\citenamefont {{Paz}},
  \citenamefont {{Lares}}, \citenamefont {{Ceccarelli}}, \citenamefont
  {{Padilla}},\ and\ \citenamefont {{Lambas}}}]{Paz2013}%
  \BibitemOpen
  \bibfield  {author} {\bibinfo {author} {\bibfnamefont {D.}~\bibnamefont
  {{Paz}}}, \bibinfo {author} {\bibfnamefont {M.}~\bibnamefont {{Lares}}},
  \bibinfo {author} {\bibfnamefont {L.}~\bibnamefont {{Ceccarelli}}}, \bibinfo
  {author} {\bibfnamefont {N.}~\bibnamefont {{Padilla}}},\ and\ \bibinfo
  {author} {\bibfnamefont {D.~G.}\ \bibnamefont {{Lambas}}},\ }\href
  {https://doi.org/10.1093/mnras/stt1836} {\bibfield  {journal} {\bibinfo
  {journal} {\mnras}\ }\textbf {\bibinfo {volume} {436}},\ \bibinfo {pages}
  {3480} (\bibinfo {year} {2013})},\ \Eprint {https://arxiv.org/abs/1306.5799}
  {arXiv:1306.5799 [astro-ph.CO]} \BibitemShut {NoStop}%
\bibitem [{\citenamefont {{Hamaus}}\ \emph
  {et~al.}(2014{\natexlab{a}})\citenamefont {{Hamaus}}, \citenamefont
  {{Sutter}},\ and\ \citenamefont {{Wandelt}}}]{Hamaus2014a}%
  \BibitemOpen
  \bibfield  {author} {\bibinfo {author} {\bibfnamefont {N.}~\bibnamefont
  {{Hamaus}}}, \bibinfo {author} {\bibfnamefont {P.~M.}\ \bibnamefont
  {{Sutter}}},\ and\ \bibinfo {author} {\bibfnamefont {B.~D.}\ \bibnamefont
  {{Wandelt}}},\ }\href {https://doi.org/10.48550/arXiv.1409.7621} {\bibfield
  {journal} {\bibinfo  {journal} {arXiv e-prints}\ ,\ \bibinfo {eid}
  {arXiv:1409.7621}} (\bibinfo {year} {2014}{\natexlab{a}})},\ \Eprint
  {https://arxiv.org/abs/1409.7621} {arXiv:1409.7621 [astro-ph.CO]}
  \BibitemShut {NoStop}%
\bibitem [{\citenamefont {{Hamaus}}\ \emph
  {et~al.}(2014{\natexlab{b}})\citenamefont {{Hamaus}}, \citenamefont
  {{Sutter}},\ and\ \citenamefont {{Wandelt}}}]{Hamaus2014b}%
  \BibitemOpen
  \bibfield  {author} {\bibinfo {author} {\bibfnamefont {N.}~\bibnamefont
  {{Hamaus}}}, \bibinfo {author} {\bibfnamefont {P.~M.}\ \bibnamefont
  {{Sutter}}},\ and\ \bibinfo {author} {\bibfnamefont {B.~D.}\ \bibnamefont
  {{Wandelt}}},\ }\href {https://doi.org/10.1103/PhysRevLett.112.251302}
  {\bibfield  {journal} {\bibinfo  {journal} {\prl}\ }\textbf {\bibinfo
  {volume} {112}},\ \bibinfo {eid} {251302} (\bibinfo {year}
  {2014}{\natexlab{b}})},\ \Eprint {https://arxiv.org/abs/1403.5499}
  {arXiv:1403.5499 [astro-ph.CO]} \BibitemShut {NoStop}%
\bibitem [{\citenamefont {{Massara}}\ and\ \citenamefont
  {{Sheth}}(2018)}]{Massara2018}%
  \BibitemOpen
  \bibfield  {author} {\bibinfo {author} {\bibfnamefont {E.}~\bibnamefont
  {{Massara}}}\ and\ \bibinfo {author} {\bibfnamefont {R.~K.}\ \bibnamefont
  {{Sheth}}},\ }\href {https://doi.org/10.48550/arXiv.1811.03132} {\bibfield
  {journal} {\bibinfo  {journal} {arXiv e-prints}\ ,\ \bibinfo {eid}
  {arXiv:1811.03132}} (\bibinfo {year} {2018})},\ \Eprint
  {https://arxiv.org/abs/1811.03132} {arXiv:1811.03132 [astro-ph.CO]}
  \BibitemShut {NoStop}%
\bibitem [{\citenamefont {{Feng}}\ \emph {et~al.}(2016)\citenamefont {{Feng}},
  \citenamefont {{Chu}}, \citenamefont {{Seljak}},\ and\ \citenamefont
  {{McDonald}}}]{Feng2016}%
  \BibitemOpen
  \bibfield  {author} {\bibinfo {author} {\bibfnamefont {Y.}~\bibnamefont
  {{Feng}}}, \bibinfo {author} {\bibfnamefont {M.-Y.}\ \bibnamefont {{Chu}}},
  \bibinfo {author} {\bibfnamefont {U.}~\bibnamefont {{Seljak}}},\ and\
  \bibinfo {author} {\bibfnamefont {P.}~\bibnamefont {{McDonald}}},\ }\href
  {https://doi.org/10.1093/mnras/stw2123} {\bibfield  {journal} {\bibinfo
  {journal} {\mnras}\ }\textbf {\bibinfo {volume} {463}},\ \bibinfo {pages}
  {2273} (\bibinfo {year} {2016})},\ \Eprint {https://arxiv.org/abs/1603.00476}
  {arXiv:1603.00476 [astro-ph.CO]} \BibitemShut {NoStop}%
\bibitem [{\citenamefont {{Bayer}}\ \emph
  {et~al.}(2021{\natexlab{b}})\citenamefont {{Bayer}}, \citenamefont
  {{Banerjee}},\ and\ \citenamefont {{Feng}}}]{Bayer2021b}%
  \BibitemOpen
  \bibfield  {author} {\bibinfo {author} {\bibfnamefont {A.~E.}\ \bibnamefont
  {{Bayer}}}, \bibinfo {author} {\bibfnamefont {A.}~\bibnamefont
  {{Banerjee}}},\ and\ \bibinfo {author} {\bibfnamefont {Y.}~\bibnamefont
  {{Feng}}},\ }\href {https://doi.org/10.1088/1475-7516/2021/01/016} {\bibfield
   {journal} {\bibinfo  {journal} {\jcap}\ }\textbf {\bibinfo {volume}
  {2021}},\ \bibinfo {eid} {016} (\bibinfo {year} {2021}{\natexlab{b}})},\
  \Eprint {https://arxiv.org/abs/2007.13394} {arXiv:2007.13394 [astro-ph.CO]}
  \BibitemShut {NoStop}%
\bibitem [{\citenamefont {{Berlind}}\ and\ \citenamefont
  {{Weinberg}}(2002)}]{Berlind2002}%
  \BibitemOpen
  \bibfield  {author} {\bibinfo {author} {\bibfnamefont {A.~A.}\ \bibnamefont
  {{Berlind}}}\ and\ \bibinfo {author} {\bibfnamefont {D.~H.}\ \bibnamefont
  {{Weinberg}}},\ }\href {https://doi.org/10.1086/341469} {\bibfield  {journal}
  {\bibinfo  {journal} {\apj}\ }\textbf {\bibinfo {volume} {575}},\ \bibinfo
  {pages} {587} (\bibinfo {year} {2002})},\ \Eprint
  {https://arxiv.org/abs/astro-ph/0109001} {arXiv:astro-ph/0109001 [astro-ph]}
  \BibitemShut {NoStop}%
\bibitem [{\citenamefont {{Cooray}}\ and\ \citenamefont
  {{Sheth}}(2002)}]{Cooray2002}%
  \BibitemOpen
  \bibfield  {author} {\bibinfo {author} {\bibfnamefont {A.}~\bibnamefont
  {{Cooray}}}\ and\ \bibinfo {author} {\bibfnamefont {R.}~\bibnamefont
  {{Sheth}}},\ }\href {https://doi.org/10.1016/S0370-1573(02)00276-4}
  {\bibfield  {journal} {\bibinfo  {journal} {\physrep}\ }\textbf {\bibinfo
  {volume} {372}},\ \bibinfo {pages} {1} (\bibinfo {year} {2002})},\ \Eprint
  {https://arxiv.org/abs/astro-ph/0206508} {arXiv:astro-ph/0206508 [astro-ph]}
  \BibitemShut {NoStop}%
\bibitem [{\citenamefont {{Wechsler}}\ and\ \citenamefont
  {{Tinker}}(2018)}]{Wechsler2018}%
  \BibitemOpen
  \bibfield  {author} {\bibinfo {author} {\bibfnamefont {R.~H.}\ \bibnamefont
  {{Wechsler}}}\ and\ \bibinfo {author} {\bibfnamefont {J.~L.}\ \bibnamefont
  {{Tinker}}},\ }\href {https://doi.org/10.1146/annurev-astro-081817-051756}
  {\bibfield  {journal} {\bibinfo  {journal} {\araa}\ }\textbf {\bibinfo
  {volume} {56}},\ \bibinfo {pages} {435} (\bibinfo {year} {2018})},\ \Eprint
  {https://arxiv.org/abs/1804.03097} {arXiv:1804.03097 [astro-ph.GA]}
  \BibitemShut {NoStop}%
\bibitem [{\citenamefont {{Cranmer}}\ \emph {et~al.}(2020)\citenamefont
  {{Cranmer}}, \citenamefont {{Brehmer}},\ and\ \citenamefont
  {{Louppe}}}]{Cranmer2020}%
  \BibitemOpen
  \bibfield  {author} {\bibinfo {author} {\bibfnamefont {K.}~\bibnamefont
  {{Cranmer}}}, \bibinfo {author} {\bibfnamefont {J.}~\bibnamefont
  {{Brehmer}}},\ and\ \bibinfo {author} {\bibfnamefont {G.}~\bibnamefont
  {{Louppe}}},\ }\href {https://doi.org/10.1073/pnas.1912789117} {\bibfield
  {journal} {\bibinfo  {journal} {Proceedings of the National Academy of
  Science}\ }\textbf {\bibinfo {volume} {117}},\ \bibinfo {pages} {30055}
  (\bibinfo {year} {2020})},\ \Eprint {https://arxiv.org/abs/1911.01429}
  {arXiv:1911.01429 [stat.ML]} \BibitemShut {NoStop}%
\bibitem [{\citenamefont {{Hahn}}\ \emph {et~al.}(2022)\citenamefont {{Hahn}},
  \citenamefont {{Eickenberg}}, \citenamefont {{Ho}}, \citenamefont {{Hou}},
  \citenamefont {{Lemos}}, \citenamefont {{Massara}}, \citenamefont {{Modi}},
  \citenamefont {{Moradinezhad Dizgah}}, \citenamefont {{R{\'e}galdo-Saint
  Blancard}},\ and\ \citenamefont {{Abidi}}}]{Hahn2022}%
  \BibitemOpen
  \bibfield  {author} {\bibinfo {author} {\bibfnamefont {C.}~\bibnamefont
  {{Hahn}}}, \bibinfo {author} {\bibfnamefont {M.}~\bibnamefont
  {{Eickenberg}}}, \bibinfo {author} {\bibfnamefont {S.}~\bibnamefont {{Ho}}},
  \bibinfo {author} {\bibfnamefont {J.}~\bibnamefont {{Hou}}}, \bibinfo
  {author} {\bibfnamefont {P.}~\bibnamefont {{Lemos}}}, \bibinfo {author}
  {\bibfnamefont {E.}~\bibnamefont {{Massara}}}, \bibinfo {author}
  {\bibfnamefont {C.}~\bibnamefont {{Modi}}}, \bibinfo {author} {\bibfnamefont
  {A.}~\bibnamefont {{Moradinezhad Dizgah}}}, \bibinfo {author} {\bibfnamefont
  {B.}~\bibnamefont {{R{\'e}galdo-Saint Blancard}}},\ and\ \bibinfo {author}
  {\bibfnamefont {M.~M.}\ \bibnamefont {{Abidi}}},\ }\href
  {https://doi.org/10.48550/arXiv.2211.00723} {\bibfield  {journal} {\bibinfo
  {journal} {arXiv e-prints}\ ,\ \bibinfo {eid} {arXiv:2211.00723}} (\bibinfo
  {year} {2022})},\ \Eprint {https://arxiv.org/abs/2211.00723}
  {arXiv:2211.00723 [astro-ph.CO]} \BibitemShut {NoStop}%
\bibitem [{\citenamefont {{Hahn}}\ \emph {et~al.}(2023)\citenamefont {{Hahn}},
  \citenamefont {{Eickenberg}}, \citenamefont {{Ho}}, \citenamefont {{Hou}},
  \citenamefont {{Lemos}}, \citenamefont {{Massara}}, \citenamefont {{Modi}},
  \citenamefont {{Moradinezhad Dizgah}}, \citenamefont {{R{\'e}galdo-Saint
  Blancard}},\ and\ \citenamefont {{Abidi}}}]{Hahn2023}%
  \BibitemOpen
  \bibfield  {author} {\bibinfo {author} {\bibfnamefont {C.}~\bibnamefont
  {{Hahn}}}, \bibinfo {author} {\bibfnamefont {M.}~\bibnamefont
  {{Eickenberg}}}, \bibinfo {author} {\bibfnamefont {S.}~\bibnamefont {{Ho}}},
  \bibinfo {author} {\bibfnamefont {J.}~\bibnamefont {{Hou}}}, \bibinfo
  {author} {\bibfnamefont {P.}~\bibnamefont {{Lemos}}}, \bibinfo {author}
  {\bibfnamefont {E.}~\bibnamefont {{Massara}}}, \bibinfo {author}
  {\bibfnamefont {C.}~\bibnamefont {{Modi}}}, \bibinfo {author} {\bibfnamefont
  {A.}~\bibnamefont {{Moradinezhad Dizgah}}}, \bibinfo {author} {\bibfnamefont
  {B.}~\bibnamefont {{R{\'e}galdo-Saint Blancard}}},\ and\ \bibinfo {author}
  {\bibfnamefont {M.~M.}\ \bibnamefont {{Abidi}}},\ }\href
  {https://doi.org/10.1088/1475-7516/2023/04/010} {\bibfield  {journal}
  {\bibinfo  {journal} {\jcap}\ }\textbf {\bibinfo {volume} {2023}},\ \bibinfo
  {eid} {010} (\bibinfo {year} {2023})},\ \Eprint
  {https://arxiv.org/abs/2211.00660} {arXiv:2211.00660 [astro-ph.CO]}
  \BibitemShut {NoStop}%
\bibitem [{\citenamefont {{Philcox}}\ and\ \citenamefont
  {{Ivanov}}(2022)}]{Philcox2022}%
  \BibitemOpen
  \bibfield  {author} {\bibinfo {author} {\bibfnamefont {O.~H.~E.}\
  \bibnamefont {{Philcox}}}\ and\ \bibinfo {author} {\bibfnamefont {M.~M.}\
  \bibnamefont {{Ivanov}}},\ }\href
  {https://doi.org/10.1103/PhysRevD.105.043517} {\bibfield  {journal} {\bibinfo
   {journal} {\prd}\ }\textbf {\bibinfo {volume} {105}},\ \bibinfo {eid}
  {043517} (\bibinfo {year} {2022})},\ \Eprint
  {https://arxiv.org/abs/2112.04515} {arXiv:2112.04515 [astro-ph.CO]}
  \BibitemShut {NoStop}%
\bibitem [{\citenamefont {Lewis}\ \emph {et~al.}(2000)\citenamefont {Lewis},
  \citenamefont {Challinor},\ and\ \citenamefont {Lasenby}}]{Lewis:1999bs}%
  \BibitemOpen
  \bibfield  {author} {\bibinfo {author} {\bibfnamefont {A.}~\bibnamefont
  {Lewis}}, \bibinfo {author} {\bibfnamefont {A.}~\bibnamefont {Challinor}},\
  and\ \bibinfo {author} {\bibfnamefont {A.}~\bibnamefont {Lasenby}},\ }\href
  {https://doi.org/10.1086/309179} {\bibfield  {journal} {\bibinfo  {journal}
  {\apj}\ }\textbf {\bibinfo {volume} {538}},\ \bibinfo {pages} {473} (\bibinfo
  {year} {2000})},\ \Eprint {https://arxiv.org/abs/astro-ph/9911177}
  {arXiv:astro-ph/9911177 [astro-ph]} \BibitemShut {NoStop}%
\bibitem [{\citenamefont {Howlett}\ \emph {et~al.}(2012)\citenamefont
  {Howlett}, \citenamefont {Lewis}, \citenamefont {Hall},\ and\ \citenamefont
  {Challinor}}]{Howlett:2012mh}%
  \BibitemOpen
  \bibfield  {author} {\bibinfo {author} {\bibfnamefont {C.}~\bibnamefont
  {Howlett}}, \bibinfo {author} {\bibfnamefont {A.}~\bibnamefont {Lewis}},
  \bibinfo {author} {\bibfnamefont {A.}~\bibnamefont {Hall}},\ and\ \bibinfo
  {author} {\bibfnamefont {A.}~\bibnamefont {Challinor}},\ }\href
  {https://doi.org/10.1088/1475-7516/2012/04/027} {\bibfield  {journal}
  {\bibinfo  {journal} {\jcap}\ }\textbf {\bibinfo {volume} {1204}},\ \bibinfo
  {pages} {027} (\bibinfo {year} {2012})},\ \Eprint
  {https://arxiv.org/abs/1201.3654} {arXiv:1201.3654 [astro-ph.CO]}
  \BibitemShut {NoStop}%
\bibitem [{\citenamefont {{Blas}}\ \emph {et~al.}(2011)\citenamefont {{Blas}},
  \citenamefont {{Lesgourgues}},\ and\ \citenamefont {{Tram}}}]{Blas2011}%
  \BibitemOpen
  \bibfield  {author} {\bibinfo {author} {\bibfnamefont {D.}~\bibnamefont
  {{Blas}}}, \bibinfo {author} {\bibfnamefont {J.}~\bibnamefont
  {{Lesgourgues}}},\ and\ \bibinfo {author} {\bibfnamefont {T.}~\bibnamefont
  {{Tram}}},\ }\href {https://doi.org/10.1088/1475-7516/2011/07/034} {\bibfield
   {journal} {\bibinfo  {journal} {\jcap}\ }\textbf {\bibinfo {volume}
  {2011}},\ \bibinfo {eid} {034} (\bibinfo {year} {2011})},\ \Eprint
  {https://arxiv.org/abs/1104.2933} {arXiv:1104.2933 [astro-ph.CO]}
  \BibitemShut {NoStop}%
\bibitem [{\citenamefont {{Zennaro}}\ \emph {et~al.}(2016)\citenamefont
  {{Zennaro}}, \citenamefont {{Bel}}, \citenamefont {{Villaescusa-Navarro}},
  \citenamefont {{Carbone}}, \citenamefont {{Sefusatti}},\ and\ \citenamefont
  {{Guzzo}}}]{Zennaro2016}%
  \BibitemOpen
  \bibfield  {author} {\bibinfo {author} {\bibfnamefont {M.}~\bibnamefont
  {{Zennaro}}}, \bibinfo {author} {\bibfnamefont {J.}~\bibnamefont {{Bel}}},
  \bibinfo {author} {\bibfnamefont {F.}~\bibnamefont {{Villaescusa-Navarro}}},
  \bibinfo {author} {\bibfnamefont {C.}~\bibnamefont {{Carbone}}}, \bibinfo
  {author} {\bibfnamefont {E.}~\bibnamefont {{Sefusatti}}},\ and\ \bibinfo
  {author} {\bibfnamefont {L.}~\bibnamefont {{Guzzo}}},\ }\href@noop {}
  {\bibinfo {title} {{REPS: REscaled Power Spectra for initial conditions with
  massive neutrinos}}},\ \bibinfo {howpublished} {Astrophysics Source Code
  Library, record ascl:1612.022} (\bibinfo {year} {2016}),\ \Eprint
  {https://arxiv.org/abs/1612.022} {ascl:1612.022} \BibitemShut {NoStop}%
\bibitem [{\citenamefont {{Zennaro}}\ \emph {et~al.}(2017)\citenamefont
  {{Zennaro}}, \citenamefont {{Bel}}, \citenamefont {{Villaescusa-Navarro}},
  \citenamefont {{Carbone}}, \citenamefont {{Sefusatti}},\ and\ \citenamefont
  {{Guzzo}}}]{Zennaro2017}%
  \BibitemOpen
  \bibfield  {author} {\bibinfo {author} {\bibfnamefont {M.}~\bibnamefont
  {{Zennaro}}}, \bibinfo {author} {\bibfnamefont {J.}~\bibnamefont {{Bel}}},
  \bibinfo {author} {\bibfnamefont {F.}~\bibnamefont {{Villaescusa-Navarro}}},
  \bibinfo {author} {\bibfnamefont {C.}~\bibnamefont {{Carbone}}}, \bibinfo
  {author} {\bibfnamefont {E.}~\bibnamefont {{Sefusatti}}},\ and\ \bibinfo
  {author} {\bibfnamefont {L.}~\bibnamefont {{Guzzo}}},\ }\href
  {https://doi.org/10.1093/mnras/stw3340} {\bibfield  {journal} {\bibinfo
  {journal} {\mnras}\ }\textbf {\bibinfo {volume} {466}},\ \bibinfo {pages}
  {3244} (\bibinfo {year} {2017})},\ \Eprint {https://arxiv.org/abs/1605.05283}
  {arXiv:1605.05283 [astro-ph.CO]} \BibitemShut {NoStop}%
\bibitem [{\citenamefont {{Villaescusa-Navarro}}\ \emph
  {et~al.}(2020)\citenamefont {{Villaescusa-Navarro}}, \citenamefont {{Hahn}},
  \citenamefont {{Massara}}, \citenamefont {{Banerjee}}, \citenamefont
  {{Delgado}} \emph {et~al.}}]{Quijote_sims}%
  \BibitemOpen
  \bibfield  {author} {\bibinfo {author} {\bibfnamefont {F.}~\bibnamefont
  {{Villaescusa-Navarro}}}, \bibinfo {author} {\bibfnamefont {C.}~\bibnamefont
  {{Hahn}}}, \bibinfo {author} {\bibfnamefont {E.}~\bibnamefont {{Massara}}},
  \bibinfo {author} {\bibfnamefont {A.}~\bibnamefont {{Banerjee}}}, \bibinfo
  {author} {\bibfnamefont {A.~M.}\ \bibnamefont {{Delgado}}}, \emph {et~al.},\
  }\href {https://doi.org/10.3847/1538-4365/ab9d82} {\bibfield  {journal}
  {\bibinfo  {journal} {\apjs}\ }\textbf {\bibinfo {volume} {250}},\ \bibinfo
  {eid} {2} (\bibinfo {year} {2020})},\ \Eprint
  {https://arxiv.org/abs/1909.05273} {arXiv:1909.05273 [astro-ph.CO]}
  \BibitemShut {NoStop}%
\bibitem [{\citenamefont {{Akiba}}\ \emph {et~al.}(2019)\citenamefont
  {{Akiba}}, \citenamefont {{Sano}}, \citenamefont {{Yanase}}, \citenamefont
  {{Ohta}},\ and\ \citenamefont {{Koyama}}}]{Akiba2019}%
  \BibitemOpen
  \bibfield  {author} {\bibinfo {author} {\bibfnamefont {T.}~\bibnamefont
  {{Akiba}}}, \bibinfo {author} {\bibfnamefont {S.}~\bibnamefont {{Sano}}},
  \bibinfo {author} {\bibfnamefont {T.}~\bibnamefont {{Yanase}}}, \bibinfo
  {author} {\bibfnamefont {T.}~\bibnamefont {{Ohta}}},\ and\ \bibinfo {author}
  {\bibfnamefont {M.}~\bibnamefont {{Koyama}}},\ }\href@noop {} {\bibfield
  {journal} {\bibinfo  {journal} {arXiv e-prints}\ ,\ \bibinfo {eid}
  {arXiv:1907.10902}} (\bibinfo {year} {2019})},\ \Eprint
  {https://arxiv.org/abs/1907.10902} {arXiv:1907.10902 [cs.LG]} \BibitemShut
  {NoStop}%
\bibitem [{\citenamefont {{Behroozi}}\ \emph {et~al.}(2012)\citenamefont
  {{Behroozi}}, \citenamefont {{Wechsler}},\ and\ \citenamefont
  {{Wu}}}]{Behroozi2012}%
  \BibitemOpen
  \bibfield  {author} {\bibinfo {author} {\bibfnamefont {P.}~\bibnamefont
  {{Behroozi}}}, \bibinfo {author} {\bibfnamefont {R.}~\bibnamefont
  {{Wechsler}}},\ and\ \bibinfo {author} {\bibfnamefont {H.-Y.}\ \bibnamefont
  {{Wu}}},\ }\href@noop {} {\bibinfo {title} {{Rockstar: Phase-space halo
  finder}}},\ \bibinfo {howpublished} {Astrophysics Source Code Library, record
  ascl:1210.008} (\bibinfo {year} {2012}),\ \Eprint
  {https://arxiv.org/abs/1210.008} {ascl:1210.008} \BibitemShut {NoStop}%
\bibitem [{\citenamefont {{Behroozi}}\ \emph {et~al.}(2013)\citenamefont
  {{Behroozi}}, \citenamefont {{Wechsler}},\ and\ \citenamefont
  {{Wu}}}]{Behroozi2013}%
  \BibitemOpen
  \bibfield  {author} {\bibinfo {author} {\bibfnamefont {P.~S.}\ \bibnamefont
  {{Behroozi}}}, \bibinfo {author} {\bibfnamefont {R.~H.}\ \bibnamefont
  {{Wechsler}}},\ and\ \bibinfo {author} {\bibfnamefont {H.-Y.}\ \bibnamefont
  {{Wu}}},\ }\href {https://doi.org/10.1088/0004-637X/762/2/109} {\bibfield
  {journal} {\bibinfo  {journal} {\apj}\ }\textbf {\bibinfo {volume} {762}},\
  \bibinfo {eid} {109} (\bibinfo {year} {2013})},\ \Eprint
  {https://arxiv.org/abs/1110.4372} {arXiv:1110.4372 [astro-ph.CO]}
  \BibitemShut {NoStop}%
\bibitem [{\citenamefont {{Zheng}}\ \emph {et~al.}(2005)\citenamefont
  {{Zheng}}, \citenamefont {{Berlind}}, \citenamefont {{Weinberg}},
  \citenamefont {{Benson}}, \citenamefont {{Baugh}}, \citenamefont {{Cole}},
  \citenamefont {{Dav{\'e}}}, \citenamefont {{Frenk}}, \citenamefont {{Katz}},\
  and\ \citenamefont {{Lacey}}}]{Zheng2005}%
  \BibitemOpen
  \bibfield  {author} {\bibinfo {author} {\bibfnamefont {Z.}~\bibnamefont
  {{Zheng}}}, \bibinfo {author} {\bibfnamefont {A.~A.}\ \bibnamefont
  {{Berlind}}}, \bibinfo {author} {\bibfnamefont {D.~H.}\ \bibnamefont
  {{Weinberg}}}, \bibinfo {author} {\bibfnamefont {A.~J.}\ \bibnamefont
  {{Benson}}}, \bibinfo {author} {\bibfnamefont {C.~M.}\ \bibnamefont
  {{Baugh}}}, \bibinfo {author} {\bibfnamefont {S.}~\bibnamefont {{Cole}}},
  \bibinfo {author} {\bibfnamefont {R.}~\bibnamefont {{Dav{\'e}}}}, \bibinfo
  {author} {\bibfnamefont {C.~S.}\ \bibnamefont {{Frenk}}}, \bibinfo {author}
  {\bibfnamefont {N.}~\bibnamefont {{Katz}}},\ and\ \bibinfo {author}
  {\bibfnamefont {C.~G.}\ \bibnamefont {{Lacey}}},\ }\href
  {https://doi.org/10.1086/466510} {\bibfield  {journal} {\bibinfo  {journal}
  {\apj}\ }\textbf {\bibinfo {volume} {633}},\ \bibinfo {pages} {791} (\bibinfo
  {year} {2005})},\ \Eprint {https://arxiv.org/abs/astro-ph/0408564}
  {arXiv:astro-ph/0408564 [astro-ph]} \BibitemShut {NoStop}%
\bibitem [{\citenamefont {{Zhu}}\ \emph {et~al.}(2006)\citenamefont {{Zhu}},
  \citenamefont {{Zheng}}, \citenamefont {{Lin}}, \citenamefont {{Jing}},
  \citenamefont {{Kang}},\ and\ \citenamefont {{Gao}}}]{Zhu2006}%
  \BibitemOpen
  \bibfield  {author} {\bibinfo {author} {\bibfnamefont {G.}~\bibnamefont
  {{Zhu}}}, \bibinfo {author} {\bibfnamefont {Z.}~\bibnamefont {{Zheng}}},
  \bibinfo {author} {\bibfnamefont {W.~P.}\ \bibnamefont {{Lin}}}, \bibinfo
  {author} {\bibfnamefont {Y.~P.}\ \bibnamefont {{Jing}}}, \bibinfo {author}
  {\bibfnamefont {X.}~\bibnamefont {{Kang}}},\ and\ \bibinfo {author}
  {\bibfnamefont {L.}~\bibnamefont {{Gao}}},\ }\href
  {https://doi.org/10.1086/501501} {\bibfield  {journal} {\bibinfo  {journal}
  {\apjl}\ }\textbf {\bibinfo {volume} {639}},\ \bibinfo {pages} {L5} (\bibinfo
  {year} {2006})},\ \Eprint {https://arxiv.org/abs/astro-ph/0601120}
  {arXiv:astro-ph/0601120 [astro-ph]} \BibitemShut {NoStop}%
\bibitem [{\citenamefont {{Zentner}}\ \emph {et~al.}(2014)\citenamefont
  {{Zentner}}, \citenamefont {{Hearin}},\ and\ \citenamefont {{van den
  Bosch}}}]{Zentner2014}%
  \BibitemOpen
  \bibfield  {author} {\bibinfo {author} {\bibfnamefont {A.~R.}\ \bibnamefont
  {{Zentner}}}, \bibinfo {author} {\bibfnamefont {A.~P.}\ \bibnamefont
  {{Hearin}}},\ and\ \bibinfo {author} {\bibfnamefont {F.~C.}\ \bibnamefont
  {{van den Bosch}}},\ }\href {https://doi.org/10.1093/mnras/stu1383}
  {\bibfield  {journal} {\bibinfo  {journal} {\mnras}\ }\textbf {\bibinfo
  {volume} {443}},\ \bibinfo {pages} {3044} (\bibinfo {year} {2014})},\ \Eprint
  {https://arxiv.org/abs/1311.1818} {arXiv:1311.1818 [astro-ph.CO]}
  \BibitemShut {NoStop}%
\bibitem [{\citenamefont {{Pujol}}\ and\ \citenamefont
  {{Gazta{\~n}aga}}(2014)}]{Pujol2014}%
  \BibitemOpen
  \bibfield  {author} {\bibinfo {author} {\bibfnamefont {A.}~\bibnamefont
  {{Pujol}}}\ and\ \bibinfo {author} {\bibfnamefont {E.}~\bibnamefont
  {{Gazta{\~n}aga}}},\ }\href {https://doi.org/10.1093/mnras/stu1001}
  {\bibfield  {journal} {\bibinfo  {journal} {\mnras}\ }\textbf {\bibinfo
  {volume} {442}},\ \bibinfo {pages} {1930} (\bibinfo {year} {2014})},\ \Eprint
  {https://arxiv.org/abs/1306.5761} {arXiv:1306.5761 [astro-ph.CO]}
  \BibitemShut {NoStop}%
\bibitem [{\citenamefont {{Reid}}\ \emph {et~al.}(2014)\citenamefont {{Reid}},
  \citenamefont {{Seo}}, \citenamefont {{Leauthaud}}, \citenamefont
  {{Tinker}},\ and\ \citenamefont {{White}}}]{Reid2014}%
  \BibitemOpen
  \bibfield  {author} {\bibinfo {author} {\bibfnamefont {B.~A.}\ \bibnamefont
  {{Reid}}}, \bibinfo {author} {\bibfnamefont {H.-J.}\ \bibnamefont {{Seo}}},
  \bibinfo {author} {\bibfnamefont {A.}~\bibnamefont {{Leauthaud}}}, \bibinfo
  {author} {\bibfnamefont {J.~L.}\ \bibnamefont {{Tinker}}},\ and\ \bibinfo
  {author} {\bibfnamefont {M.}~\bibnamefont {{White}}},\ }\href
  {https://doi.org/10.1093/mnras/stu1391} {\bibfield  {journal} {\bibinfo
  {journal} {\mnras}\ }\textbf {\bibinfo {volume} {444}},\ \bibinfo {pages}
  {476} (\bibinfo {year} {2014})},\ \Eprint {https://arxiv.org/abs/1404.3742}
  {arXiv:1404.3742 [astro-ph.CO]} \BibitemShut {NoStop}%
\bibitem [{\citenamefont {{Lin}}\ \emph {et~al.}(2016)\citenamefont {{Lin}},
  \citenamefont {{Mandelbaum}}, \citenamefont {{Huang}}, \citenamefont
  {{Huang}}, \citenamefont {{Dalal}}, \citenamefont {{Diemer}}, \citenamefont
  {{Jian}},\ and\ \citenamefont {{Kravtsov}}}]{Lin2016}%
  \BibitemOpen
  \bibfield  {author} {\bibinfo {author} {\bibfnamefont {Y.-T.}\ \bibnamefont
  {{Lin}}}, \bibinfo {author} {\bibfnamefont {R.}~\bibnamefont {{Mandelbaum}}},
  \bibinfo {author} {\bibfnamefont {Y.-H.}\ \bibnamefont {{Huang}}}, \bibinfo
  {author} {\bibfnamefont {H.-J.}\ \bibnamefont {{Huang}}}, \bibinfo {author}
  {\bibfnamefont {N.}~\bibnamefont {{Dalal}}}, \bibinfo {author} {\bibfnamefont
  {B.}~\bibnamefont {{Diemer}}}, \bibinfo {author} {\bibfnamefont {H.-Y.}\
  \bibnamefont {{Jian}}},\ and\ \bibinfo {author} {\bibfnamefont
  {A.}~\bibnamefont {{Kravtsov}}},\ }\href
  {https://doi.org/10.3847/0004-637X/819/2/119} {\bibfield  {journal} {\bibinfo
   {journal} {\apj}\ }\textbf {\bibinfo {volume} {819}},\ \bibinfo {eid} {119}
  (\bibinfo {year} {2016})},\ \Eprint {https://arxiv.org/abs/1504.07632}
  {arXiv:1504.07632 [astro-ph.GA]} \BibitemShut {NoStop}%
\bibitem [{\citenamefont {{Kobayashi}}\ \emph {et~al.}(2022)\citenamefont
  {{Kobayashi}}, \citenamefont {{Nishimichi}}, \citenamefont {{Takada}},\ and\
  \citenamefont {{Miyatake}}}]{Kobayashi2022}%
  \BibitemOpen
  \bibfield  {author} {\bibinfo {author} {\bibfnamefont {Y.}~\bibnamefont
  {{Kobayashi}}}, \bibinfo {author} {\bibfnamefont {T.}~\bibnamefont
  {{Nishimichi}}}, \bibinfo {author} {\bibfnamefont {M.}~\bibnamefont
  {{Takada}}},\ and\ \bibinfo {author} {\bibfnamefont {H.}~\bibnamefont
  {{Miyatake}}},\ }\href {https://doi.org/10.1103/PhysRevD.105.083517}
  {\bibfield  {journal} {\bibinfo  {journal} {\prd}\ }\textbf {\bibinfo
  {volume} {105}},\ \bibinfo {eid} {083517} (\bibinfo {year} {2022})},\ \Eprint
  {https://arxiv.org/abs/2110.06969} {arXiv:2110.06969 [astro-ph.CO]}
  \BibitemShut {NoStop}%
\bibitem [{\citenamefont {{Berlind}}\ \emph {et~al.}(2003)\citenamefont
  {{Berlind}}, \citenamefont {{Weinberg}}, \citenamefont {{Benson}},
  \citenamefont {{Baugh}}, \citenamefont {{Cole}}, \citenamefont {{Dav{\'e}}},
  \citenamefont {{Frenk}}, \citenamefont {{Jenkins}}, \citenamefont {{Katz}},\
  and\ \citenamefont {{Lacey}}}]{Berlind2003}%
  \BibitemOpen
  \bibfield  {author} {\bibinfo {author} {\bibfnamefont {A.~A.}\ \bibnamefont
  {{Berlind}}}, \bibinfo {author} {\bibfnamefont {D.~H.}\ \bibnamefont
  {{Weinberg}}}, \bibinfo {author} {\bibfnamefont {A.~J.}\ \bibnamefont
  {{Benson}}}, \bibinfo {author} {\bibfnamefont {C.~M.}\ \bibnamefont
  {{Baugh}}}, \bibinfo {author} {\bibfnamefont {S.}~\bibnamefont {{Cole}}},
  \bibinfo {author} {\bibfnamefont {R.}~\bibnamefont {{Dav{\'e}}}}, \bibinfo
  {author} {\bibfnamefont {C.~S.}\ \bibnamefont {{Frenk}}}, \bibinfo {author}
  {\bibfnamefont {A.}~\bibnamefont {{Jenkins}}}, \bibinfo {author}
  {\bibfnamefont {N.}~\bibnamefont {{Katz}}},\ and\ \bibinfo {author}
  {\bibfnamefont {C.~G.}\ \bibnamefont {{Lacey}}},\ }\href
  {https://doi.org/10.1086/376517} {\bibfield  {journal} {\bibinfo  {journal}
  {\apj}\ }\textbf {\bibinfo {volume} {593}},\ \bibinfo {pages} {1} (\bibinfo
  {year} {2003})},\ \Eprint {https://arxiv.org/abs/astro-ph/0212357}
  {arXiv:astro-ph/0212357 [astro-ph]} \BibitemShut {NoStop}%
\bibitem [{\citenamefont {{Yoshikawa}}\ \emph {et~al.}(2003)\citenamefont
  {{Yoshikawa}}, \citenamefont {{Jing}},\ and\ \citenamefont
  {{B{\"o}rner}}}]{Yoshikawa2003}%
  \BibitemOpen
  \bibfield  {author} {\bibinfo {author} {\bibfnamefont {K.}~\bibnamefont
  {{Yoshikawa}}}, \bibinfo {author} {\bibfnamefont {Y.~P.}\ \bibnamefont
  {{Jing}}},\ and\ \bibinfo {author} {\bibfnamefont {G.}~\bibnamefont
  {{B{\"o}rner}}},\ }\href {https://doi.org/10.1086/375148} {\bibfield
  {journal} {\bibinfo  {journal} {\apj}\ }\textbf {\bibinfo {volume} {590}},\
  \bibinfo {pages} {654} (\bibinfo {year} {2003})},\ \Eprint
  {https://arxiv.org/abs/astro-ph/0303053} {arXiv:astro-ph/0303053 [astro-ph]}
  \BibitemShut {NoStop}%
\bibitem [{\citenamefont {{van den Bosch}}\ \emph {et~al.}(2005)\citenamefont
  {{van den Bosch}}, \citenamefont {{Weinmann}}, \citenamefont {{Yang}},
  \citenamefont {{Mo}}, \citenamefont {{Li}},\ and\ \citenamefont
  {{Jing}}}]{vandenBosch2005}%
  \BibitemOpen
  \bibfield  {author} {\bibinfo {author} {\bibfnamefont {F.~C.}\ \bibnamefont
  {{van den Bosch}}}, \bibinfo {author} {\bibfnamefont {S.~M.}\ \bibnamefont
  {{Weinmann}}}, \bibinfo {author} {\bibfnamefont {X.}~\bibnamefont {{Yang}}},
  \bibinfo {author} {\bibfnamefont {H.~J.}\ \bibnamefont {{Mo}}}, \bibinfo
  {author} {\bibfnamefont {C.}~\bibnamefont {{Li}}},\ and\ \bibinfo {author}
  {\bibfnamefont {Y.~P.}\ \bibnamefont {{Jing}}},\ }\href
  {https://doi.org/10.1111/j.1365-2966.2005.09260.x} {\bibfield  {journal}
  {\bibinfo  {journal} {\mnras}\ }\textbf {\bibinfo {volume} {361}},\ \bibinfo
  {pages} {1203} (\bibinfo {year} {2005})},\ \Eprint
  {https://arxiv.org/abs/astro-ph/0502466} {arXiv:astro-ph/0502466 [astro-ph]}
  \BibitemShut {NoStop}%
\bibitem [{\citenamefont {{Guo}}\ \emph {et~al.}(2015)\citenamefont {{Guo}},
  \citenamefont {{Zheng}}, \citenamefont {{Zehavi}}, \citenamefont {{Dawson}},
  \citenamefont {{Skibba}}, \citenamefont {{Tinker}}, \citenamefont
  {{Weinberg}}, \citenamefont {{White}},\ and\ \citenamefont
  {{Schneider}}}]{Guo2015}%
  \BibitemOpen
  \bibfield  {author} {\bibinfo {author} {\bibfnamefont {H.}~\bibnamefont
  {{Guo}}}, \bibinfo {author} {\bibfnamefont {Z.}~\bibnamefont {{Zheng}}},
  \bibinfo {author} {\bibfnamefont {I.}~\bibnamefont {{Zehavi}}}, \bibinfo
  {author} {\bibfnamefont {K.}~\bibnamefont {{Dawson}}}, \bibinfo {author}
  {\bibfnamefont {R.~A.}\ \bibnamefont {{Skibba}}}, \bibinfo {author}
  {\bibfnamefont {J.~L.}\ \bibnamefont {{Tinker}}}, \bibinfo {author}
  {\bibfnamefont {D.~H.}\ \bibnamefont {{Weinberg}}}, \bibinfo {author}
  {\bibfnamefont {M.}~\bibnamefont {{White}}},\ and\ \bibinfo {author}
  {\bibfnamefont {D.~P.}\ \bibnamefont {{Schneider}}},\ }\href
  {https://doi.org/10.1093/mnras/stu2120} {\bibfield  {journal} {\bibinfo
  {journal} {\mnras}\ }\textbf {\bibinfo {volume} {446}},\ \bibinfo {pages}
  {578} (\bibinfo {year} {2015})},\ \Eprint {https://arxiv.org/abs/1407.4811}
  {arXiv:1407.4811 [astro-ph.CO]} \BibitemShut {NoStop}%
\bibitem [{\citenamefont {{Zhai}}\ \emph {et~al.}(2023)\citenamefont {{Zhai}},
  \citenamefont {{Tinker}}, \citenamefont {{Banerjee}}, \citenamefont
  {{DeRose}}, \citenamefont {{Guo}}, \citenamefont {{Mao}}, \citenamefont
  {{McLaughlin}}, \citenamefont {{Storey-Fisher}},\ and\ \citenamefont
  {{Wechsler}}}]{Zhai2023}%
  \BibitemOpen
  \bibfield  {author} {\bibinfo {author} {\bibfnamefont {Z.}~\bibnamefont
  {{Zhai}}}, \bibinfo {author} {\bibfnamefont {J.~L.}\ \bibnamefont
  {{Tinker}}}, \bibinfo {author} {\bibfnamefont {A.}~\bibnamefont
  {{Banerjee}}}, \bibinfo {author} {\bibfnamefont {J.}~\bibnamefont
  {{DeRose}}}, \bibinfo {author} {\bibfnamefont {H.}~\bibnamefont {{Guo}}},
  \bibinfo {author} {\bibfnamefont {Y.-Y.}\ \bibnamefont {{Mao}}}, \bibinfo
  {author} {\bibfnamefont {S.}~\bibnamefont {{McLaughlin}}}, \bibinfo {author}
  {\bibfnamefont {K.}~\bibnamefont {{Storey-Fisher}}},\ and\ \bibinfo {author}
  {\bibfnamefont {R.~H.}\ \bibnamefont {{Wechsler}}},\ }\href
  {https://doi.org/10.3847/1538-4357/acc65b} {\bibfield  {journal} {\bibinfo
  {journal} {\apj}\ }\textbf {\bibinfo {volume} {948}},\ \bibinfo {eid} {99}
  (\bibinfo {year} {2023})},\ \Eprint {https://arxiv.org/abs/2203.08999}
  {arXiv:2203.08999 [astro-ph.CO]} \BibitemShut {NoStop}%
\bibitem [{\citenamefont {{Carlson}}\ and\ \citenamefont
  {{White}}(2010)}]{Carlson2010}%
  \BibitemOpen
  \bibfield  {author} {\bibinfo {author} {\bibfnamefont {J.}~\bibnamefont
  {{Carlson}}}\ and\ \bibinfo {author} {\bibfnamefont {M.}~\bibnamefont
  {{White}}},\ }\href {https://doi.org/10.1088/0067-0049/190/2/311} {\bibfield
  {journal} {\bibinfo  {journal} {\apjs}\ }\textbf {\bibinfo {volume} {190}},\
  \bibinfo {pages} {311} (\bibinfo {year} {2010})},\ \Eprint
  {https://arxiv.org/abs/1003.3178} {arXiv:1003.3178 [astro-ph.CO]}
  \BibitemShut {NoStop}%
\bibitem [{\citenamefont {{Sutter}}\ \emph {et~al.}(2015)\citenamefont
  {{Sutter}}, \citenamefont {{Lavaux}}, \citenamefont {{Hamaus}}, \citenamefont
  {{Pisani}}, \citenamefont {{Wandelt}}, \citenamefont {{Warren}},
  \citenamefont {{Villaescusa-Navarro}}, \citenamefont {{Zivick}},
  \citenamefont {{Mao}},\ and\ \citenamefont {{Thompson}}}]{Sutter2015}%
  \BibitemOpen
  \bibfield  {author} {\bibinfo {author} {\bibfnamefont {P.~M.}\ \bibnamefont
  {{Sutter}}}, \bibinfo {author} {\bibfnamefont {G.}~\bibnamefont {{Lavaux}}},
  \bibinfo {author} {\bibfnamefont {N.}~\bibnamefont {{Hamaus}}}, \bibinfo
  {author} {\bibfnamefont {A.}~\bibnamefont {{Pisani}}}, \bibinfo {author}
  {\bibfnamefont {B.~D.}\ \bibnamefont {{Wandelt}}}, \bibinfo {author}
  {\bibfnamefont {M.}~\bibnamefont {{Warren}}}, \bibinfo {author}
  {\bibfnamefont {F.}~\bibnamefont {{Villaescusa-Navarro}}}, \bibinfo {author}
  {\bibfnamefont {P.}~\bibnamefont {{Zivick}}}, \bibinfo {author}
  {\bibfnamefont {Q.}~\bibnamefont {{Mao}}},\ and\ \bibinfo {author}
  {\bibfnamefont {B.~B.}\ \bibnamefont {{Thompson}}},\ }\href
  {https://doi.org/10.1016/j.ascom.2014.10.002} {\bibfield  {journal} {\bibinfo
   {journal} {Astronomy and Computing}\ }\textbf {\bibinfo {volume} {9}},\
  \bibinfo {pages} {1} (\bibinfo {year} {2015})},\ \Eprint
  {https://arxiv.org/abs/1406.1191} {arXiv:1406.1191 [astro-ph.CO]}
  \BibitemShut {NoStop}%
\bibitem [{\citenamefont {{Neyrinck}}(2008)}]{Neyrinck2008}%
  \BibitemOpen
  \bibfield  {author} {\bibinfo {author} {\bibfnamefont {M.~C.}\ \bibnamefont
  {{Neyrinck}}},\ }\href {https://doi.org/10.1111/j.1365-2966.2008.13180.x}
  {\bibfield  {journal} {\bibinfo  {journal} {\mnras}\ }\textbf {\bibinfo
  {volume} {386}},\ \bibinfo {pages} {2101} (\bibinfo {year} {2008})},\ \Eprint
  {https://arxiv.org/abs/0712.3049} {arXiv:0712.3049 [astro-ph]} \BibitemShut
  {NoStop}%
\bibitem [{\citenamefont {{Colberg}}\ \emph {et~al.}(2008)\citenamefont
  {{Colberg}}, \citenamefont {{Pearce}}, \citenamefont {{Foster}},
  \citenamefont {{Platen}}, \citenamefont {{Brunino}} \emph
  {et~al.}}]{Colberg2008}%
  \BibitemOpen
  \bibfield  {author} {\bibinfo {author} {\bibfnamefont {J.~M.}\ \bibnamefont
  {{Colberg}}}, \bibinfo {author} {\bibfnamefont {F.}~\bibnamefont {{Pearce}}},
  \bibinfo {author} {\bibfnamefont {C.}~\bibnamefont {{Foster}}}, \bibinfo
  {author} {\bibfnamefont {E.}~\bibnamefont {{Platen}}}, \bibinfo {author}
  {\bibfnamefont {R.}~\bibnamefont {{Brunino}}}, \emph {et~al.},\ }\href
  {https://doi.org/10.1111/j.1365-2966.2008.13307.x} {\bibfield  {journal}
  {\bibinfo  {journal} {\mnras}\ }\textbf {\bibinfo {volume} {387}},\ \bibinfo
  {pages} {933} (\bibinfo {year} {2008})},\ \Eprint
  {https://arxiv.org/abs/0803.0918} {arXiv:0803.0918 [astro-ph]} \BibitemShut
  {NoStop}%
\bibitem [{\citenamefont {{van de Weygaert}}\ and\ \citenamefont
  {{Schaap}}(2009)}]{vandeWeygaert2009}%
  \BibitemOpen
  \bibfield  {author} {\bibinfo {author} {\bibfnamefont {R.}~\bibnamefont {{van
  de Weygaert}}}\ and\ \bibinfo {author} {\bibfnamefont {W.}~\bibnamefont
  {{Schaap}}},\ }in\ \href {https://doi.org/10.1007/978-3-540-44767-2_11}
  {\emph {\bibinfo {booktitle} {Data Analysis in Cosmology}}},\ Vol.\ \bibinfo
  {volume} {665},\ \bibinfo {editor} {edited by\ \bibinfo {editor}
  {\bibfnamefont {V.~J.}\ \bibnamefont {{Mart{\'\i}nez}}}, \bibinfo {editor}
  {\bibfnamefont {E.}~\bibnamefont {{Saar}}}, \bibinfo {editor} {\bibfnamefont
  {E.}~\bibnamefont {{Mart{\'\i}nez-Gonz{\'a}lez}}},\ and\ \bibinfo {editor}
  {\bibfnamefont {M.~J.}\ \bibnamefont {{Pons-Border{\'\i}a}}}}\ (\bibinfo
  {year} {2009})\ pp.\ \bibinfo {pages} {291--413}\BibitemShut {NoStop}%
\bibitem [{\citenamefont {{Cautun}}\ \emph {et~al.}(2018)\citenamefont
  {{Cautun}}, \citenamefont {{Paillas}}, \citenamefont {{Cai}}, \citenamefont
  {{Bose}}, \citenamefont {{Armijo}}, \citenamefont {{Li}},\ and\ \citenamefont
  {{Padilla}}}]{Cautun2018}%
  \BibitemOpen
  \bibfield  {author} {\bibinfo {author} {\bibfnamefont {M.}~\bibnamefont
  {{Cautun}}}, \bibinfo {author} {\bibfnamefont {E.}~\bibnamefont {{Paillas}}},
  \bibinfo {author} {\bibfnamefont {Y.-C.}\ \bibnamefont {{Cai}}}, \bibinfo
  {author} {\bibfnamefont {S.}~\bibnamefont {{Bose}}}, \bibinfo {author}
  {\bibfnamefont {J.}~\bibnamefont {{Armijo}}}, \bibinfo {author}
  {\bibfnamefont {B.}~\bibnamefont {{Li}}},\ and\ \bibinfo {author}
  {\bibfnamefont {N.}~\bibnamefont {{Padilla}}},\ }\href
  {https://doi.org/10.1093/mnras/sty463} {\bibfield  {journal} {\bibinfo
  {journal} {\mnras}\ }\textbf {\bibinfo {volume} {476}},\ \bibinfo {pages}
  {3195} (\bibinfo {year} {2018})},\ \Eprint {https://arxiv.org/abs/1710.01730}
  {arXiv:1710.01730 [astro-ph.CO]} \BibitemShut {NoStop}%
\bibitem [{\citenamefont {{Hand}}\ \emph {et~al.}(2018)\citenamefont {{Hand}},
  \citenamefont {{Feng}}, \citenamefont {{Beutler}}, \citenamefont {{Li}},
  \citenamefont {{Modi}}, \citenamefont {{Seljak}},\ and\ \citenamefont
  {{Slepian}}}]{Hand2018}%
  \BibitemOpen
  \bibfield  {author} {\bibinfo {author} {\bibfnamefont {N.}~\bibnamefont
  {{Hand}}}, \bibinfo {author} {\bibfnamefont {Y.}~\bibnamefont {{Feng}}},
  \bibinfo {author} {\bibfnamefont {F.}~\bibnamefont {{Beutler}}}, \bibinfo
  {author} {\bibfnamefont {Y.}~\bibnamefont {{Li}}}, \bibinfo {author}
  {\bibfnamefont {C.}~\bibnamefont {{Modi}}}, \bibinfo {author} {\bibfnamefont
  {U.}~\bibnamefont {{Seljak}}},\ and\ \bibinfo {author} {\bibfnamefont
  {Z.}~\bibnamefont {{Slepian}}},\ }\href
  {https://doi.org/10.3847/1538-3881/aadae0} {\bibfield  {journal} {\bibinfo
  {journal} {\aj}\ }\textbf {\bibinfo {volume} {156}},\ \bibinfo {eid} {160}
  (\bibinfo {year} {2018})},\ \Eprint {https://arxiv.org/abs/1712.05834}
  {arXiv:1712.05834 [astro-ph.IM]} \BibitemShut {NoStop}%
\bibitem [{\citenamefont {{Hand}}\ \emph {et~al.}(2019)\citenamefont {{Hand}},
  \citenamefont {{Feng}}, \citenamefont {{Beutler}}, \citenamefont {{Li}},
  \citenamefont {{Modi}}, \citenamefont {{Seljak}},\ and\ \citenamefont
  {{Slepian}}}]{Hand2019}%
  \BibitemOpen
  \bibfield  {author} {\bibinfo {author} {\bibfnamefont {N.}~\bibnamefont
  {{Hand}}}, \bibinfo {author} {\bibfnamefont {Y.}~\bibnamefont {{Feng}}},
  \bibinfo {author} {\bibfnamefont {F.}~\bibnamefont {{Beutler}}}, \bibinfo
  {author} {\bibfnamefont {Y.}~\bibnamefont {{Li}}}, \bibinfo {author}
  {\bibfnamefont {C.}~\bibnamefont {{Modi}}}, \bibinfo {author} {\bibfnamefont
  {U.}~\bibnamefont {{Seljak}}},\ and\ \bibinfo {author} {\bibfnamefont
  {Z.}~\bibnamefont {{Slepian}}},\ }\href@noop {} {\bibinfo {title} {{nbodykit:
  Massively parallel, large-scale structure toolkit}}},\ \bibinfo
  {howpublished} {Astrophysics Source Code Library, record ascl:1904.027}
  (\bibinfo {year} {2019}),\ \Eprint {https://arxiv.org/abs/1904.027}
  {ascl:1904.027} \BibitemShut {NoStop}%
\bibitem [{\citenamefont {{Feldman}}\ \emph {et~al.}(1994)\citenamefont
  {{Feldman}}, \citenamefont {{Kaiser}},\ and\ \citenamefont
  {{Peacock}}}]{Feldman1994}%
  \BibitemOpen
  \bibfield  {author} {\bibinfo {author} {\bibfnamefont {H.~A.}\ \bibnamefont
  {{Feldman}}}, \bibinfo {author} {\bibfnamefont {N.}~\bibnamefont
  {{Kaiser}}},\ and\ \bibinfo {author} {\bibfnamefont {J.~A.}\ \bibnamefont
  {{Peacock}}},\ }\href {https://doi.org/10.1086/174036} {\bibfield  {journal}
  {\bibinfo  {journal} {\apj}\ }\textbf {\bibinfo {volume} {426}},\ \bibinfo
  {pages} {23} (\bibinfo {year} {1994})},\ \Eprint
  {https://arxiv.org/abs/astro-ph/9304022} {arXiv:astro-ph/9304022 [astro-ph]}
  \BibitemShut {NoStop}%
\bibitem [{\citenamefont {{Reid}}\ \emph {et~al.}(2016)\citenamefont {{Reid}},
  \citenamefont {{Ho}}, \citenamefont {{Padmanabhan}}, \citenamefont
  {{Percival}}, \citenamefont {{Tinker}} \emph {et~al.}}]{Reid2016}%
  \BibitemOpen
  \bibfield  {author} {\bibinfo {author} {\bibfnamefont {B.}~\bibnamefont
  {{Reid}}}, \bibinfo {author} {\bibfnamefont {S.}~\bibnamefont {{Ho}}},
  \bibinfo {author} {\bibfnamefont {N.}~\bibnamefont {{Padmanabhan}}}, \bibinfo
  {author} {\bibfnamefont {W.~J.}\ \bibnamefont {{Percival}}}, \bibinfo
  {author} {\bibfnamefont {J.}~\bibnamefont {{Tinker}}}, \emph {et~al.},\
  }\href {https://doi.org/10.1093/mnras/stv2382} {\bibfield  {journal}
  {\bibinfo  {journal} {\mnras}\ }\textbf {\bibinfo {volume} {455}},\ \bibinfo
  {pages} {1553} (\bibinfo {year} {2016})},\ \Eprint
  {https://arxiv.org/abs/1509.06529} {arXiv:1509.06529 [astro-ph.CO]}
  \BibitemShut {NoStop}%
\bibitem [{\citenamefont {{Cousinou}}\ \emph {et~al.}(2019)\citenamefont
  {{Cousinou}}, \citenamefont {{Pisani}}, \citenamefont {{Tilquin}},
  \citenamefont {{Hamaus}}, \citenamefont {{Hawken}},\ and\ \citenamefont
  {{Escoffier}}}]{Cousinou2019}%
  \BibitemOpen
  \bibfield  {author} {\bibinfo {author} {\bibfnamefont {M.~C.}\ \bibnamefont
  {{Cousinou}}}, \bibinfo {author} {\bibfnamefont {A.}~\bibnamefont
  {{Pisani}}}, \bibinfo {author} {\bibfnamefont {A.}~\bibnamefont {{Tilquin}}},
  \bibinfo {author} {\bibfnamefont {N.}~\bibnamefont {{Hamaus}}}, \bibinfo
  {author} {\bibfnamefont {A.~J.}\ \bibnamefont {{Hawken}}},\ and\ \bibinfo
  {author} {\bibfnamefont {S.}~\bibnamefont {{Escoffier}}},\ }\href
  {https://doi.org/10.1016/j.ascom.2019.03.001} {\bibfield  {journal} {\bibinfo
   {journal} {Astronomy and Computing}\ }\textbf {\bibinfo {volume} {27}},\
  \bibinfo {eid} {53} (\bibinfo {year} {2019})},\ \Eprint
  {https://arxiv.org/abs/1805.07181} {arXiv:1805.07181 [astro-ph.CO]}
  \BibitemShut {NoStop}%
\bibitem [{\citenamefont {{Pisani}}\ \emph
  {et~al.}(2015{\natexlab{b}})\citenamefont {{Pisani}}, \citenamefont
  {{Sutter}},\ and\ \citenamefont {{Wandelt}}}]{Pisani2015b}%
  \BibitemOpen
  \bibfield  {author} {\bibinfo {author} {\bibfnamefont {A.}~\bibnamefont
  {{Pisani}}}, \bibinfo {author} {\bibfnamefont {P.~M.}\ \bibnamefont
  {{Sutter}}},\ and\ \bibinfo {author} {\bibfnamefont {B.~D.}\ \bibnamefont
  {{Wandelt}}},\ }\href {https://doi.org/10.48550/arXiv.1506.07982} {\bibfield
  {journal} {\bibinfo  {journal} {arXiv e-prints}\ ,\ \bibinfo {eid}
  {arXiv:1506.07982}} (\bibinfo {year} {2015}{\natexlab{b}})},\ \Eprint
  {https://arxiv.org/abs/1506.07982} {arXiv:1506.07982 [astro-ph.CO]}
  \BibitemShut {NoStop}%
\bibitem [{\citenamefont {{Ivanov}}\ \emph
  {et~al.}(2020{\natexlab{b}})\citenamefont {{Ivanov}}, \citenamefont
  {{Simonovi{\'c}}},\ and\ \citenamefont {{Zaldarriaga}}}]{Ivanov2020a}%
  \BibitemOpen
  \bibfield  {author} {\bibinfo {author} {\bibfnamefont {M.~M.}\ \bibnamefont
  {{Ivanov}}}, \bibinfo {author} {\bibfnamefont {M.}~\bibnamefont
  {{Simonovi{\'c}}}},\ and\ \bibinfo {author} {\bibfnamefont {M.}~\bibnamefont
  {{Zaldarriaga}}},\ }\href {https://doi.org/10.1088/1475-7516/2020/05/042}
  {\bibfield  {journal} {\bibinfo  {journal} {\jcap}\ }\textbf {\bibinfo
  {volume} {2020}},\ \bibinfo {eid} {042} (\bibinfo {year}
  {2020}{\natexlab{b}})},\ \Eprint {https://arxiv.org/abs/1909.05277}
  {arXiv:1909.05277 [astro-ph.CO]} \BibitemShut {NoStop}%
\bibitem [{\citenamefont {{Cranmer}}\ \emph {et~al.}(2015)\citenamefont
  {{Cranmer}}, \citenamefont {{Pavez}},\ and\ \citenamefont
  {{Louppe}}}]{Cranmer2015}%
  \BibitemOpen
  \bibfield  {author} {\bibinfo {author} {\bibfnamefont {K.}~\bibnamefont
  {{Cranmer}}}, \bibinfo {author} {\bibfnamefont {J.}~\bibnamefont {{Pavez}}},\
  and\ \bibinfo {author} {\bibfnamefont {G.}~\bibnamefont {{Louppe}}},\ }\href
  {https://doi.org/10.48550/arXiv.1506.02169} {\bibfield  {journal} {\bibinfo
  {journal} {arXiv e-prints}\ ,\ \bibinfo {eid} {arXiv:1506.02169}} (\bibinfo
  {year} {2015})},\ \Eprint {https://arxiv.org/abs/1506.02169}
  {arXiv:1506.02169 [stat.AP]} \BibitemShut {NoStop}%
\bibitem [{\citenamefont {{Hermans}}\ \emph {et~al.}(2019)\citenamefont
  {{Hermans}}, \citenamefont {{Begy}},\ and\ \citenamefont
  {{Louppe}}}]{Hermans2019}%
  \BibitemOpen
  \bibfield  {author} {\bibinfo {author} {\bibfnamefont {J.}~\bibnamefont
  {{Hermans}}}, \bibinfo {author} {\bibfnamefont {V.}~\bibnamefont {{Begy}}},\
  and\ \bibinfo {author} {\bibfnamefont {G.}~\bibnamefont {{Louppe}}},\ }\href
  {https://doi.org/10.48550/arXiv.1903.04057} {\bibfield  {journal} {\bibinfo
  {journal} {arXiv e-prints}\ ,\ \bibinfo {eid} {arXiv:1903.04057}} (\bibinfo
  {year} {2019})},\ \Eprint {https://arxiv.org/abs/1903.04057}
  {arXiv:1903.04057 [stat.ML]} \BibitemShut {NoStop}%
\bibitem [{\citenamefont {{Durkan}}\ \emph {et~al.}(2020)\citenamefont
  {{Durkan}}, \citenamefont {{Murray}},\ and\ \citenamefont
  {{Papamakarios}}}]{Durkan2020}%
  \BibitemOpen
  \bibfield  {author} {\bibinfo {author} {\bibfnamefont {C.}~\bibnamefont
  {{Durkan}}}, \bibinfo {author} {\bibfnamefont {I.}~\bibnamefont {{Murray}}},\
  and\ \bibinfo {author} {\bibfnamefont {G.}~\bibnamefont {{Papamakarios}}},\
  }\href {https://doi.org/10.48550/arXiv.2002.03712} {\bibfield  {journal}
  {\bibinfo  {journal} {arXiv e-prints}\ ,\ \bibinfo {eid} {arXiv:2002.03712}}
  (\bibinfo {year} {2020})},\ \Eprint {https://arxiv.org/abs/2002.03712}
  {arXiv:2002.03712 [stat.ML]} \BibitemShut {NoStop}%
\bibitem [{\citenamefont {{Delaunoy}}\ \emph {et~al.}(2022)\citenamefont
  {{Delaunoy}}, \citenamefont {{Hermans}}, \citenamefont {{Rozet}},
  \citenamefont {{Wehenkel}},\ and\ \citenamefont {{Louppe}}}]{Delaunoy2022}%
  \BibitemOpen
  \bibfield  {author} {\bibinfo {author} {\bibfnamefont {A.}~\bibnamefont
  {{Delaunoy}}}, \bibinfo {author} {\bibfnamefont {J.}~\bibnamefont
  {{Hermans}}}, \bibinfo {author} {\bibfnamefont {F.}~\bibnamefont {{Rozet}}},
  \bibinfo {author} {\bibfnamefont {A.}~\bibnamefont {{Wehenkel}}},\ and\
  \bibinfo {author} {\bibfnamefont {G.}~\bibnamefont {{Louppe}}},\ }\href
  {https://doi.org/10.48550/arXiv.2208.13624} {\bibfield  {journal} {\bibinfo
  {journal} {arXiv e-prints}\ ,\ \bibinfo {eid} {arXiv:2208.13624}} (\bibinfo
  {year} {2022})},\ \Eprint {https://arxiv.org/abs/2208.13624}
  {arXiv:2208.13624 [stat.ML]} \BibitemShut {NoStop}%
\bibitem [{\citenamefont {{Miller}}\ \emph {et~al.}(2022)\citenamefont
  {{Miller}}, \citenamefont {{Weniger}},\ and\ \citenamefont
  {{Forr{\'e}}}}]{Miller2022}%
  \BibitemOpen
  \bibfield  {author} {\bibinfo {author} {\bibfnamefont {B.~K.}\ \bibnamefont
  {{Miller}}}, \bibinfo {author} {\bibfnamefont {C.}~\bibnamefont
  {{Weniger}}},\ and\ \bibinfo {author} {\bibfnamefont {P.}~\bibnamefont
  {{Forr{\'e}}}},\ }\href {https://doi.org/10.48550/arXiv.2210.06170}
  {\bibfield  {journal} {\bibinfo  {journal} {arXiv e-prints}\ ,\ \bibinfo
  {eid} {arXiv:2210.06170}} (\bibinfo {year} {2022})},\ \Eprint
  {https://arxiv.org/abs/2210.06170} {arXiv:2210.06170 [stat.ML]} \BibitemShut
  {NoStop}%
\bibitem [{\citenamefont {{Foreman-Mackey}}\ \emph
  {et~al.}(2013{\natexlab{a}})\citenamefont {{Foreman-Mackey}}, \citenamefont
  {{Conley}}, \citenamefont {{Meierjurgen Farr}}, \citenamefont {{Hogg}},
  \citenamefont {{Lang}}, \citenamefont {{Marshall}}, \citenamefont
  {{Price-Whelan}}, \citenamefont {{Sanders}},\ and\ \citenamefont
  {{Zuntz}}}]{Foreman-Mackey2013a}%
  \BibitemOpen
  \bibfield  {author} {\bibinfo {author} {\bibfnamefont {D.}~\bibnamefont
  {{Foreman-Mackey}}}, \bibinfo {author} {\bibfnamefont {A.}~\bibnamefont
  {{Conley}}}, \bibinfo {author} {\bibfnamefont {W.}~\bibnamefont {{Meierjurgen
  Farr}}}, \bibinfo {author} {\bibfnamefont {D.~W.}\ \bibnamefont {{Hogg}}},
  \bibinfo {author} {\bibfnamefont {D.}~\bibnamefont {{Lang}}}, \bibinfo
  {author} {\bibfnamefont {P.}~\bibnamefont {{Marshall}}}, \bibinfo {author}
  {\bibfnamefont {A.}~\bibnamefont {{Price-Whelan}}}, \bibinfo {author}
  {\bibfnamefont {J.}~\bibnamefont {{Sanders}}},\ and\ \bibinfo {author}
  {\bibfnamefont {J.}~\bibnamefont {{Zuntz}}},\ }\href@noop {} {\bibinfo
  {title} {{emcee: The MCMC Hammer}}},\ \bibinfo {howpublished} {Astrophysics
  Source Code Library, record ascl:1303.002} (\bibinfo {year}
  {2013}{\natexlab{a}}),\ \Eprint {https://arxiv.org/abs/1303.002}
  {ascl:1303.002} \BibitemShut {NoStop}%
\bibitem [{\citenamefont {{Foreman-Mackey}}\ \emph
  {et~al.}(2013{\natexlab{b}})\citenamefont {{Foreman-Mackey}}, \citenamefont
  {{Hogg}}, \citenamefont {{Lang}},\ and\ \citenamefont
  {{Goodman}}}]{Foreman-Mackey2013b}%
  \BibitemOpen
  \bibfield  {author} {\bibinfo {author} {\bibfnamefont {D.}~\bibnamefont
  {{Foreman-Mackey}}}, \bibinfo {author} {\bibfnamefont {D.~W.}\ \bibnamefont
  {{Hogg}}}, \bibinfo {author} {\bibfnamefont {D.}~\bibnamefont {{Lang}}},\
  and\ \bibinfo {author} {\bibfnamefont {J.}~\bibnamefont {{Goodman}}},\ }\href
  {https://doi.org/10.1086/670067} {\bibfield  {journal} {\bibinfo  {journal}
  {\pasp}\ }\textbf {\bibinfo {volume} {125}},\ \bibinfo {pages} {306}
  (\bibinfo {year} {2013}{\natexlab{b}})},\ \Eprint
  {https://arxiv.org/abs/1202.3665} {arXiv:1202.3665 [astro-ph.IM]}
  \BibitemShut {NoStop}%
\bibitem [{\citenamefont {Tejero-Cantero}\ \emph {et~al.}(2020)\citenamefont
  {Tejero-Cantero}, \citenamefont {Boelts}, \citenamefont {Deistler},
  \citenamefont {Lueckmann}, \citenamefont {Durkan}, \citenamefont
  {Gonçalves}, \citenamefont {Greenberg},\ and\ \citenamefont
  {Macke}}]{tejero-cantero2020sbi}%
  \BibitemOpen
  \bibfield  {author} {\bibinfo {author} {\bibfnamefont {A.}~\bibnamefont
  {Tejero-Cantero}}, \bibinfo {author} {\bibfnamefont {J.}~\bibnamefont
  {Boelts}}, \bibinfo {author} {\bibfnamefont {M.}~\bibnamefont {Deistler}},
  \bibinfo {author} {\bibfnamefont {J.-M.}\ \bibnamefont {Lueckmann}}, \bibinfo
  {author} {\bibfnamefont {C.}~\bibnamefont {Durkan}}, \bibinfo {author}
  {\bibfnamefont {P.~J.}\ \bibnamefont {Gonçalves}}, \bibinfo {author}
  {\bibfnamefont {D.~S.}\ \bibnamefont {Greenberg}},\ and\ \bibinfo {author}
  {\bibfnamefont {J.~H.}\ \bibnamefont {Macke}},\ }\href
  {https://doi.org/10.21105/joss.02505} {\bibfield  {journal} {\bibinfo
  {journal} {Journal of Open Source Software}\ }\textbf {\bibinfo {volume}
  {5}},\ \bibinfo {pages} {2505} (\bibinfo {year} {2020})}\BibitemShut
  {NoStop}%
\bibitem [{\citenamefont {{Heavens}}\ \emph {et~al.}(2000)\citenamefont
  {{Heavens}}, \citenamefont {{Jimenez}},\ and\ \citenamefont
  {{Lahav}}}]{Heavens2000}%
  \BibitemOpen
  \bibfield  {author} {\bibinfo {author} {\bibfnamefont {A.~F.}\ \bibnamefont
  {{Heavens}}}, \bibinfo {author} {\bibfnamefont {R.}~\bibnamefont
  {{Jimenez}}},\ and\ \bibinfo {author} {\bibfnamefont {O.}~\bibnamefont
  {{Lahav}}},\ }\href {https://doi.org/10.1046/j.1365-8711.2000.03692.x}
  {\bibfield  {journal} {\bibinfo  {journal} {\mnras}\ }\textbf {\bibinfo
  {volume} {317}},\ \bibinfo {pages} {965} (\bibinfo {year} {2000})},\ \Eprint
  {https://arxiv.org/abs/astro-ph/9911102} {arXiv:astro-ph/9911102 [astro-ph]}
  \BibitemShut {NoStop}%
\bibitem [{\citenamefont {{Alsing}}\ and\ \citenamefont
  {{Wandelt}}(2019)}]{Alsing2019}%
  \BibitemOpen
  \bibfield  {author} {\bibinfo {author} {\bibfnamefont {J.}~\bibnamefont
  {{Alsing}}}\ and\ \bibinfo {author} {\bibfnamefont {B.}~\bibnamefont
  {{Wandelt}}},\ }\href {https://doi.org/10.1093/mnras/stz1900} {\bibfield
  {journal} {\bibinfo  {journal} {\mnras}\ }\textbf {\bibinfo {volume} {488}},\
  \bibinfo {pages} {5093} (\bibinfo {year} {2019})},\ \Eprint
  {https://arxiv.org/abs/1903.01473} {arXiv:1903.01473 [astro-ph.CO]}
  \BibitemShut {NoStop}%
\bibitem [{\citenamefont {{d'Amico}}\ \emph {et~al.}(2020)\citenamefont
  {{d'Amico}}, \citenamefont {{Gleyzes}}, \citenamefont {{Kokron}},
  \citenamefont {{Markovic}}, \citenamefont {{Senatore}}, \citenamefont
  {{Zhang}}, \citenamefont {{Beutler}},\ and\ \citenamefont
  {{Gil-Mar{\'\i}n}}}]{dAmico2020}%
  \BibitemOpen
  \bibfield  {author} {\bibinfo {author} {\bibfnamefont {G.}~\bibnamefont
  {{d'Amico}}}, \bibinfo {author} {\bibfnamefont {J.}~\bibnamefont
  {{Gleyzes}}}, \bibinfo {author} {\bibfnamefont {N.}~\bibnamefont {{Kokron}}},
  \bibinfo {author} {\bibfnamefont {K.}~\bibnamefont {{Markovic}}}, \bibinfo
  {author} {\bibfnamefont {L.}~\bibnamefont {{Senatore}}}, \bibinfo {author}
  {\bibfnamefont {P.}~\bibnamefont {{Zhang}}}, \bibinfo {author} {\bibfnamefont
  {F.}~\bibnamefont {{Beutler}}},\ and\ \bibinfo {author} {\bibfnamefont
  {H.}~\bibnamefont {{Gil-Mar{\'\i}n}}},\ }\href
  {https://doi.org/10.1088/1475-7516/2020/05/005} {\bibfield  {journal}
  {\bibinfo  {journal} {\jcap}\ }\textbf {\bibinfo {volume} {2020}},\ \bibinfo
  {eid} {005} (\bibinfo {year} {2020})},\ \Eprint
  {https://arxiv.org/abs/1909.05271} {arXiv:1909.05271 [astro-ph.CO]}
  \BibitemShut {NoStop}%
\bibitem [{\citenamefont {{Chen}}\ \emph {et~al.}(2022)\citenamefont {{Chen}},
  \citenamefont {{Vlah}},\ and\ \citenamefont {{White}}}]{Chen2022}%
  \BibitemOpen
  \bibfield  {author} {\bibinfo {author} {\bibfnamefont {S.-F.}\ \bibnamefont
  {{Chen}}}, \bibinfo {author} {\bibfnamefont {Z.}~\bibnamefont {{Vlah}}},\
  and\ \bibinfo {author} {\bibfnamefont {M.}~\bibnamefont {{White}}},\ }\href
  {https://doi.org/10.1088/1475-7516/2022/02/008} {\bibfield  {journal}
  {\bibinfo  {journal} {\jcap}\ }\textbf {\bibinfo {volume} {2022}},\ \bibinfo
  {eid} {008} (\bibinfo {year} {2022})},\ \Eprint
  {https://arxiv.org/abs/2110.05530} {arXiv:2110.05530 [astro-ph.CO]}
  \BibitemShut {NoStop}%
\bibitem [{\citenamefont {{Philcox}}\ \emph {et~al.}(2020)\citenamefont
  {{Philcox}}, \citenamefont {{Ivanov}}, \citenamefont {{Simonovi{\'c}}},\ and\
  \citenamefont {{Zaldarriaga}}}]{Philcox2020}%
  \BibitemOpen
  \bibfield  {author} {\bibinfo {author} {\bibfnamefont {O.~H.~E.}\
  \bibnamefont {{Philcox}}}, \bibinfo {author} {\bibfnamefont {M.~M.}\
  \bibnamefont {{Ivanov}}}, \bibinfo {author} {\bibfnamefont {M.}~\bibnamefont
  {{Simonovi{\'c}}}},\ and\ \bibinfo {author} {\bibfnamefont {M.}~\bibnamefont
  {{Zaldarriaga}}},\ }\href {https://doi.org/10.1088/1475-7516/2020/05/032}
  {\bibfield  {journal} {\bibinfo  {journal} {\jcap}\ }\textbf {\bibinfo
  {volume} {2020}},\ \bibinfo {eid} {032} (\bibinfo {year} {2020})},\ \Eprint
  {https://arxiv.org/abs/2002.04035} {arXiv:2002.04035 [astro-ph.CO]}
  \BibitemShut {NoStop}%
\bibitem [{\citenamefont {{Philcox}}(2021)}]{Philcox2021}%
  \BibitemOpen
  \bibfield  {author} {\bibinfo {author} {\bibfnamefont {O.~H.~E.}\
  \bibnamefont {{Philcox}}},\ }\href
  {https://doi.org/10.1103/PhysRevD.103.103504} {\bibfield  {journal} {\bibinfo
   {journal} {\prd}\ }\textbf {\bibinfo {volume} {103}},\ \bibinfo {eid}
  {103504} (\bibinfo {year} {2021})},\ \Eprint
  {https://arxiv.org/abs/2012.09389} {arXiv:2012.09389 [astro-ph.CO]}
  \BibitemShut {NoStop}%
\bibitem [{\citenamefont {{Chudaykin}}\ \emph {et~al.}(2020)\citenamefont
  {{Chudaykin}}, \citenamefont {{Ivanov}}, \citenamefont {{Philcox}},\ and\
  \citenamefont {{Simonovi{\'c}}}}]{Chudaykin2020}%
  \BibitemOpen
  \bibfield  {author} {\bibinfo {author} {\bibfnamefont {A.}~\bibnamefont
  {{Chudaykin}}}, \bibinfo {author} {\bibfnamefont {M.~M.}\ \bibnamefont
  {{Ivanov}}}, \bibinfo {author} {\bibfnamefont {O.~H.~E.}\ \bibnamefont
  {{Philcox}}},\ and\ \bibinfo {author} {\bibfnamefont {M.}~\bibnamefont
  {{Simonovi{\'c}}}},\ }\href {https://doi.org/10.1103/PhysRevD.102.063533}
  {\bibfield  {journal} {\bibinfo  {journal} {\prd}\ }\textbf {\bibinfo
  {volume} {102}},\ \bibinfo {eid} {063533} (\bibinfo {year} {2020})},\ \Eprint
  {https://arxiv.org/abs/2004.10607} {arXiv:2004.10607 [astro-ph.CO]}
  \BibitemShut {NoStop}%
\bibitem [{\citenamefont {{Hermans}}\ \emph {et~al.}(2021)\citenamefont
  {{Hermans}}, \citenamefont {{Delaunoy}}, \citenamefont {{Rozet}},
  \citenamefont {{Wehenkel}}, \citenamefont {{Begy}},\ and\ \citenamefont
  {{Louppe}}}]{Hermans2021}%
  \BibitemOpen
  \bibfield  {author} {\bibinfo {author} {\bibfnamefont {J.}~\bibnamefont
  {{Hermans}}}, \bibinfo {author} {\bibfnamefont {A.}~\bibnamefont
  {{Delaunoy}}}, \bibinfo {author} {\bibfnamefont {F.}~\bibnamefont {{Rozet}}},
  \bibinfo {author} {\bibfnamefont {A.}~\bibnamefont {{Wehenkel}}}, \bibinfo
  {author} {\bibfnamefont {V.}~\bibnamefont {{Begy}}},\ and\ \bibinfo {author}
  {\bibfnamefont {G.}~\bibnamefont {{Louppe}}},\ }\href
  {https://doi.org/10.48550/arXiv.2110.06581} {\bibfield  {journal} {\bibinfo
  {journal} {arXiv e-prints}\ ,\ \bibinfo {eid} {arXiv:2110.06581}} (\bibinfo
  {year} {2021})},\ \Eprint {https://arxiv.org/abs/2110.06581}
  {arXiv:2110.06581 [stat.ML]} \BibitemShut {NoStop}%
\bibitem [{\citenamefont {{Anscombe}}(1948)}]{Anscombe1948}%
  \BibitemOpen
  \bibfield  {author} {\bibinfo {author} {\bibfnamefont {F.~J.}\ \bibnamefont
  {{Anscombe}}},\ }\href {https://doi.org/10.1093/biomet/35.3-4.246} {\bibfield
   {journal} {\bibinfo  {journal} {Biometrika}\ }\textbf {\bibinfo {volume}
  {35}},\ \bibinfo {pages} {246} (\bibinfo {year} {1948})},\ \Eprint
  {https://arxiv.org/abs/https://academic.oup.com/biomet/article-pdf/35/3-4/246/785684/35-3-4-246.pdf}
  {https://academic.oup.com/biomet/article-pdf/35/3-4/246/785684/35-3-4-246.pdf}
  \BibitemShut {NoStop}%
\bibitem [{\citenamefont {{Ivezi{\'c}}}\ \emph {et~al.}(2019)\citenamefont
  {{Ivezi{\'c}}}, \citenamefont {{Kahn}}, \citenamefont {{Tyson}},
  \citenamefont {{Abel}}, \citenamefont {{Acosta}} \emph
  {et~al.}}]{Ivezic2019}%
  \BibitemOpen
  \bibfield  {author} {\bibinfo {author} {\bibfnamefont {{\v{Z}}.}~\bibnamefont
  {{Ivezi{\'c}}}}, \bibinfo {author} {\bibfnamefont {S.~M.}\ \bibnamefont
  {{Kahn}}}, \bibinfo {author} {\bibfnamefont {J.~A.}\ \bibnamefont {{Tyson}}},
  \bibinfo {author} {\bibfnamefont {B.}~\bibnamefont {{Abel}}}, \bibinfo
  {author} {\bibfnamefont {E.}~\bibnamefont {{Acosta}}}, \emph {et~al.},\
  }\href {https://doi.org/10.3847/1538-4357/ab042c} {\bibfield  {journal}
  {\bibinfo  {journal} {\apj}\ }\textbf {\bibinfo {volume} {873}},\ \bibinfo
  {eid} {111} (\bibinfo {year} {2019})},\ \Eprint
  {https://arxiv.org/abs/0805.2366} {arXiv:0805.2366 [astro-ph]} \BibitemShut
  {NoStop}%
\bibitem [{\citenamefont {{Dey}}\ \emph {et~al.}(2019)\citenamefont {{Dey}},
  \citenamefont {{Schlegel}}, \citenamefont {{Lang}} \emph {et~al.}}]{Dey2019}%
  \BibitemOpen
  \bibfield  {author} {\bibinfo {author} {\bibfnamefont {A.}~\bibnamefont
  {{Dey}}}, \bibinfo {author} {\bibfnamefont {D.~J.}\ \bibnamefont
  {{Schlegel}}}, \bibinfo {author} {\bibfnamefont {D.}~\bibnamefont {{Lang}}},
  \emph {et~al.},\ }\href {https://doi.org/10.3847/1538-3881/ab089d} {\bibfield
   {journal} {\bibinfo  {journal} {\aj}\ }\textbf {\bibinfo {volume} {157}},\
  \bibinfo {eid} {168} (\bibinfo {year} {2019})},\ \Eprint
  {https://arxiv.org/abs/1804.08657} {arXiv:1804.08657 [astro-ph.IM]}
  \BibitemShut {NoStop}%
\bibitem [{\citenamefont {{Laureijs}}\ \emph {et~al.}(2011)\citenamefont
  {{Laureijs}}, \citenamefont {{Amiaux}}, \citenamefont {{Arduini}},
  \citenamefont {{Augu{\`e}res}}, \citenamefont {{Brinchmann}}, \citenamefont
  {{Cole}} \emph {et~al.}}]{Laureijs2011}%
  \BibitemOpen
  \bibfield  {author} {\bibinfo {author} {\bibfnamefont {R.}~\bibnamefont
  {{Laureijs}}}, \bibinfo {author} {\bibfnamefont {J.}~\bibnamefont
  {{Amiaux}}}, \bibinfo {author} {\bibfnamefont {S.}~\bibnamefont {{Arduini}}},
  \bibinfo {author} {\bibfnamefont {J.~L.}\ \bibnamefont {{Augu{\`e}res}}},
  \bibinfo {author} {\bibfnamefont {J.}~\bibnamefont {{Brinchmann}}}, \bibinfo
  {author} {\bibfnamefont {R.}~\bibnamefont {{Cole}}}, \emph {et~al.},\ }\href
  {https://doi.org/10.48550/arXiv.1110.3193} {\bibfield  {journal} {\bibinfo
  {journal} {arXiv e-prints}\ ,\ \bibinfo {eid} {arXiv:1110.3193}} (\bibinfo
  {year} {2011})},\ \Eprint {https://arxiv.org/abs/1110.3193} {arXiv:1110.3193
  [astro-ph.CO]} \BibitemShut {NoStop}%
\bibitem [{\citenamefont {{Dor{\'e}}}\ \emph {et~al.}(2014)\citenamefont
  {{Dor{\'e}}}, \citenamefont {{Bock}} \emph {et~al.}}]{Dore2014}%
  \BibitemOpen
  \bibfield  {author} {\bibinfo {author} {\bibfnamefont {O.}~\bibnamefont
  {{Dor{\'e}}}}, \bibinfo {author} {\bibfnamefont {J.}~\bibnamefont {{Bock}}},
  \emph {et~al.},\ }\href {https://doi.org/10.48550/arXiv.1412.4872} {\bibfield
   {journal} {\bibinfo  {journal} {arXiv e-prints}\ ,\ \bibinfo {eid}
  {arXiv:1412.4872}} (\bibinfo {year} {2014})},\ \Eprint
  {https://arxiv.org/abs/1412.4872} {arXiv:1412.4872 [astro-ph.CO]}
  \BibitemShut {NoStop}%
\bibitem [{\citenamefont {{Takada}}\ \emph {et~al.}(2014)\citenamefont
  {{Takada}}, \citenamefont {{Ellis}}, \citenamefont {{Chiba}}, \citenamefont
  {{Greene}}, \citenamefont {{Aihara}} \emph {et~al.}}]{Takada2014}%
  \BibitemOpen
  \bibfield  {author} {\bibinfo {author} {\bibfnamefont {M.}~\bibnamefont
  {{Takada}}}, \bibinfo {author} {\bibfnamefont {R.~S.}\ \bibnamefont
  {{Ellis}}}, \bibinfo {author} {\bibfnamefont {M.}~\bibnamefont {{Chiba}}},
  \bibinfo {author} {\bibfnamefont {J.~E.}\ \bibnamefont {{Greene}}}, \bibinfo
  {author} {\bibfnamefont {H.}~\bibnamefont {{Aihara}}}, \emph {et~al.},\
  }\href {https://doi.org/10.1093/pasj/pst019} {\bibfield  {journal} {\bibinfo
  {journal} {\pasj}\ }\textbf {\bibinfo {volume} {66}},\ \bibinfo {eid} {R1}
  (\bibinfo {year} {2014})},\ \Eprint {https://arxiv.org/abs/1206.0737}
  {arXiv:1206.0737 [astro-ph.CO]} \BibitemShut {NoStop}%
\bibitem [{\citenamefont {{Spergel}}\ \emph {et~al.}(2015)\citenamefont
  {{Spergel}}, \citenamefont {{Gehrels}}, \citenamefont {{Baltay}},
  \citenamefont {{Bennett}}, \citenamefont {{Breckinridge}} \emph
  {et~al.}}]{Spergel2015}%
  \BibitemOpen
  \bibfield  {author} {\bibinfo {author} {\bibfnamefont {D.}~\bibnamefont
  {{Spergel}}}, \bibinfo {author} {\bibfnamefont {N.}~\bibnamefont
  {{Gehrels}}}, \bibinfo {author} {\bibfnamefont {C.}~\bibnamefont {{Baltay}}},
  \bibinfo {author} {\bibfnamefont {D.}~\bibnamefont {{Bennett}}}, \bibinfo
  {author} {\bibfnamefont {J.}~\bibnamefont {{Breckinridge}}}, \emph {et~al.},\
  }\href {https://doi.org/10.48550/arXiv.1503.03757} {\bibfield  {journal}
  {\bibinfo  {journal} {arXiv e-prints}\ ,\ \bibinfo {eid} {arXiv:1503.03757}}
  (\bibinfo {year} {2015})},\ \Eprint {https://arxiv.org/abs/1503.03757}
  {arXiv:1503.03757 [astro-ph.IM]} \BibitemShut {NoStop}%
\bibitem [{\citenamefont {{Navarro}}\ \emph {et~al.}(1997)\citenamefont
  {{Navarro}}, \citenamefont {{Frenk}},\ and\ \citenamefont
  {{White}}}]{Navarro1997}%
  \BibitemOpen
  \bibfield  {author} {\bibinfo {author} {\bibfnamefont {J.~F.}\ \bibnamefont
  {{Navarro}}}, \bibinfo {author} {\bibfnamefont {C.~S.}\ \bibnamefont
  {{Frenk}}},\ and\ \bibinfo {author} {\bibfnamefont {S.~D.~M.}\ \bibnamefont
  {{White}}},\ }\href {https://doi.org/10.1086/304888} {\bibfield  {journal}
  {\bibinfo  {journal} {\apj}\ }\textbf {\bibinfo {volume} {490}},\ \bibinfo
  {pages} {493} (\bibinfo {year} {1997})},\ \Eprint
  {https://arxiv.org/abs/astro-ph/9611107} {arXiv:astro-ph/9611107 [astro-ph]}
  \BibitemShut {NoStop}%
\bibitem [{\citenamefont {{Duffy}}\ \emph {et~al.}(2008)\citenamefont
  {{Duffy}}, \citenamefont {{Schaye}}, \citenamefont {{Kay}},\ and\
  \citenamefont {{Dalla Vecchia}}}]{Duffy2008}%
  \BibitemOpen
  \bibfield  {author} {\bibinfo {author} {\bibfnamefont {A.~R.}\ \bibnamefont
  {{Duffy}}}, \bibinfo {author} {\bibfnamefont {J.}~\bibnamefont {{Schaye}}},
  \bibinfo {author} {\bibfnamefont {S.~T.}\ \bibnamefont {{Kay}}},\ and\
  \bibinfo {author} {\bibfnamefont {C.}~\bibnamefont {{Dalla Vecchia}}},\
  }\href {https://doi.org/10.1111/j.1745-3933.2008.00537.x} {\bibfield
  {journal} {\bibinfo  {journal} {\mnras}\ }\textbf {\bibinfo {volume} {390}},\
  \bibinfo {pages} {L64} (\bibinfo {year} {2008})},\ \Eprint
  {https://arxiv.org/abs/0804.2486} {arXiv:0804.2486 [astro-ph]} \BibitemShut
  {NoStop}%
\bibitem [{\citenamefont {{Robotham}}\ and\ \citenamefont
  {{Howlett}}(2018)}]{Robotham2018}%
  \BibitemOpen
  \bibfield  {author} {\bibinfo {author} {\bibfnamefont {A.~S.~G.}\
  \bibnamefont {{Robotham}}}\ and\ \bibinfo {author} {\bibfnamefont
  {C.}~\bibnamefont {{Howlett}}},\ }\href
  {https://doi.org/10.3847/2515-5172/aacc70} {\bibfield  {journal} {\bibinfo
  {journal} {Research Notes of the American Astronomical Society}\ }\textbf
  {\bibinfo {volume} {2}},\ \bibinfo {eid} {55} (\bibinfo {year} {2018})},\
  \Eprint {https://arxiv.org/abs/1805.09550} {arXiv:1805.09550 [astro-ph.CO]}
  \BibitemShut {NoStop}%
\bibitem [{\citenamefont {{Hearin}}\ \emph
  {et~al.}(2016{\natexlab{a}})\citenamefont {{Hearin}}, \citenamefont
  {{Zentner}}, \citenamefont {{van den Bosch}}, \citenamefont {{Campbell}},\
  and\ \citenamefont {{Tollerud}}}]{Hearin2016a}%
  \BibitemOpen
  \bibfield  {author} {\bibinfo {author} {\bibfnamefont {A.~P.}\ \bibnamefont
  {{Hearin}}}, \bibinfo {author} {\bibfnamefont {A.~R.}\ \bibnamefont
  {{Zentner}}}, \bibinfo {author} {\bibfnamefont {F.~C.}\ \bibnamefont {{van
  den Bosch}}}, \bibinfo {author} {\bibfnamefont {D.}~\bibnamefont
  {{Campbell}}},\ and\ \bibinfo {author} {\bibfnamefont {E.}~\bibnamefont
  {{Tollerud}}},\ }\href {https://doi.org/10.1093/mnras/stw840} {\bibfield
  {journal} {\bibinfo  {journal} {\mnras}\ }\textbf {\bibinfo {volume} {460}},\
  \bibinfo {pages} {2552} (\bibinfo {year} {2016}{\natexlab{a}})},\ \Eprint
  {https://arxiv.org/abs/1512.03050} {arXiv:1512.03050 [astro-ph.CO]}
  \BibitemShut {NoStop}%
\bibitem [{\citenamefont {{Hearin}}\ \emph
  {et~al.}(2016{\natexlab{b}})\citenamefont {{Hearin}}, \citenamefont
  {{Tollerud}}, \citenamefont {{Robitaille}}, \citenamefont {{Droettboom}},
  \citenamefont {{Zentner}} \emph {et~al.}}]{Hearin2016b}%
  \BibitemOpen
  \bibfield  {author} {\bibinfo {author} {\bibfnamefont {A.}~\bibnamefont
  {{Hearin}}}, \bibinfo {author} {\bibfnamefont {E.}~\bibnamefont
  {{Tollerud}}}, \bibinfo {author} {\bibfnamefont {T.}~\bibnamefont
  {{Robitaille}}}, \bibinfo {author} {\bibfnamefont {M.}~\bibnamefont
  {{Droettboom}}}, \bibinfo {author} {\bibfnamefont {A.}~\bibnamefont
  {{Zentner}}}, \emph {et~al.},\ }\href@noop {} {\bibinfo {title} {{Halotools:
  Galaxy-Halo connection models}}},\ \bibinfo {howpublished} {Astrophysics
  Source Code Library, record ascl:1604.005} (\bibinfo {year}
  {2016}{\natexlab{b}}),\ \Eprint {https://arxiv.org/abs/1604.005}
  {ascl:1604.005} \BibitemShut {NoStop}%
\end{thebibliography}
\end{document}